\documentclass[%
 reprint,prd,
preprintnumbers,
 amsmath,amssymb,nofootinbib,
 aps, groupedaddress, superscriptaddress,
]{revtex4-1}

\pdfoutput=1

\usepackage[dvipsnames]{xcolor}
\usepackage{graphicx}
\usepackage{dcolumn}
\usepackage{bm}
\usepackage{hyperref}
\usepackage{multirow}
\usepackage{soul}
\usepackage{color}
\usepackage{lineno}
\usepackage[normalem]{ulem}
\usepackage{fontawesome}

\definecolor{myblue}{rgb}{0.152941176,0.549019608,0.670588235}
\definecolor{newred}{cmyk}{0,1,1,0.2}
\definecolor{newblue}{cmyk}{1,1,0,0.1}

\definecolor{linkcolor}{rgb}{0.6,0,0}
\definecolor{citecolor}{rgb}{0,0,0.75}
\definecolor{urlcolor}{rgb}{0.12,0.46,0.7}

\hypersetup{   
    pdftoolbar=true,        
    pdfmenubar=true,        
    pdffitwindow=false,
    pdfkeywords={BBN} {lepton asymmetry} {Neff} {Early Universe},
    pdfnewwindow=true,
    urlcolor=urlcolor, 
    linkcolor=linkcolor,
    citecolor=citecolor,
    colorlinks=true,
    linktocpage=true
}

\def\equationautorefname~#1\null{Eq.\,(#1)\null}
\newcommand{\appendixref}[1]{\hyperref[#1]{appendix~\ref{#1}}}

\newcommand{\github}{\href{https://github.com/vallima/PRyMordial}{\faGithub}}

\begin{document}

\title{The QCD Axion: Some Like It Hot}

\author{Federico Bianchini}
\email{fbianc@stanford.edu}
\affiliation{
SLAC National Accelerator Laboratory, 2575 Sand Hill Road, Menlo Park, CA, 94025, USA}
\affiliation{Kavli Institute for Particle Astrophysics and Cosmology, Stanford University, 452 Lomita Mall, Stanford, CA, 94305, USA}
\affiliation{
Department of Physics, Stanford University, 382 Via Pueblo Mall, Stanford, CA, 94305, USA}

\author{Giovanni Grilli di Cortona}
\email{grillidc@lnf.infn.it}
\affiliation{
 INFN Sezione di Roma, Piazzale Aldo Moro 2, I-00185 Rome, Italy
}
\author{Mauro Valli}
\email{vallima@roma1.infn.it}
\affiliation{
 INFN Sezione di Roma, Piazzale Aldo Moro 2, I-00185 Rome, Italy
}

\date{\today}

\begin{abstract}
We compare the QCD axion phase-space distribution from unitarized next-to-leading order chiral perturbation theory with the one extracted from pion-scattering data. We derive a robust bound by confronting momentum-dependent Boltzmann equations against up-to-date observations of the Cosmic Microwave Background, of the Baryonic Acoustic Oscillations and of primordial abundances. These datasets imply $m_{a} \leq \, 0.16 $~eV for the 95\% credible interval, i.e. $\sim$30\% stronger bound than what previously found. We present forecasts using dedicated likelihoods for future cosmological surveys and the sphaleron rate from unquenched lattice QCD.
\end{abstract}

\maketitle

\textbf{Introduction.}~The Peccei-Quinn mechanism~\cite{Peccei:1977hh, Peccei:1977ur} predicts the existence of a pseudo-Goldstone boson -- the axion $a$~\cite{Weinberg:1977ma, Wilczek:1977pj} -- which elegantly solves the strong CP problem~\cite{Baker:2006ts,Pendlebury:2015lrz,Abel:2020pzs} by dynamically relaxing QCD $\theta$ vacua to the minimum-energy state~\cite{Vafa:1984xg,Dvali:2022fdv}.
The axion may serve as cold dark matter, produced via the misalignment mechanism~\cite{Preskill:1982cy,Abbott:1982af,Dine:1982ah,Davis:1986xc}. Furthermore, the minimal coupling of the axion to the gluon field strength $G$ opens up the possibility of generating an abundance of new light species from the thermal bath of the Early Universe~\cite{Turner:1986tb, Kolb:1990vq,Berezhiani:1992rk, Chang:1993gm,Baumann:2016wac,DEramo:2022nvb}.

In fact, a reheating temperature well above the QCD crossover $T_c\sim 150$ MeV~\cite{Aoki:2006br,Borsanyi:2010bp,Bazavov:2011nk} would imply an axion in thermal equilibrium with 
the Standard Model (SM), satisfying a production rate $\Gamma_{a} \sim T^3 / f_{a}^2 
\gtrsim H$, with $f_a=5.7\times 10^6\, \mathrm{GeV}(\mathrm{eV}/m_a)$~\cite{diCortona:2015ldu,Gorghetto:2018ocs} being the axion decay constant. 
As a consequence, a bound on the mass of the QCD axion can be simply extracted from the upper limit on the extra radiation $\Delta N_{\rm eff} \equiv  \rho_{a}/\rho_{\nu}$ allowed in the Early Universe by the time of recombination~\cite{Aghanim:2018eyx}:
\begin{equation}
\label{eq:DeltaNeff}
\Delta N_{\rm eff}\big|_{T_{\rm CMB}} \simeq \frac{4}{7}  \left(\frac{11}{4}\frac{g_{\star s}(T_{\rm CMB})}{g_{\star s}(T_{\rm dec})}\right)^{4/3} \lesssim 0.5 \,~\footnote{The constraint corresponds to the 95\% credible interval from the \textit{Planck} Collaboration with no prior on helium-4.} \, ,
\end{equation}
where $g_{\star s}$ are the entropic degrees of freedom of the SM thermal bath (see, e.g., Appendix~D of \cite{DEramo:2021lgb}),
$T_{\rm CMB}~\simeq~0.26$~eV, and the decoupling temperature of the axion is $T_{\rm dec}~\sim~\mathcal{O}({\rm MeV}) \times ({\rm eV}/m_{a})^2$.
\begin{figure}[!t!]
  \centering
  \includegraphics[width=0.9\columnwidth]{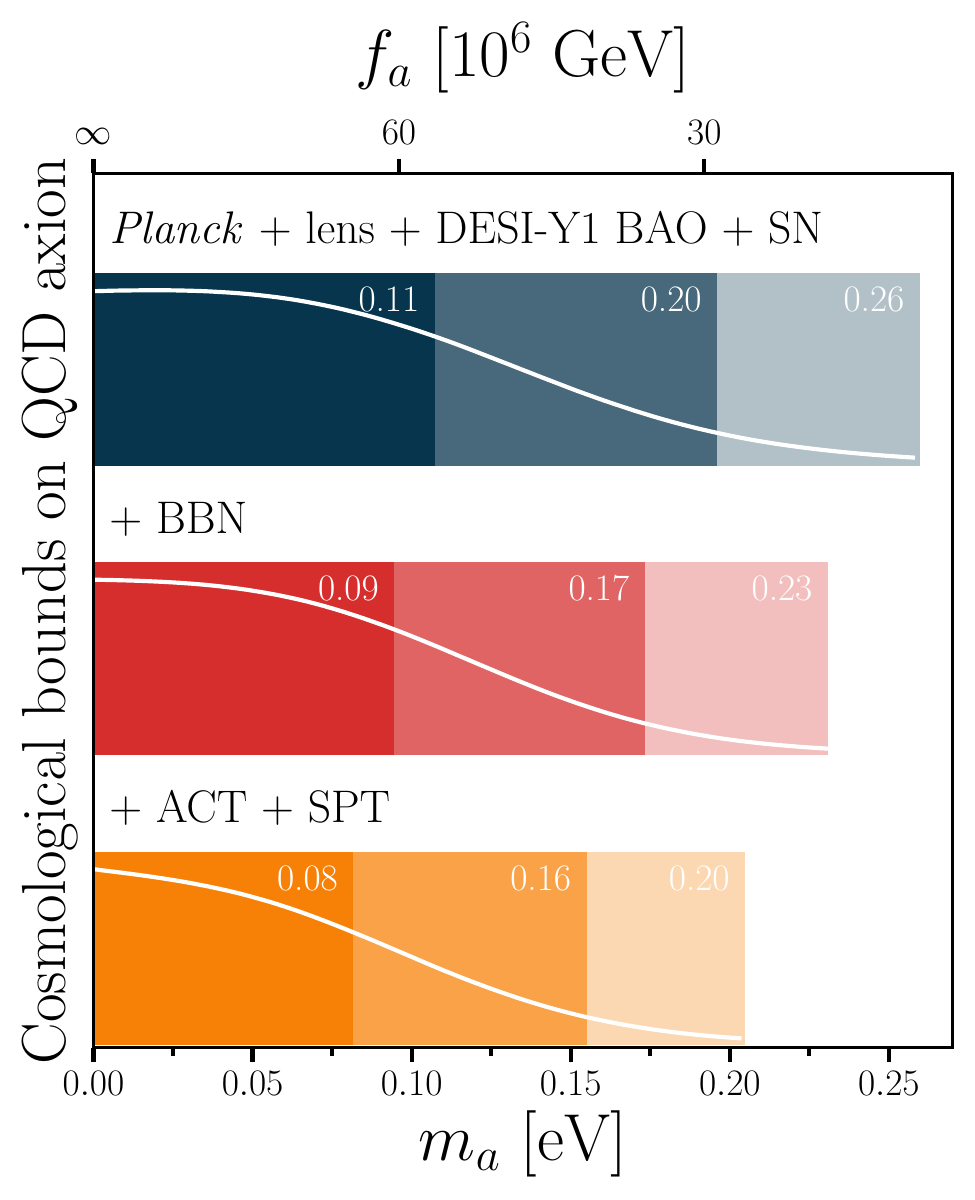}
  \caption{QCD axion mass posterior distribution from present cosmological data; 68\%, 95\%, 99.7\% credible intervals are shown from darker to lighter color shade.}
  \label{fig:fig1}
\end{figure}
According to Eq.~\eqref{eq:DeltaNeff}, with $g_{\star s}$ monotonically decreasing with the temperature, a lower $T_{\mathrm{dec}}$ (or larger $m_a$) results in an increase in $\Delta N_{\mathrm{eff}}$. Consequently, a QCD axion which decouples at high temperatures from the SM bath  -- say at $T_{\rm dec} \gtrsim T_{\rm EW} \sim \mathcal{O}(100)$~GeV -- under quite general assumptions would leave an indelible cosmological imprint at the level of $\Delta N_{\rm eff} \sim \mathcal{O}(0.01)$~\cite{Baumann:2016wac}.
This could potentially be within the reach of future surveys~\cite{Abazajian:2019eic,CMB-HD:2022bsz}. Most notably, an axion produced below $T_c$ is under the spotlight of current cosmological observations{, that rule out QCD axions with $f_a \lesssim \mathcal{O}(10^7)$ GeV. This result must be confronted with the current astrophysical bounds. In particular, the QCD axion can be constrained by the observed neutrino signal from SN1987A, excluding $f_a \lesssim 2 \times 10^8$ GeV~\cite{Carenza:2019pxu}. Furthermore, the observation of the cooling of neutron stars provides information on the axion-nucleon coupling. The rapid cooling of the neutron star in Cassiopeia A excludes $f_a \lesssim 3\times 10^7$ GeV~\cite{Leinson:2021ety}. However, astrophysical constraints are reasonably expected to be more susceptible to theoretical and observational systematics than cosmological ones. This is because astrophysical bounds are often model-dependent and largely rely on the assumption that uncertainties in stellar physics -- an active area of research -- are well understood~\cite{Raffelt:2006cw,MillerBertolami:2014rka,Ayala:2014pea, Chang:2018rso, Bar:2019ifz,Carenza:2020cis, Buschmann:2021juv,Dolan:2022kul,Lella:2023bfb}.} In this work we conduct an in-depth study of the \textit{hot QCD axion}, namely on its model-independent impact on the cosmological history of the Universe.

Many cosmological analyses focusing on the QCD axion~\cite{Hannestad:2005df,Melchiorri:2007cd,Hannestad:2007dd,Hannestad:2008js,Hannestad:2010yi,Archidiacono:2013cha,Giusarma:2014zza,DiValentino:2015zta,DiValentino:2015wba,Archidiacono:2015mda,Ferreira:2020bpb,Arias-Aragon:2020shv,DEramo:2021psx,DiValentino:2022edq} primarily rely on scattering amplitudes evaluated at leading order in a chiral expansion of the axion-pion effective Lagrangian~\cite{Georgi:1986df}, and a Bose-Einstein distribution for the axion in the SM bath assuming fully-established thermal equilibrium~\cite{Masso:2002np}. 
Interestingly, the regime of validity of these two ingredients have been overlooked until recently. 
In Refs.~\cite{DiLuzio:2021vjd,DiLuzio:2022gsc}, an explicit next-to-leading-order (NLO) computation of the axion-pion rate in chiral perturbation theory ($\chi$PT) invalidated previous results, raising questions about the robustness of the bound on the QCD axion from cosmology.
Furthermore, Ref.~\cite{Notari:2022zxo} developed a phenomenological approach to constrain the QCD axion mass without relying on a thermal distribution for its phase space $\mathcal{F}_{a}(E,t)$,  where $E = \sqrt{|\mathbf{k}|^2+m_{a}^2}$. 
In particular, in a Universe that reheats at a temperature around (or above) the QCD crossover with no initial axion abundance~\footnote{This is the proper initial condition to claim a conservative bound on $m_{a}$ in light of an experimental sensitivity to $T_{\rm dec} \lesssim T_{c}$.}, the assumption $\mathcal{F}_{a} = \mathcal{F}_{a}^{\rm eq} = 1/(\exp(E/T)-1)$ is indeed not justified. 
Departures from $\mathcal{F}_{a}^{\rm eq}$ should be investigated by solving the set of  momentum-dependent Boltzmann equations with a collision term characterized by the imaginary part of the axion self-energy~\cite{Graf:2010tv,Salvio:2013iaa,Notari:2022zxo}:
\begin{eqnarray}
\label{eq:axionpsd}
\frac{\partial\mathcal{F}_{a}}{\partial t}  & - & H \, |\mathbf{k}|  \frac{\partial \mathcal{F}_{a}}{\partial |\mathbf{k}|} = \, \Gamma_{a} \, (\mathcal{F}^{\rm \, eq}_{a}-\mathcal{F}_{a}) \ , \\
\Pi_{a}^{\rm R} & = & i \, \int d^4 x \,  e^{i \, x k} \, \langle \, [ \, \mathcal{Q}(x), \, \mathcal{Q}(0)] \Theta(t) \,  \rangle \ , \nonumber
\end{eqnarray}
where $\Gamma_{a} \equiv \textrm{Im}(\Pi_{a}^{\rm R})/(E f_{a}^2)$, $\mathcal{Q} \equiv \alpha_{s}/(8\pi) \, G \tilde{G}$, $ \langle \dots \rangle$ implies the trace over the thermal density matrix, and $\Theta(t)$ is the Heaviside function~\cite{Laine:2016hma}.

In this work, we follow the approach of Ref.~\cite{Notari:2022zxo} and improve upon their result by adopting the recently released one-year Dark Energy Spectroscopic Instrument (DESI-Y1) measurements on baryonic acoustic oscillation (BAO)~\cite{DESI:2024mwx}, by using up-to-date information from the Big Bang Nucleosynthesis (BBN) era, and by including also high-resolution observations of the damping tail of the Cosmic Microwave Background (CMB) power spectra from ground-based surveys \cite{ACT:2020gnv, Balkenhol23}.
Fig.~\ref{fig:fig1} shows how these improvements yield an overall 30\% tighter constraint on the QCD axion mass compared to the result obtained in Ref.~\cite{Notari:2022zxo}.
We support our findings by quantifying the theoretical systematic uncertainties of our computation. 
For this reason, we compare for the first time the non-thermal distribution of hot axions for different masses $m_a$ by solving Eq.~\eqref{eq:axionpsd} with NLO $\chi$PT unitarized rates~\cite{DiLuzio:2022tbb}, with the distribution obtained from the phenomenological approach of~\cite{Notari:2022zxo}. 
We conclude with dedicated forecasts on the QCD axion mass based on recent progress on non-perturbative effects in $\Pi_{a}^{\rm R}$ at $T \gtrsim T_{c}$ from lattice QCD~\cite{Bonanno:2023thi,Bonanno:2023xfv}, and state-of-the-art likelihoods for the sensitivity of next-generation cosmological surveys like the Simons Observatory (SO)~\cite{SimonsObservatory:2018koc}, CMB-S4~\cite{abazajian2019cmbs4,CMB-S4:2022ght}, and DESI with its expected five-year data taking (DESI-Y5)~\cite{DESI:2016fyo}. The forecasts are shown in Fig.~\ref{fig:fig3}.

~

\textbf{Hot axions from pions.}~While the model-building landscape related to the QCD axion is rather vast~\cite{DiLuzio:2020wdo}, in this work we conservatively take the model-independent interaction between $a$ and the QCD topological charge $\mathcal{Q}$ to be the main phenomenological driver. In such a minimal setup, with the extra requirement of a reheating temperature being $\gtrsim T_{c}$,  the axion abundance can be efficiently built up in the thermal bath from $\pi \pi \to \pi a$ scatterings, which dominate the two-point function of Eq.~\eqref{eq:axionpsd} for $T \lesssim T_{c}$. Then, the optical theorem and the chiral Lagrangian allow one to evaluate the thermalization rate $\Gamma_{a}$, a task performed at LO in $\chi$PT already three decades ago~\cite{Chang:1993gm}. 
Such a LO result has been extensively adopted in the community, despite the breakdown of $\chi$PT can be estimated with the appearance of resonant structures around 500~MeV~\cite{Aydemir:2012nz}. 
This becomes problematic when considering that the mean energy of a pion in the SM bath at a temperature where it can still be considered semi-relativistic is three times larger, namely $\langle E \rangle \simeq \rho_{\pi}/n_{\pi} = \pi^{4}T/(30\,\zeta_3) $. 
This implies a center of mass energy $\sqrt{s}$ for $\pi\pi$ scattering outside the regime where $\chi$PT is reliable~\cite{Schenk:1993ru}. 
This reasoning recently inherited further credit from the computation of the axion production rate in Refs.~\cite{DiLuzio:2021vjd,DiLuzio:2022gsc}, that pointed out  $\Gamma^{\rm (NLO)}_{a}/\Gamma^{\rm(LO)}_{a}  \sim  \mathcal{O}(1)$ for the $\pi^\pm\pi^0$ channel already at $T \sim 70$~MeV.
In order to overcome this limitation in the evaluation of $\Gamma_{a}$ at $T \lesssim T_{c}$, two different approaches have been recently explored:
\begin{itemize}
\item[(I)] Ref.~\cite{DiLuzio:2022tbb} applied the Inverse Amplitude Method (IAM) originally developed in~\cite{Truong:1988zp} and recently reviewed in~\cite{Salas-Bernardez:2020hua} to evaluate unitary corrections to axion-pion scattering at NLO in $\chi$PT via the phase shifts induced by pion final-state interactions~\cite{PhysRev.88.1163};
\item[(II)] Ref.~\cite{Notari:2022zxo} took advantage of the symmetries of the chiral Lagrangian to show that the amplitude for $a \pi \to \pi \pi $ can be directly related to the experimentally measured pion-pion scattering up to corrections $\mathcal{O}(m_{\pi}^2 / s) $ at all orders in the chiral expansion.
\end{itemize}

\begin{figure}[!t!]
  \centering
  \includegraphics[width=\columnwidth]{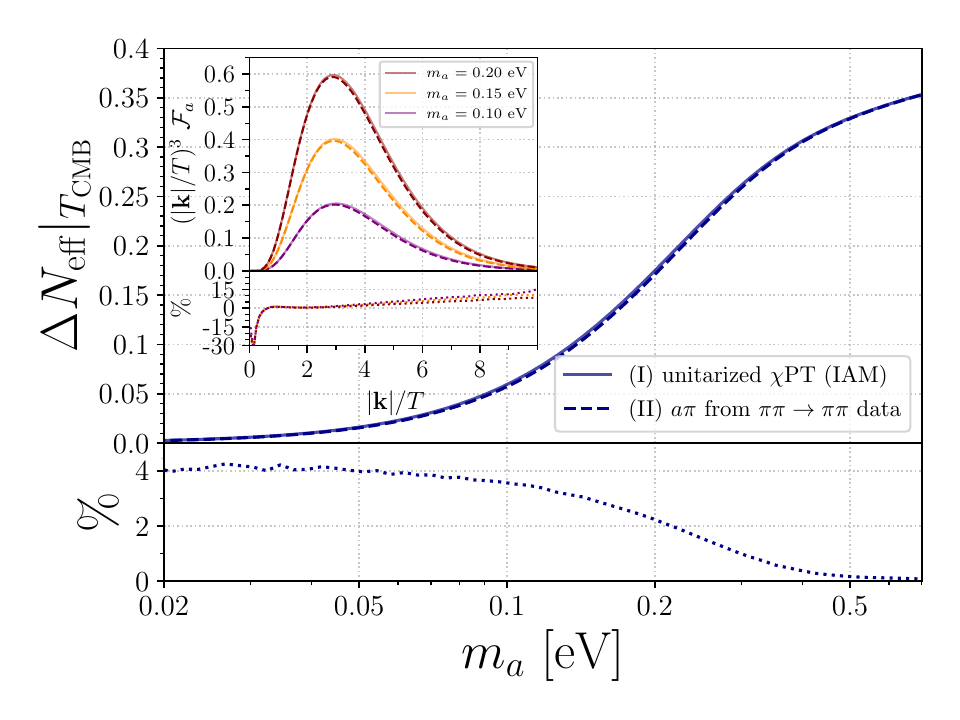}
  \caption{Extra relativistic degrees of freedom at recombination due to the hot QCD axion produced via pion scattering in the two different approaches described in the text. In the insert we compare the two methods also on the axion phase space.}
  \label{fig:fig2}
\end{figure}

\begin{figure}
  \centering
  \includegraphics[width=\columnwidth]{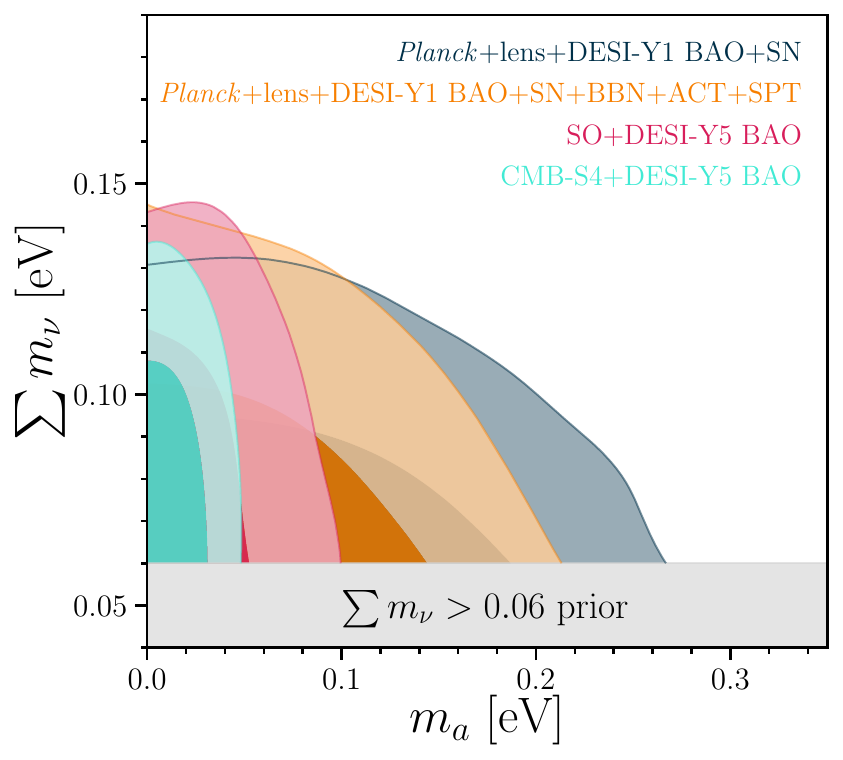}
  \caption{68\% and 95\% credible region for the axion mass and the sum of neutrino masses from current cosmological data (blue and orange contours) and future CMB/LSS surveys (magenta and turquoise contours). See text for details on the forecast.}
  \label{fig:fig3}
\end{figure}

A direct comparison of the two methods through the solution of Eq.~\eqref{eq:axionpsd} has been missing. We fill this gap in Fig.~\ref{fig:fig2}, where we show in the inset the axion phase-space distribution from approach (I) and (II) for three different mass values within the range allowed by current experiments. 
In Appendix~\ref{app:A} we report the details to reproduce our computation of $\mathcal{F}_{a}$. 
The percentage difference between the two phase-space distributions for the same $m_a$ shows a mismatch above 10\% only at the tails, suggesting a negligible impact on the cosmological data analysis. 
This is confirmed by examining Fig.~\ref{fig:fig2}, which shows $\Delta N_{\rm eff}$ as a function of the axion mass. 
For $m_a \gtrsim 0.1$ eV, the difference in terms of energy density is below 4\%.

In Appendix~\ref{app:B} we demonstrate that the effect of spectral distortions~\cite{Ma:1995ey,Lesgourgues:2011rh} in linear cosmological perturbations is minimal. 
This is because an analysis with the axion species described by a Bose-Einstein distribution and with the underlying $\Delta N_{\rm eff}(m_{a})$ of Fig.~\ref{fig:fig2} leads to the same results.
We conclude that as long as a precise evaluation of $\Delta N_{\rm eff}(m_{a})$ is considered, and axions are not treated merely as dark radiation, the cosmological bounds on $m_a$ in Fig.~\ref{fig:fig1} remain unaffected by the choice of method (I) over (II) for the determination of $\Gamma_{a}$.\footnote{Nevertheless, for ease of comparison with previous work, we present cosmological constraints on $m_a$ based on $\mathcal{F}_{a}$ extracted via approach (II).} 

~

\textbf{The BBN-CMB bound.} Similarly to massive neutrinos in cosmology~\cite{Lesgourgues:2006nd}, axions manifest themselves as extra radiation in the Early Universe, altering the expansion rate and affecting the matter-radiation equality and Silk damping scales. Furthermore, they leave an important imprint on the large-scale structure (LSS) when they become non-relativistic. 
For this reason, the QCD axion can be constrained by a set of cosmological data. These include measurements at low- and high-$\ell$ of the CMB temperature and polarization power spectra, exquisitely measured by the \textit{Planck} Collaboration~\cite{Planck:2018vyg}, up-to-date measurements from DESI-Y1~\cite{DESI:2024mwx}, which probe the large-scale structure of the universe and the expansion history by measuring the scale of the BAO peaks with unprecedented accuracy. 
In order to exploit the statistical power of the rich cosmological dataset of Refs.~\cite{Aghanim:2019ame,Alam:2016hwk,Beutler:2011hx,Ross:2014qpa,Alam:2020sor,Scolnic:2017caz,ACT:2023dou,Carron_2022}, we implement the hot QCD axion as a non-cold dark-matter species in the CMB Boltzmann-solver code $\texttt{CLASS}$~\cite{2011arXiv1104.2932L,2011JCAP07034B} using the set of phase-space distributions $\mathcal{F}_{a}$ at the core of Fig.~\ref{fig:fig2}. 
We compare theory predictions obtained in an extended $\Lambda$CDM cosmology with massive neutrinos (satisfying a lower bound of $\sum m_{\nu} > 0.06~$eV as hinted by $\nu$-oscillation global fits~\cite{Capozzi:2017ipn,deSalas:2020pgw,Gonzalez-Garcia:2021dve}) and the QCD axion against CMB and LSS data via a Monte Carlo Markov Chain (MCMC) analysis, using the general-purpose Bayesian package $\texttt{Cobaya}$~\cite{Torrado:2020dgo} and compute 95\% credible regions with $\texttt{GetDist}$~\cite{Lewis:2019xzd} (see Appendix~\ref{app:B} for more details).

In Fig.~\ref{fig:fig1} we show the marginalized posterior distribution of $m_a$ constrained by CMB and LSS data (blue band): the 95\% probability of the QCD axion mass posterior yields $m_a \leq 0.20$~eV. 
This constraint is stronger than that of Ref.~\cite{Notari:2022zxo} due to the inclusion of the recent DESI dataset, as well as the combined analysis of ACT and \textit{Planck} for CMB lensing \cite{Carron_2022,ACT:2023dou}.\footnote{We verified that restricting the exact dataset of~\cite{Notari:2022zxo} we reproduce the 95\% credible interval $m_a \leq 0.24$~eV.} 
The bound does not significantly depend on the theoretical prior assumed for the mass fraction of helium-4, $Y_{P}$, despite the expected degeneracy with $N_{\rm eff}$~\cite{Schoneberg:2019wmt}, which affects the free-electron fraction at recombination~\cite{Lee:2020obi} and hence Silk damping~\cite{Steigman:2010pa}. 
It is the inclusion of LSS data and the careful treatment of the axion as hot dark matter which breaks the degeneracy with $Y_{P}$.
As we verified in Appendix~\ref{app:B}, allowing $Y_P$ to be a free parameter of the fit would only degrade the constraint by a few percent. For the combined \textit{Planck} + lens + DESI-Y1 BAO + SN dataset, the $2\sigma$  upper bound on $m_a$ goes from 0.20 eV (when $Y_P$ is set to BBN theory) to 0.22 eV (when $Y_P$ is free), a degradation of approximately 10\%. 

The primordial helium-4 mass fraction is not only relevant for CMB physics. 
It constitutes a key observable to learn about the Early Universe~\cite{Sarkar:1995dd,Olive:1999ij,Pospelov:2010hj}. 
It is measured at the percent level~\cite{2020ApJ89677H} or more~\cite{Kurichin:2021ppm} in metal-poor systems, while being predicted at the permil level in standard BBN as an outcome of weak interactions going out-of-equilibrium~\cite{Pitrou18}. 
The relative number density of primordial deuterium, $D/H$, also features outstanding observational inference from fits to quasar absorption spectra~\cite{Riemer-Sorensen:2017pey,Cooke:2017cwo}, and can be predicted conservatively if large systematics in the treatment of nuclear rates like $DD$ fusion~\cite{Pitrou:2021vqr} are not dismissed a priori~\cite{Burns:2022hkq,Yeh:2022heq}.

In this work, we use the new publicly released package~\texttt{PRyMordial}~\cite{Burns:2023sgx} -- dedicated to the study of the physics of the Early Universe -- to provide an up-to-date prediction of helium-4 and a conservative evaluation of the relative abundance of deuterium, where the key thermonuclear rates are extracted from Ref.~\cite{Xu:2013fha}. 
We perform a set of 1300 Monte Carlo runs with \texttt{PRyMordial} to predict $X = \{Y_{P},D/H\}$ beyond $\Lambda$CDM as a function of $N_{\rm eff} = 3.044+ \Delta N_{\rm eff}$ and the cosmic baryon density $\Omega_{b}$. For each run, we extract the mean $X_{\rm th}$ and the standard deviation $\delta X_{\rm th}$ after marginalizing over thermonuclear-rate uncertainties and the neutron lifetime (whose most recent average includes only ultracold-neutron determinations~\cite{PDG:2022pth}). 
Using the result in Fig.~\ref{fig:fig2}, we feed the updated $Y_{P}(\Omega_b,m_{a})$ to \texttt{CLASS} for the computation of CMB power spectra and we implement in \texttt{Cobaya} a primordial element abundance Gaussian likelihood (see Appendix~\ref{app:B} for further details).
Adopting the measurements on light primordial abundances recommended by the Particle Data Group~\cite{PDG:2022pth}, we obtain as a main result that BBN theory combined with observations are able to narrow the QCD axion bound at 95\% probability down to $m_{a} \leq 0.17$~eV, providing a 15\% improvement from blue to red in Fig.~\ref{fig:fig1}. 

We conclude our discussion on the current cosmological constraints on the hot QCD axion, emphasizing the importance of ground-based experiments.
ACT~\cite{ACT:2020gnv} and SPT~\cite{Balkenhol23} accurately map out the CMB temperature and polarization anisotropies at angular scales smaller than those measured by \textit{Planck}. 
Including these datasets in our analysis, we obtain the marginalized posterior shown as the orange band in Fig.~\ref{fig:fig1}, ruling out axions with $m_{a}\geq0.16$~eV for the 95\% credible interval. Notice also that analyzing the cosmological setup with the addition of ACT and SPT but without the BBN likelihood would yield $m_{a}\leq0.17$~eV at 95\% probability.

In Fig.~\ref{fig:fig3}, we present a summary of our findings, illustrating the two-dimensional joint posterior distribution of $m_a$ with the sum of neutrino masses. 
While the constraint on neutrinos is saturated by the statistical weight of LSS data, the bound on the QCD axion mass is improved by the addition of BBN + ACT + SPT. Figure~\ref{fig:fig3} also shows that improving our knowledge on $\sum m_{\nu}$ would impact the cosmological constraint on the QCD axion.\footnote{It is well known that the amount of lensing inferred from the smoothing of the acoustic peaks in the \textit{Planck} temperature and polarization power spectra, exceeds the  $\Lambda$CDM predictions by a 2 to 3$\sigma$ level \cite[usually referred to as $A_L$ or $A_{\rm lens}$ tension in literature,][]{Aghanim:2018eyx,motloch20}. As a direct consequence, constraints on the sum of the neutrino masses are artificially strengthened~\cite{bianchini20a}.
In our baseline cosmological setup we find $\sum m_{\nu} < 0.118~$eV for the 95\% credible interval. The ground-based CMB surveys do not observe a similar excess of acoustic-peak smearing \cite{ACT:2020gnv,Balkenhol23}. When ACT and SPT are combined with \textit{Planck}, we then obtain a slightly relaxed bound, i.e. $\sum m_{\nu} < 0.125~$eV at 95\% probability, bearing in mind that neutrinos decouple from the thermal bath later than the QCD axion~\cite{Lesgourgues:1999wu,Arias:2023wyg}.}

~

\textbf{Forecasting with sphalerons.}~As mentioned in the introduction, Eq.~\eqref{eq:DeltaNeff} provides a beacon for future searches: $\Delta N_{\rm eff} \sim \mathcal{O}(0.01)$ points to an axion decoupling temperature $T_c\lesssim T_\mathrm{dec}\lesssim \mathcal{O}(1\,\mathrm{GeV})$, i.e. in between the confined phase of QCD and the quark-gluon plasma. 
While the axion production rate from gluon-gluon scattering in~\cite{Salvio:2013iaa} has been extended by Ref.~\cite{DEramo:2021lgb} down to $\mathcal{O}$(GeV), Ref.~\cite{Notari:2022zxo} observed that already at those temperatures, the main contribution to $\Pi_{a}^R$ should be of non-perturbative nature due to {\it strong sphalerons}, QCD topological configurations responsible of quark chirality-violating transitions at finite temperature~\cite{McLerran:1990de,Giudice:1993bb,Fukushima:2008xe,Moore:2010jd}.

In this study, we forecast the sensitivity of future cosmological surveys to the QCD axion mass , taking advantage of the first-principle non-perturbative computation of the real-time correlator of $\mathcal{Q}$ in Eq.~\eqref{eq:axionpsd} beyond the pure Yang-Mills limit, as recently obtained by the 2+1 lattice QCD simulations above chiral crossover of Refs.~\cite{Bonanno:2023thi,Bonanno:2023xfv}. Adopting the best-fit results~\footnote{We checked that extrapolating the strong sphaleron rate down to 150~MeV with either of the two fits present in~\cite{Bonanno:2023thi} leads to the same results in our study.} on the sphaleron rate in~\cite{Bonanno:2023thi}, we derive the momentum-averaged one using the expected size of sphaleron configurations (see App.~E of~\cite{Notari:2022zxo}). In doing so, we obtain for $150~$MeV~$\lesssim T \lesssim 600~$MeV a conservative estimate of the thermalization rate $\overline{\Gamma}_{a}$ governing the (integrated) Boltzmann equation of the yield $Y_{a} \equiv n_a/s_{\gamma}\,$:
\begin{eqnarray}
 \label{eq:gammasph}
  \frac{dY_{a}}{dt} & = &  \, \frac{\overline{\Gamma}_a}{H} \, (Y^{\rm eq}_a - Y_{a})   \ ,  \\ 
   \overline{\Gamma}_{a}  & \simeq &  \kappa \, \left(\frac{T^3}{f_{a}^2} \right) \, \left(c_0 + c_1 \, \frac{T_{c}}{T} + c_2 \, \frac{T_{c}^2}{T^2} \right) \, , \nonumber
\end{eqnarray}
$s_{\gamma}$ being the total entropy density of the thermal bath, $\kappa = 0.1$, and $c_{0,1,2} = \{0.04,-0.28,1.5\}$.

Trading time for temperature using entropy conservation, we solve Eq.~\eqref{eq:gammasph} from 600~MeV down to $T_{c}$, and set the initial condition for the momentum-dependent Boltzmann equations in Eqs.~\eqref{eq:axionpsd} using the approximation: $ \mathcal{F}_{a}(E,T_{c}) \simeq \mathcal{F}^{\rm eq}_{a}(E,\overline{T})$ and determining $\overline{T}$ by $Y_{a}(T_{c}) = Y^{\rm eq}_{a}(\overline{T})$. 
Considering an optimistic forecast, we assume that the Universe underwent reheating above the electroweak crossover, suggesting for the initial condition of Eq.~\eqref{eq:gammasph} $Y_{a}(600\,{\rm MeV}) \simeq Y^{\rm eq}_{a}(T_{\rm EW})$.  

We produce our forecasts with state-of-the-art likelihoods using the dedicated routine in \texttt{Cobaya} for next-gen CMB/LSS experiments and adopting the expected sensitivities reported in~\cite{SimonsObservatory:2018koc,CMB-S4:2022ght,DESI:2016fyo} respectively for SO, CMB-S4 and DESI-Y5 (BAO), as well as a fiducial $\Lambda$CDM cosmology with three degenerate neutrinos of $m_{\nu} = 0.02$~eV. 
We forecast two different future scenarios: SO+DESI-Y5 and CMB-S4+DESI-Y5. 
We perform an MCMC analysis for each of the cases assuming the same priors used in the real data analysis (see Appendix~\ref{app:B}). 
In the cosmological Boltzmann-solver \texttt{CLASS} we include the QCD axion as hot dark matter with phase-space distribution imprinted by: thermal equilibrium at high temperature, strong-sphaleron transitions according to the averaged rate estimated for $1 \lesssim T/T_{c} \lesssim 4$, and energy-dependent scattering with pions for $T \lesssim T_{c}$.

In Fig.~\ref{fig:fig3} we show the joint $m_{a}$-$\sum m_{\nu}$ 68\% and 95\% credible region from our forecasts. 
At the 95\% probability, SO+DESI-Y5 will be sensitive to $m_a \lesssim 0.08$ eV, while CMB-S4+DESI-Y5 could exclude axions with $m_{a} \geq 0.04~$eV, competitive with the current supernovae disfavoured region~\cite{Lella:2023bfb}.
While futuristic CMB and LSS proposals like~\cite{Schlegel:2022vrv,MacInnis:2023vif} may further improve those projections, foreseeable progress in BBN physics might play a marginal role. Assuming a 50\% refined inference for helium-4 and deuterium observations, and an order-of-magnitude more precise $D/H$ prediction -- matching the present one for $Y_P$ -- we added BBN to both forecasts via importance sampling and improved of a few percent the bound on $m_a$ only for SO+DESI-Y5.\footnote{In Appendix~\ref{app:B} we show two forecasts obtained for a fiducial cosmology that assumes three equally massive neutrinos with total mass $\sum m_\nu=0.06$ eV and a QCD axion mass $m_a=0.05$, $0.1$ eV.}


~

\textbf{Conclusions.}~We determined a conservative bound on $m_a$ from up-to-date measurements of the CMB, including ground-based telescopes, and abundances from BBN. We treated the axion as hot dark matter evaluating its phase-space distribution beyond the thermal approximation -- i.e. including spectral distortions from momentum-dependent scattering -- finding $m_a\leq0.16$ eV for the 95\% credible interval, shown by the orange band in Fig.~\ref{fig:fig1}. 
We also reported the two dimensional joint posterior distributions of $m_a$ and $\sum m_\nu$ in Fig.~\ref{fig:fig3} from current cosmological probes and provided a forecast for future cosmological surveys including the effect of the non-perturbative production rate expected from strong sphalerons extracted from 2+1 lattice QCD at the physical point. 

While Fig.~\ref{fig:fig3} captures an exciting prospect, we point out that the future for a cosmological detection of the QCD axion via primeval hot modes might be even brighter thanks to dedicated studies of the Lyman-$\alpha$ forest~\cite{Irsic:2023equ}, advancements in the full-shape analysis of LSS~\cite{Chudaykin:2020aoj,DAmico:2020kxu}, and progress in 21~cm intensity mapping~\cite{Karkare:2022bai}.

{\bf Acknowledgements.} We thank Luca Di Luzio, Peter Graham, Maxim Pospelov, Neelima Sehgal, Luca Silvestrini and Wei Xue for discussions. G.G.d.C. is grateful to SCGP, F.B. to INFN Rome and Sapienza, M.V. to SITP and SLAC, for the hospitality during the completion of the work. F.B. would like to thank Emmanuel Schaan and Kimmy Wu for stylistic suggestions on Fig.~\ref{fig:fig1}. F.B. acknowledges support by the Department of Energy, Contract DE-AC02-76SF00515.
Some of the computing for this project was performed on the Sherlock cluster. We would like to thank Stanford University and the Stanford Research Computing Center for providing computational resources and support that contributed to these research results. We also acknowledge support by COST (European Cooperation in Science and Technology) via the COST Action COSMIC WISPers CA21106.

\appendix
\onecolumngrid

\onecolumngrid
\section{Axion Thermalization Rate}
\label{app:A}
We report more details about the numerical computation of the thermalization rate for hot axion production and the evaluation of the axion phase-space distribution. We follow the strategy originally worked out in detail in~\cite{1976Ap&SS..39..429Y,Hannestad:1995rs} and more recently applied to the case of the QCD axion in Ref.~\cite{Notari:2022zxo}. The momentum-dependent thermalization rate $\Gamma_{a}$ can be explicitly written for the process $ \pi \pi\to \pi a$ as follows:
\begin{equation}
   \mathcal{F}_{a}^{\rm eq} \, \Gamma_a = \frac{1}{2\,E}\int \biggl(\prod_{i=1}^3 \frac{d^3\mathbf{k}_i}{(2\pi)^3\,2\,E_i}\biggr) \mathcal{F}_1^\mathrm{eq}\mathcal{F}_2^\mathrm{eq}(1+\mathcal{F}_3^\mathrm{eq})(2\pi)^4 \delta^{(4)}(k_1^\mu+k_2^\mu-k_3^\mu-k^\mu) \, |\mathcal{M}|^2,
\end{equation}
where $|\mathcal{M}|^2$ is the modulus squared of the scattering amplitude (summed over initial and final spins) and $k_i^\mu$ with $i=1,\,2,\,3$ are the 4-momenta of the incoming pions and the outgoing one, respectively, while $k$ is the 4-momentum of the axion.
Following similar steps carried out in the literature for the collision term of the momentum-dependent Boltzmann equations of neutrinos~\cite{1976Ap&SS..39..429Y,Hannestad:1995rs}, we can rewrite the above as: 
\begin{equation}
    \Gamma_a = \frac{\exp(E/T)-1}{16 \pi^4 \,E}\int_0^\infty \biggl(\prod_{i=1}^2 \frac{d|\mathbf{k}_i| |\mathbf{k}_i|^2}{E_i}\biggr) \int_{-1}^{1} d c_\alpha \int_{-1}^{1} d c_\beta \,\mathcal{F}_1^\mathrm{eq}\mathcal{F}_2^\mathrm{eq}(1+\mathcal{F}_3^\mathrm{eq})\frac{\Theta(E_1+E_2-E) \Theta(1-c_{\gamma_0}^2)}{|h'(\gamma_0)|} |\mathcal{M}|^2,
\end{equation}
where $h(\gamma) = k_3^2 - m_3^2$ and $\gamma_{0}$ are the zeros of $h(\gamma)$. The scattering angles $\alpha$, $\beta$ and $\gamma$ are defined as:
\begin{equation}
    c_\alpha = \cos\alpha = \frac{\mathbf{k}_1\cdot\mathbf{k}_2}{|\mathbf{k}_1||\mathbf{k}_2|}, \qquad c_\beta = \cos\beta = \frac{\mathbf{k}_1\cdot\mathbf{k}}{|\mathbf{k}_1||\mathbf{k}|}, \qquad \frac{\mathbf{k}_2\cdot\mathbf{k}}{|\mathbf{k}_2||\mathbf{k}|} = \cos\alpha \cos \beta + \sin\alpha \sin\beta \cos\gamma \ ,
    \label{eq:destruction_rate}
\end{equation}
and $\mathcal{F}_3^\mathrm{eq}$ is now a function of $|\mathbf{{k}_1}|$ and $|\mathbf{k}_2|$ via energy conservation. Also the amplitude $|\mathcal{M}|^2$, which depends on the Mandelstam variables $s,t,u$ of the $2 \to 2$ process, can be expressed as a function of the 3-momenta $|\mathbf{{k}_1}|$ and $|\mathbf{{k}_2}|$. 

For the numerical analysis, we used $f_\pi=92.3$ MeV, an average pion mass $m_\pi = 138$ MeV and the isospin-suppression factor: $(m_d-m_u)/(m_d+m_u)\simeq 0.47$~\cite{ParticleDataGroup:2022pth}, where $m_u$ and $m_d$ are the up- and down-quark masses. 
The multi-dimensional integration was performed with use of the Python library \href{https://zenodo.org/record/8175999}{\texttt{vegas}}~\cite{Lepage:2020tgj}.
The missing ingredient of equation~\eqref{eq:destruction_rate} -- the modulus square of the amplitude -- is described in the following for the two different methods outlined in the text, see also~\cite{Notari:2022zxo,DiLuzio:2022gsc} for more details.  

\subsection{Inverse Amplitude Method} 

The amplitude $\mathcal{M}$ can be conveniently decomposed in a basis with definite total isospin $I$; in doing so, one can obtain the following relations involving Clebsch-Gordan weights~\cite{DiLuzio:2022gsc}:
\begin{eqnarray}
    \mathcal{M}_{+0} &=& -\frac{A_1+A_2'}{\sqrt{2}} = -\mathcal{M}_{-0},\qquad \mathcal{M}_{\pm \mp} = - \frac{\sqrt{2}A_0 + A_2}{\sqrt{6}},\\
    \mathcal{M}_{0+} &=& \frac{A_1-A_2'}{\sqrt{2}} = -\mathcal{M}_{0-},\qquad \quad \mathcal{M}_{00} = -\frac{A_0-\sqrt{2}A_2}{\sqrt{3}}.\nonumber
\end{eqnarray}
Notice that the amplitudes in the defined total isospin basis $A_2$ and $A_2'$ are different because the axion coupling to pions violates isospin, and $\mathcal{M}_{+-}=\mathcal{M}_{-+}$ because of charge-conjugation symmetry.

We can now further project the amplitudes with defined total isospin into a basis of states with well-defined total angular momentum $J$, obtaining
\begin{equation}
    A_I(s,\cos\theta) = \sum_{J=0}^\infty (2 J + 1)P_J(\cos\theta) A_{IJ}(s),
\end{equation}
where $\cos\theta$ is the scattering angle in the center of mass frame, and $P_J(\cos\theta)$ are Legendre polynomials.
We can then apply the inverse amplitude method~\cite{Salas-Bernardez:2020hua} at the next-to-leading-order expressing perturbatively $A_{IJ}$ as:
\begin{equation}
    A_{IJ}(s) = \frac{A_{IJ}^{(2)}(s)}{1-A_{IJ}^{(4)}(s)/A_{IJ}^{(2)}(s)},
\end{equation}
where by $A_{IJ}^{(2n)}$ we denote the amplitudes calculated in $\chi$PT up to $\mathcal{O}(p^{2n})$ from the partial-wave decomposition:
\begin{equation}
    A_{IJ}^{(2n)}(s)=\frac{1}{2}\int_{-1}^{+1} d\cos\theta \,P_J(\cos\theta) A_I^{(2n)}(s,\cos\theta).
\end{equation}
In our computation we include the contributions coming from S-wave ($J=0$, $I=0,\,2$) and P-wave ($J=1$, $I=1$), and adopt the analytical expression of the amplitudes for the $\pi\pi\to \pi a$ reported in~\cite{DiLuzio:2022gsc}. For our analysis we used the following low-energy constants: $\bar{\ell}_1=-0.525$~\cite{Dobado:1996ps}, $\bar{\ell}_2 = 5.425$~\cite{Dobado:1996ps}, $\bar{\ell}_3 = 3.53$~\cite{FlavourLatticeAveragingGroupFLAG:2021npn}, $\bar{\ell}_4 = 4.73$~\cite{FlavourLatticeAveragingGroupFLAG:2021npn} and $\ell_7 = 0.007$~\cite{diCortona:2015ldu}.

\subsection{Phenomenological Approach} 
The amplitude from the  phenomenological partial-wave phases of $\pi\pi$ scattering is obtained by decomposing $\mathcal{M}$ in the charged basis into a basis with well-defined isospin, and then expanding each amplitude into partial waves. Following the notation of~\cite{Gasser:1983yg}, the partial-wave amplitudes are related to the real phase shifts $\delta_{IJ}$ as:
\begin{equation}
    A_{IJ}(s) = \sqrt{\frac{s}{s-4 m_\pi^2}} \left(\frac{32 \pi}{\cot\delta_{IJ}(s)-i} \right) \ ,
\end{equation}
where the $\delta_{IJ}(s)$ are parameterised as in~\cite{Garcia-Martin:2011iqs}. We include only the S- and P-wave phase shift, since other terms give a negligible contribution to the rate. We find the modulus squared of the total amplitude $\pi \pi \to \pi a$ to be given by:
\begin{equation}
    |\mathcal{M}|^2 = \frac{1}{24} \left(\frac{m_d-m_u}{m_d+m_u} \right)^2 \left(\frac{f_\pi}{f_a}\right)^2
    \left(|A_{0 0}|^2+27\,|\cos\theta\, A_{11}|^2+5\,|A_{20}|^2\right).
\end{equation}
\subsection{The Strong Sphaleron Rate} 
The axion thermalization rate at zero-momentum from strong-sphaleron transitions is taken from the results of the lattice QCD simulation at the physical point for 2+1 flavors of Ref.s~\cite{Bonanno:2023thi,Bonanno:2023xfv}, according to the Monte Carlo simulations performed in that work for five sampling temperatures: $T = $230, 300, 365, 430 and 570~MeV. We adopted the phenomenological parameterization given by the power-law fit:
\begin{equation}
\Gamma_{\rm sph}(|\mathbf{k}| = 0) = \Lambda_{0}^4 \ (T/T_{c})^{\epsilon} \ ,
\end{equation}
where $ \Lambda_{0} \simeq 142.3~$MeV,  $\epsilon \simeq 1.81$ , $T_{c} = 155~$MeV. 

We notice that the above parameterization (adopted in the range $1 \lesssim T/T_{c} \lesssim 4$) gives a slightly smaller  $\Delta N_{\rm eff}(m_{a})$ in the final computation with respect to the one obtained from the semiclassically-inspired fit reported in the same Ref.s~\cite{Bonanno:2023thi,Bonanno:2023xfv}. 
To compute the axion phase-space distribution with the inclusion of non-perturbative effects, we conservatively estimated the momentum-averaged axion rate $\overline{\Gamma}_{a}$ using the fact that $\Gamma_{\rm sph}$ should be the same within a shell of momentum $|\mathbf{k}| < |\mathbf{k_{s}}|$, with $|\mathbf{k_{s}}|$ set by the expected sphaleron size, $\mathcal{O}\left(1/(\alpha_{s} T)\right)$, namely:
\begin{equation}
 n_{a}^{\rm eq} \, \overline{\Gamma}_{a} = \frac{1}{f_{a}^2}\int \frac{d^{3} |{\bf k}|}{(2 \pi)^32 E} \, \Gamma_{\rm sph}(|\mathbf{k}|) \, \frac{\mathcal{F}^{\rm eq}_{a}(E,T)}{\left(1+ \mathcal{F}^{\rm eq}_{a}(E,T)\right)} \gtrsim \frac{\Lambda_{0}^4}{4 \pi^2 f_{a}^2} \left(\frac{T}{T_{c}}\right)^{\epsilon}\int_{0}^{3\alpha_{s} T} d|\mathbf{k}| |\mathbf{k}| \exp(-|{\bf k}|/T) \ ,
\end{equation}
see also Appendix~E of Ref.~\cite{Notari:2022zxo}. Comparing the rate saturating the inequality against expectations from dimensional analysis, i.e. $\sim T^3 /f_{a}^2$, a simple analytical form can be obtained valid for temperatures $150~{\rm MeV} \lesssim T \lesssim 600~$MeV and reported in the main text. We remind that the solution of the integrated Boltzmann equation governed by $\overline{\Gamma}_{a}$ allows one to estimate the temperature which sets the initial condition $\mathcal{F}_{a} > 0$ at $T_{c}$, adopted for our forecasts.


\section{Cosmological Analysis}
\label{app:B}
We detail the integration of hot axions into the Boltzmann-solver \texttt{CLASS} \citep{Lesgourgues:2011rh} and provide further details on our Markov Chain Monte Carlo (MCMC) analyses using \texttt{Cobaya} \citep{Torrado:2020dgo}.

\subsection{Axions As Hot Dark Matter}

Our cosmological analysis treats the axion as an additional non-cold dark matter (\texttt{ncdm}) species.
This is accomplished by coding a new phase-space distribution in the \texttt{background.c} source file implementing analogous modifications to what already highlighted in the work of  Ref.~\cite{Notari:2022zxo}.

In our numerical analysis, for a grid of masses $m_{a}$ we have approximated the numerical phase-space distribution obtained from the solution to the momentum-Boltzmann equations with a Fermi-Dirac o Bose-Einstein distribution with non-zero chemical potential.
We have verified the accuracy of this approximation over a range of scales $0.04 \lesssim |\mathbf{k}|/T \lesssim 150$ and masses $0.01 \lesssim m_a/\mathrm{eV} \lesssim 4$ and found very good agreement.
Although the phase-space distribution above solely depends on $ |\mathbf{k}|/T$ and $m_a$, \texttt{CLASS} requires us to assign a temperature $T_{\rm ncdm}$ to the non-cold dark matter species. 
This temperature essentially serves as an overall normalization of the phase-space  distribution.
We properly set this temperature to be in line with an extra-neutrino species: $T_{a} = \left[ g_{*s}(3\,\text{MeV})/g_{*s}(30\,\text{MeV})\right]^{1/3} (4/11)^{1/3} T$, where $g_{*s}(T)$ is the number of entropic degrees of freedom of the thermal bath at temperature $T$~\cite{Borsanyi:2010bp}, and we remind the reader that $T_{\nu} = (4/11)^{1/3} \, T $ in the limit of neutrino instantaneous decoupling.

When neglecting spectral distortions and modeling the QCD axion with a Bose-Einstein distribution, we exploit the definition of $\Delta N_{\rm eff} \equiv \rho_{a}/\rho_{\nu}$ to set the species temperature to:
\begin{equation}
    T_a(m_a) = \left(\frac{7}{4}\Delta N_{\rm eff}(m_a)\right)^{1/4} \left(\frac{4}{11} \right)^{1/3} T \ ,
\end{equation}
where $\Delta N_{\rm eff}(m_a)$ is an analytical fit to the numerically evaluated contribution of the QCD axion to the effective number of relativistic degrees of freedom.

\subsection{Cosmological Inference Framework}
Our baseline cosmology extends the $\Lambda$CDM model to include neutrinos and the QCD axion, which are treated as hot dark matter species, and is built on purely adiabatic scalar fluctuations.

This baseline model comprises a total of eight cosmological parameters,  including the physical density of cold dark matter ($\Omega_{\rm c}h^2$), the physical density of baryons ($\Omega_{\rm b}h^2$), the approximated angular size of the sound horizon at recombination ($100\,\theta_{\rm s}$), the optical depth at reionization ($\tau$), the amplitude of curvature perturbations at $\hat{k} = 0.05$ Mpc$^{-1}$ ($A_s$), and the spectral index ($n_s$) of the power law power spectrum of primordial scalar fluctuations.

In addition to these six standard $\Lambda$CDM parameters, we include the axion mass $m_a$  and the sum of the active neutrino masses $\sum m_{\nu}$. 
We assume these to be a degenerate combination of three equally massive neutrinos, as done in the publications of the \textit{Planck} Collaboration \citep{Lesgourgues:2006nd,Planck18res}.
The lensed CMB and CMB lensing potential power spectra are computed using the \texttt{CLASS} Boltzmann code (\texttt{v3.3.1}). 
To infer cosmological parameter constraints, we sample the posterior space using the Metropolis-Hastings sampler with adaptive covariance learning, provided in the Markov Chain Monte Carlo (MCMC) \texttt{Cobaya} package \cite{Torrado:2020dgo}.
When sampling the parameter space, we adopt the priors listed in Tab.~\ref{tab:priors}.
Note, in particular, that we assume $m_{\nu} \ge 0.02$ eV for three degenerate neutrinos to enforce the lower bound on the sum of the neutrino masses $\sum m_\nu \ge 0.06$ eV from oscillation measurements \cite{Capozzi:2017ipn,deSalas:2020pgw,Gonzalez-Garcia:2021dve}.
For each cosmological model and dataset combination, we run four \texttt{MPI} parallel chains (each one using four \texttt{OpenMP} processes) until the Gelman–Rubin statistic reaches a convergence threshold of $R-1 \leq 0.02$.

\begin{table}
\renewcommand{\arraystretch}{2}
\centering
\caption{Cosmological parameters varied in this work and their respective priors. $\mathcal{U}(a,b)$ denotes a uniform distribution between $[a,b]$, while $\mathcal{N}(\mu,\sigma)$ indicates a Gaussian distribution with mean $\mu$ and variance $\sigma^2$. Note that when analyzing mock data from future CMB surveys, we adopt a \textit{Planck}-based Gaussian prior on  $\tau \sim \mathcal{N}(0.055, 0.007)$.\\ }
\label{tab:priors}
\begin{tabular}{c|c|c|c|c|c|c|c|c|c}
\hline
\hline
\textit{Parameter} & \boldmath${\Omega_{b} h^2}$ & \boldmath$\Omega_{\rm c} h^2$ & \boldmath$100\, \theta_s$ & \boldmath$\tau$ & \boldmath$n_{s}$ & \boldmath$\ln (10^{10}A_{\rm s})$ & \boldmath$\sum m_{\rm \nu}$ {[}eV{]} & \boldmath$m_a$ {[}eV{]} & \boldmath$Y_P$ \\
\hline
\hline
\textit{Prior} & $\, \mathcal{U}(0.005,0.1) \, $ & $\, \mathcal{U}(0.001,0.99)\, $ & $\, \mathcal{U}(0.5,10)\, $ & $\, \mathcal{U}(0.01,0.8)\, $ & $\, \mathcal{U}(0.8,1.2)\, $ & $\, \mathcal{U}(1.6,3.9)\, $ & $\, \mathcal{U}(0.06,5)\, $ & $\, \mathcal{U}(0,4)\, $ & $\, \mathcal{U}(0.1,0.5)\, $ \\
\hline

\end{tabular}
\end{table}

\subsection{Cosmological Datasets}
Our cosmological analysis incorporates seven distinct types of observations: primary CMB, CMB lensing, Baryon Acoustic Oscillations (BAO), redshift space distortions (RSD), Type Ia supernova (SNeIa) distance moduli, and the observational inference of the primordial abundances of the mass fraction of helium-4, $Y_{P}$, and of the relative number density of deuterium, $D/H$.

For primary CMB, we utilize the CMB temperature and polarization anisotropies measurements as presented in the \textit{Planck} 2018 data release \cite{Aghanim:2019ame}. Specifically, we combine both the low- and high-$\ell$ temperature and polarization likelihoods derived from PR3 maps. This dataset is referred to as ‘\textit{Planck}’ in our figures. 
We supplement Planck observations with high-resolution CMB temperature and polarization measurements at intermediate and small angular scales from ground-based experiments. 
Specifically, we incorporate the Atacama Cosmology Telescope (ACT) DR4  $TT/TE/EE$ results from \cite{ACT:2020gnv} and the 2018 South Pole Telescope (SPT) SPT-3G $TT/TE/EE$ measurements from \cite{Balkenhol23}. 
We denote the combination of these two datasets as ‘ACT+SPT’.

We utilize measurements of the CMB lensing power spectrum derived from a joint analysis of \textit{Planck} PR4 \cite{Carron_2022} and ACT DR6 data \cite{ACT:2023dou}. 
This data product is referred to as ‘lens’.

In this paper, we utilize the DESI Year 1 (DESI-Y1) BAO data. 
BAO are fluctuations in the density of the visible baryonic matter of the universe, caused by acoustic density waves in the primordial plasma before the universe recombined. These oscillations leave imprints in the distribution of galaxies, providing a "standard ruler" for cosmological distance measurements.
The observables include measurements and correlations for the comoving distance over the drag epoch, $D_M / r_d$, and the distance variable, $D_H / r_d$. Here, $r_d$ denotes the sound horizon at the drag epoch, the distance sound waves traveled in the early universe until baryons decoupled from photons. For datasets with lower signal-to-noise ratios, the angle-averaged quantity, $D_V / r_d$, is provided instead.
Here, we employ the recently published DESI-Y1 BAO data \cite{DESI:2024mwx}, encompassing BAO distance measurements derived from various galaxy samples: the Bright Galaxy Sample (BGS, $0.1 < z < 0.4$), the Luminous Red Galaxy Sample (LRG, $0.4 < z < 1.1$), the Emission Line Galaxy Sample (ELG, $1.1 < z < 1.6$), the Quasar Sample (QSO, $0.8 < z < 2.1$), and the Lyman-$\alpha$ Forest Sample (Ly$\alpha$, $1.77 < z < 4.16$). These measurements span a broad range of redshifts, which is crucial for breaking degeneracies in the $\Omega_m$-$r_dH_0$ plane, leading to a more precise determination of these cosmological parameters.


The SNeIa distance moduli measurements are from the Pantheon sample \cite{Scolnic:2017caz}. We refer to this dataset as ‘SN’.

We use the measurements on light primordial abundances recommended by the Particle Data Group~\cite{PDG:2022pth}: $Y_{P}=~0.245 \pm 0.003$ and $D/H\times10^{5} = 2.547 \pm 0.025$. 
We refer to this dataset as ‘BBN’ (see next section for more details).

We also forecast the cosmological sensitivity of next-generation CMB and LSS surveys to the QCD axion mass. 
We consider the $TT/TE/EE/\phi\phi$ data vector from two representative upcoming ground-based CMB surveys, Simons Observatory (SO)  \cite{SimonsObservatory:2018koc} and CMB-S4 \citep{abazajian2019cmbs4,CMB-S4:2022ght}. 
The effective temperature and polarization noise curves are calculated after multi-frequency component-separation while CMB lensing is assumed to be reconstructed with a minimum-variance quadratic estimator. \footnote{That is actually available at \url{https://github.com/simonsobs/so_noise_models}. For temperature and polarization we use the following \texttt{LAT\_comp\_sep\_noise/v3.1.0} curves: \texttt{SO\_LAT\_Nell\_T\_atmv1\_goal\_fsky0p4\_ILC\_CMB.txt} and \texttt{SO\_LAT\_Nell\_P\_baseline\_fsky0p4\_ILC\_CMB\_E.txt}. \\ For lensing: \texttt{nlkk\_v3\_1\_0\_deproj0\_SENS2\_fsky0p4\_it\_lT30-\\3000\_lP30-5000.dat} in \texttt{LAT\_lensing\_noise/lensing\_v3\_1\_1/}. The primary CMB noise curves are at \url{https://sns.ias.edu/~jch/S4_190604d_2LAT_Tpol_default_noisecurves.tgz}. \\ Specifically we use \texttt{S4\_190604d\_2LAT\_T\_default\_noisecurves\allowbreak \_deproj0\_SENS0\_ mask\_16000\_ell\_TT\_yy.txt} together also with \texttt{S4\_190604d\_2LAT\_pol\_default\_noisecurves\_deproj0\_SENS0\_mask\allowbreak\_16000\_ell\_EE\_BB.txt} for $TT$ and $EE$ respectively. For the CMB lensing reconstruction noise we have adopted the following curves \texttt{kappa\_deproj0\_sens0\_16000\_lT30-3000\_lP30-5000.dat}, which are taken from \url{https://github.com/toshiyan/cmblensplus/tree/master/example/data}.}.
We assume a sky fraction of $f_{\rm sky} = 0.4$ for both surveys and use CMB information between $30 \le \ell \le 3000$ in temperature and between $30 \le \ell \le 5000$ for polarization and lensing.
The CMB power spectra likelihood is the one from \citet{Hamimeche:2008ai} (specifically we use the \texttt{CMBlikes} likelihood class in \texttt{Cobaya}).
Note that when producing forecasts using mock SO and CMB-S4 data, we add a \textit{Planck}-based Gaussian prior on the optical depth  $\tau \sim \mathcal{N}(0.055, 0.007)$ to account for the missing large-scale ($\ell \lesssim 30$) polarization information from the ground.

We also include future BAO data in the form of $r_s/D_V$ measurements from the expected five-year data taking of DESI \cite{DESI:2016fyo}, where $D_V$ is the volume averaged distance and $r_s$ is the sound horizon at the drag epoch when photons and baryons decouple. We consider the baseline DESI survey, covering 14000 deg$^2$ and targeting bright galaxies, luminous red galaxies, and emission-line galaxies in the redshift range $ z \in [0.15, 1.85]$ with 18 bins equally spaced redshift bin.

The fiducial cosmologies used to generate the mock CMB/BAO data vectors always assume three equally massive neutrinos with total mass $\sum m_\nu=0.06$ eV and a span over the following axion mass values: $m_a=0,\, 0.05,\, 0.1$ eV.

\subsection{BBN Analysis}
We report here the information about the BBN likelihood adopted in our analysis, including the details about the constraint on the axion mass derived solely by BBN. The BBN likelihood is modeled as a Gaussian according to:
\begin{equation}
-2 \, \log \mathcal{L}_{\rm BBN} = \sum_{Y_{P},D/H} \frac{\big( X_{\rm exp} - X_{\rm th}(\Omega_b,m_a)\big)^2}
{\delta X_{\rm  exp}^2+ \delta X^2_{\rm th}(\Omega_b,m_a)} \ .
\label{eq:bbn_likelihood}
\end{equation}
In the above expression, $X_{\rm th}(\Omega_b,m_a)$, with $X_{\rm th}$ running over the helium-4 mass fraction $Y_P$ and deuterium relative number density $D/H$, is the theoretical prediction obtained for the light primordial abundance, while $\delta X_{\rm th}(\Omega_b,m_a)$ is the associated error. 
Those are obtained from a set of 1300 Monte Carlo (MC) runs with the code $\texttt{PRyMordial}$~\cite{Burns:2023sgx} as a function of $\Omega_{b}$ and $N_{\rm eff}$ (which can be put in one-to-one correspondence with $m_a$). 
Each of the 1300 MC runs comprised $10^4$ events at fixed baryon density and extra relativistic number of degrees of freedom, where we varied 12 nuisance parameters related to the key nuclear rates for helium and deuterium production. 
For the latter we adopt log-normal distributions as detailed in Ref.~\cite{Coc:2014oia} (along the lines of the MC examples in Ref.~\cite{Burns:2023sgx}), and use as a main reference the NACREE~II database~\cite{Xu:2013fha} in order to avoid any strong theory prior on the prediction of deuterium, see the discussion in~\cite{Pitrou:2021vqr}. 
We also varied in our MC runs the neutron lifetime according to a Gaussian distribution  with mean and standard deviation recently recommended by the Particle Data Group~\cite{PDG:2022pth}, $\tau_{n} = (878.4 \pm 0.5$)~s, obtaining a marginal update of the state-of-the-art prediction on $Y_P$. 
For each MC run we then have computed mean and standard deviation from the posterior distribution of $Y_{P}$ and $D/H$ (which turned out to be well approximated by a Gaussian) in order to obtain $X_{\rm th}$ and 
$\delta X_{\rm th}$ as a function of $\Omega_b$ and $m_a$ for helium-4 and deuterium.\footnote{The corresponding numerical table can be found at the repository \github\  \href{https://github.com/vallima/PRyMordial}{https://github.com/vallima/PRyMordial}.}
\begin{figure}[t!]
\includegraphics[width=0.6\textwidth]{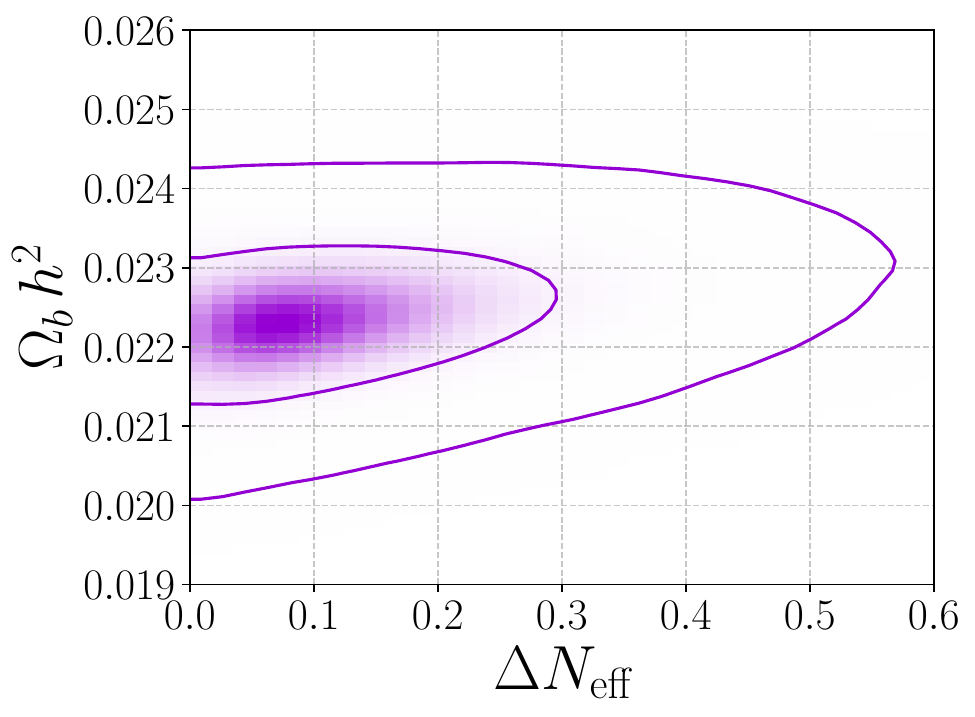}  \caption{$68\%$ and $95\%$ credible region for the cosmological baryon abundance, $\Omega_b \, h^2$, and positive extra relativistic degrees of freedom, $\Delta N_{\rm eff}$, determined using $\texttt{PRyMordial}$~\cite{Burns:2023sgx} from a fit to the measurements on light primordial abundances~\cite{PDG:2022pth}: the helium-4 mass fraction, $Y_{P} = 0.245 \pm 0.003$ and the deuterium relative number density, $D/H \times 10^{5} = 2.547 \pm 0.025$.}
\label{fig:bbn_bound}
\end{figure}

Regarding the observational inference of light primordial abundances, the study of spectroscopic emission lines in metal-poor extragalactic environments allows for the determination of the mass density fraction of helium-4. Notice that to consider the latter robust against systematics, it is important to restrict the sample of astrophysical systems under examination to the set featuring also the detection of the He $\lambda$10830 infrared emission line (relevant for parameter-degeneracy breaking~\cite{Aver:2015iza}). Moreover, it is also crucial that the underlying theoretical model describing the lines detected provides effectively an overall good fit to the observed emission spectra, see for instance Ref.s~\cite{Aver:2020fon,2020ApJ89677H}.
On the basis of those criteria, in this work we follow the recommendation of the Particle Data Group and adopt in our analysis $Y_{P} = 0.245 \pm 0.003$~\cite{PDG:2022pth}.
For what concerns deuterium, the inspection of this element in quasar absorption lines provides a clean and very plausible probe of its primordial abundance. State-of-the-art analyses from damped Lyman-$\alpha$ systems surpass percent-level precision~\cite{Cooke:2016rky,Riemer-Sorensen:2017pey,Cooke:2017cwo}, 
and a weighted average performed by the Particle Data Group on the most recent measurements yields ${D/H} \times 10^5 = 2.547 \pm 0.025$~\cite{PDG:2022pth}, adopted in our study. 

A constraint on the QCD axion mass solely from BBN can be derived from what reported in Fig.~\ref{fig:bbn_bound}. In the figure we show the 68\% and 95\% credible interval of the two-dimensional posterior distribution for the cosmological baryon abundance and a positive shift of the extra-relativistic degrees of freedom. We obtained the joint posterior distribution via a dedicated MCMC analysis of the BBN likelihood in Eq.~\eqref{eq:bbn_likelihood}, making use of the code PRyMordial and the light-element measurements taken from the Particle Data Group. Marginalizing over $\Omega_b h^2$ and trading the extra-relativistic degrees of freedom for the QCD axion mass, we obtain as BBN bound $m_a \leq 0.77~$eV for the 95\% credible interval.

\subsection{Cosmological Insights } 
\begin{figure}[b!]
  \includegraphics[width=0.45\textwidth]{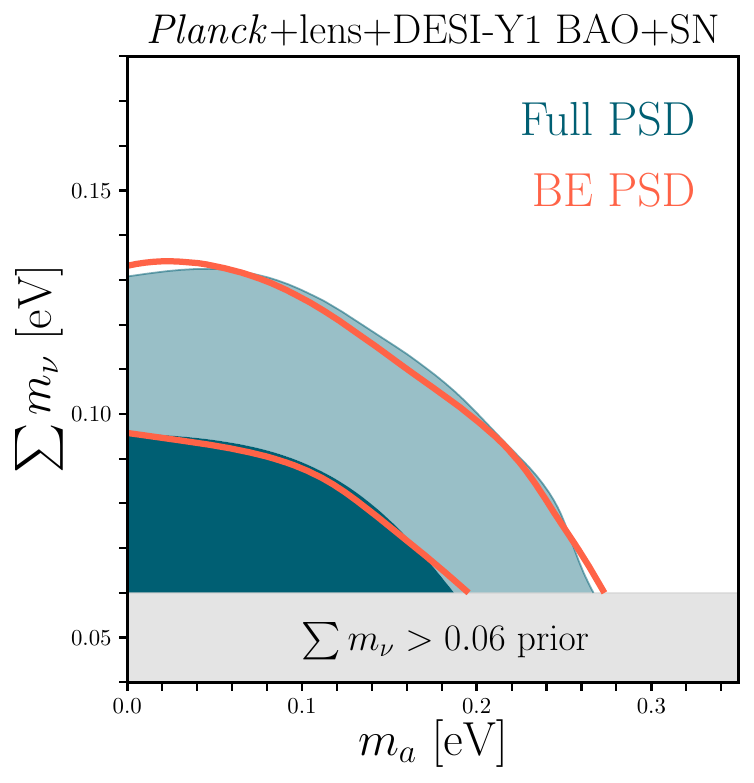}
  \caption{Comparison for the 68\% and 95\% credible region in the plane axion mass - sum of neutrino masses, inferred from the analysis of \textit{Planck} + lens + BAO + SN data under the assumption of a Bose-Einstein thermal distribution (empty red contours) matching the contribution of extra-relativistic degrees of freedom computed for the QCD axion or using directly the full numerical phase-space distribution obtained via the solution of the moment-dependent Boltzmann equations (filled blue contours).
  }
  \label{fig:fig_app1}
\end{figure}
In this final section we present some robustness tests performed at the cosmological inference level, along with some further physical insights on the outcome of our analysis.

First of all, we further investigate the role of spectral distorsions precisely computed for the QCD axion phase space on the basis of its interactions with pions. In particular, we have tested how including spectral distortions in our analysis only at the level of the background, but not at the one of linear cosmological perturbations, yields a very similar bound on the QCD axion. In Fig.~\ref{fig:fig_app1} we support this statement analyzing the baseline cosmological dataset \textit{Planck}+\textit{lens}+BAO+SN in the case where: 
\begin{itemize}
\item The analysis is carried out using the full numerical phase-space distribution computed for the QCD axion;
\item The analysis is performed approximating the full numerical phase space with a Bose-Einstein one matching the contribution of extra-relativistic degrees of freedom underlain by the exact QCD axion distribution.  
\end{itemize}
Inspecting the marginalized posterior distribution for the QCD axion mass, we observe only a $\sim 1.5\%$ relative difference in the 95\% bound for the two cases. A similar conclusion holds for the constraint on the sum of neutrino masses. This test highlights how the main effect from a precise evaluation of the phase-space distribution for the QCD axion boils down to a shift of $N_{\rm eff}$, which has a significant imprint on the cosmological background evolution.

\begin{figure}[t!]
  \includegraphics[width=0.49\textwidth]{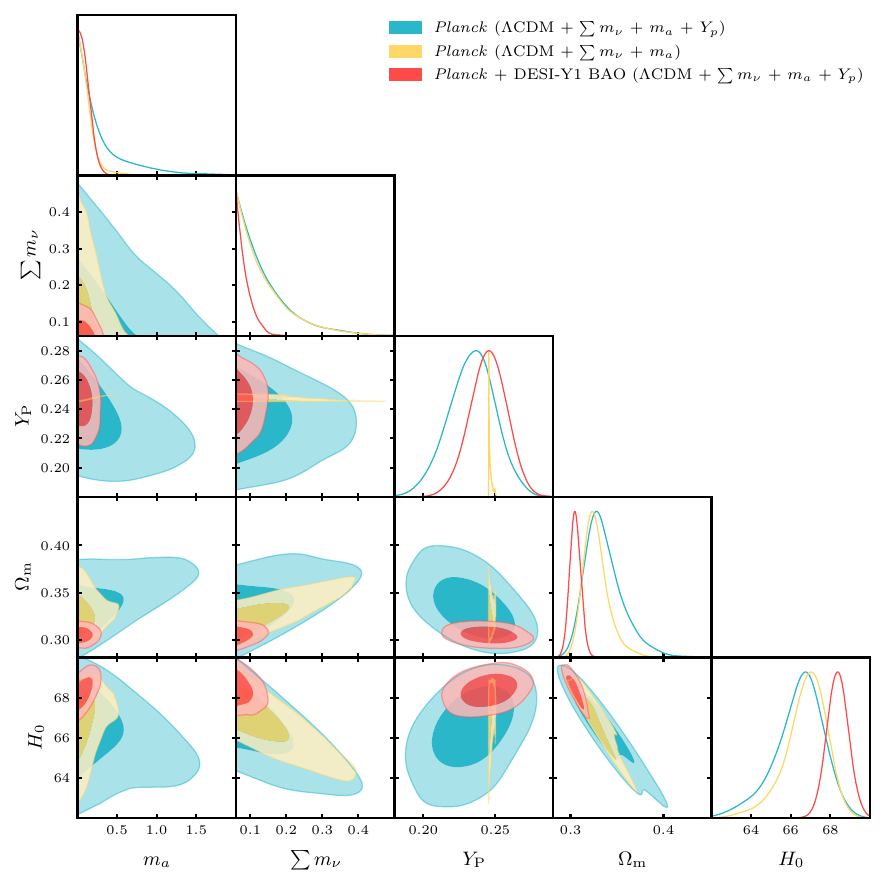}
  \includegraphics[width=0.49\textwidth]{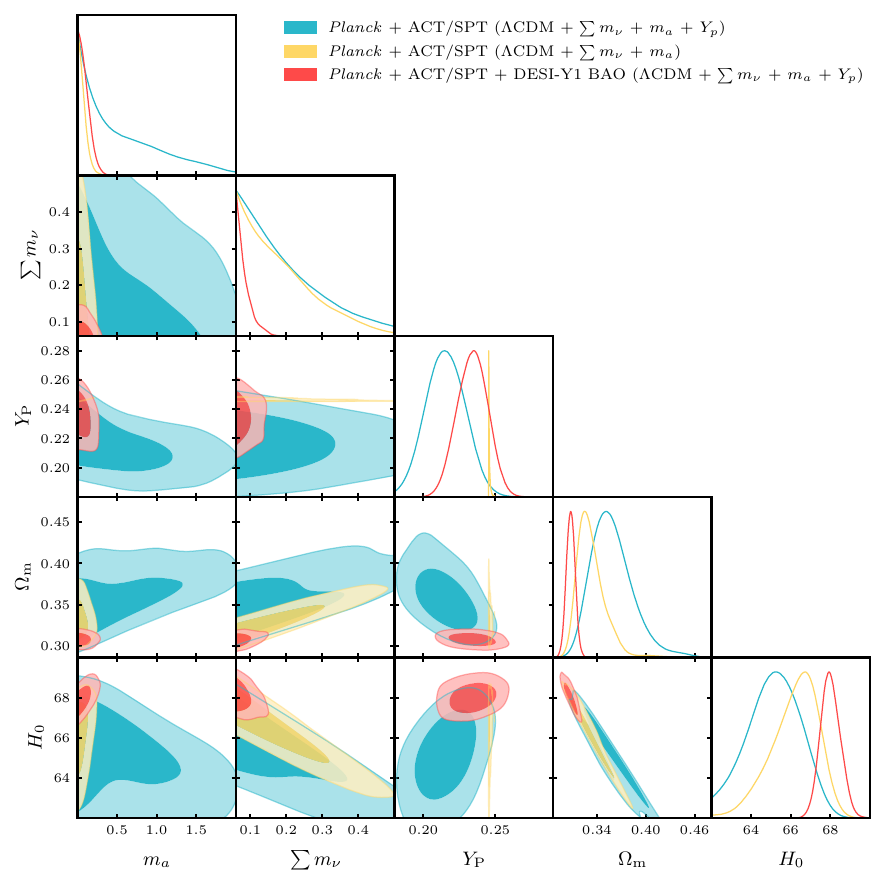}
  \caption{Triangle plots of different cosmological setups without (left panel) and with (right panel) ACT+SPT to highlight the importance of BAO and ACT+SPT versus the theoretical knowledge of helium-4 in order to constrain hot dark matter species. 
  }
  \label{fig:fig_app2}
\end{figure}

In Fig.~\ref{fig:fig_app2} we show the result of another in-depth analysis we perform here. We explore three different cosmological setups defined by the inclusion of the Planck likelihood with or without a theoretical prior from BBN theory related to the prediction of $Y_P$, and eventually with the addition of BAO from one-year data taking of DESI (left panel). We revisit these three scenarios also with the additional statistical weight provided by the cosmological likelihoods of CMB ground-based telescopes (right Figure). The highlight captured by Fig.~\ref{fig:fig_app2} is that the knowledge of $Y_P$ turns out to be crucial in our analysis when neither BAO data nor ACT+SPT measurements are included in the fit. In the absence of both set of data, in order to pin down a competitive bound on the QCD axion mass, Planck measurements need to be supplemented by a prior on $Y_P$ following from BBN theory. 

\begin{figure}[]
  \includegraphics[width=.77\textwidth]{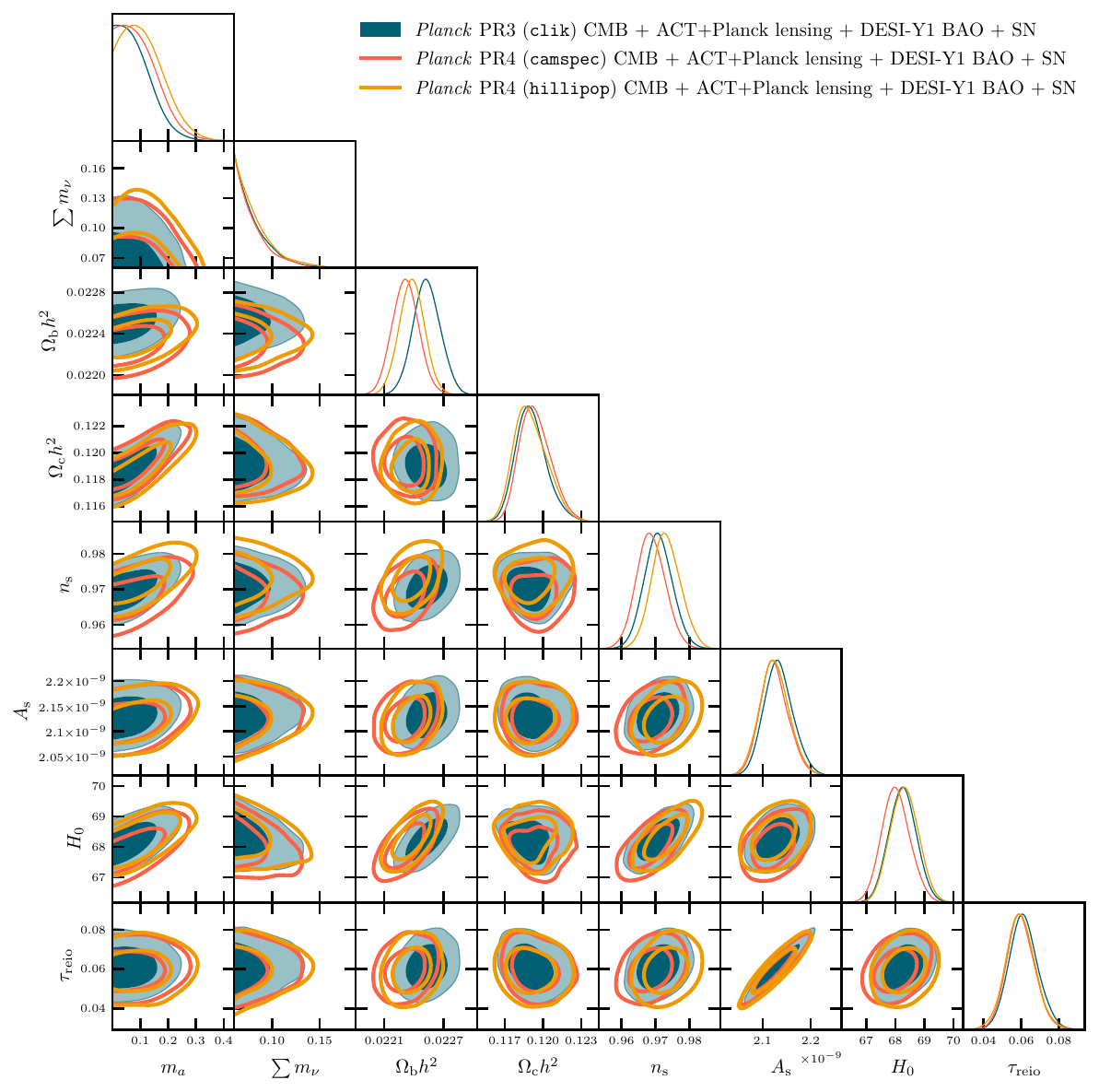}
  \caption{Triangle plots of the cosmological parameters inferred in our analysis for a cosmological dataset comprising CMB, BAO and SN data. Different colors highlight the different CMB likelihoods adopted with respect to the official release from Planck.}
  \label{fig:fig_app3}
\end{figure}

As a last important test of our analysis, we trade the official-release Planck likelihood, PR3 \texttt{clik}, for Planck PR4 \texttt{camspec}~\cite{Rosenberg:2022sdy} and also for Planck PR4 \texttt{hillipop}~\cite{Tristram:2023haj}. This exercise is instructive taking into account that:
\begin{itemize}
    \item The various CMB likelihood releases have varying lensing amplitudes, with Planck PR3 being the larger and \texttt{camspec} being consistent with unity, $A_L = 1.039 \pm 0.052$; 
    \item The $TT/TE/EE$ power spectra from the CMB likelihood \texttt{hillipop} show a slight preference for larger extra relativistic degrees of freedom, $N_{\rm eff}=3.08 \pm 0.17$, versus the baseline 2018 \textit{Planck} PR3, ($N_{\rm eff}=2.92 \pm 0.19$).
\end{itemize}
In Fig.~\ref{fig:fig_app3} we show the full triangle plot of the eight parameters inferred in our Bayesian analysis according to the dataset \textit{Planck} + lens + BAO + SN, for the three different releases of the Planck CMB likelihood. We observe that, in addition to the well-known effect on the cosmological baryon abundance, $\Omega_b h^2$, a change in the CMB likelihood also results in a slight relaxation of the constraints on the hot dark matter species that become non-relativistic first. Specifically, this pertains to the QCD axion in our study, whose marginalized posterior in the case of PR4 \texttt{hillipop} is also slightly affected by a tiny hint for $\Delta N_{\rm eff} > 0$. Quantitatively, under the aforementioned setup we find at 95\% probability: $m_a < 0.20$ eV (\texttt{clik}), $m_a < 0.22$ eV (\texttt{camspec}), $m_a < 0.25$ eV (\texttt{hillipop}), implying a degradation of the axion mass bound by approximately $\mathcal{O}(10\%)$ compared to the baseline presented in the main text.

\begin{figure}[]
\includegraphics[width=0.73\textwidth]{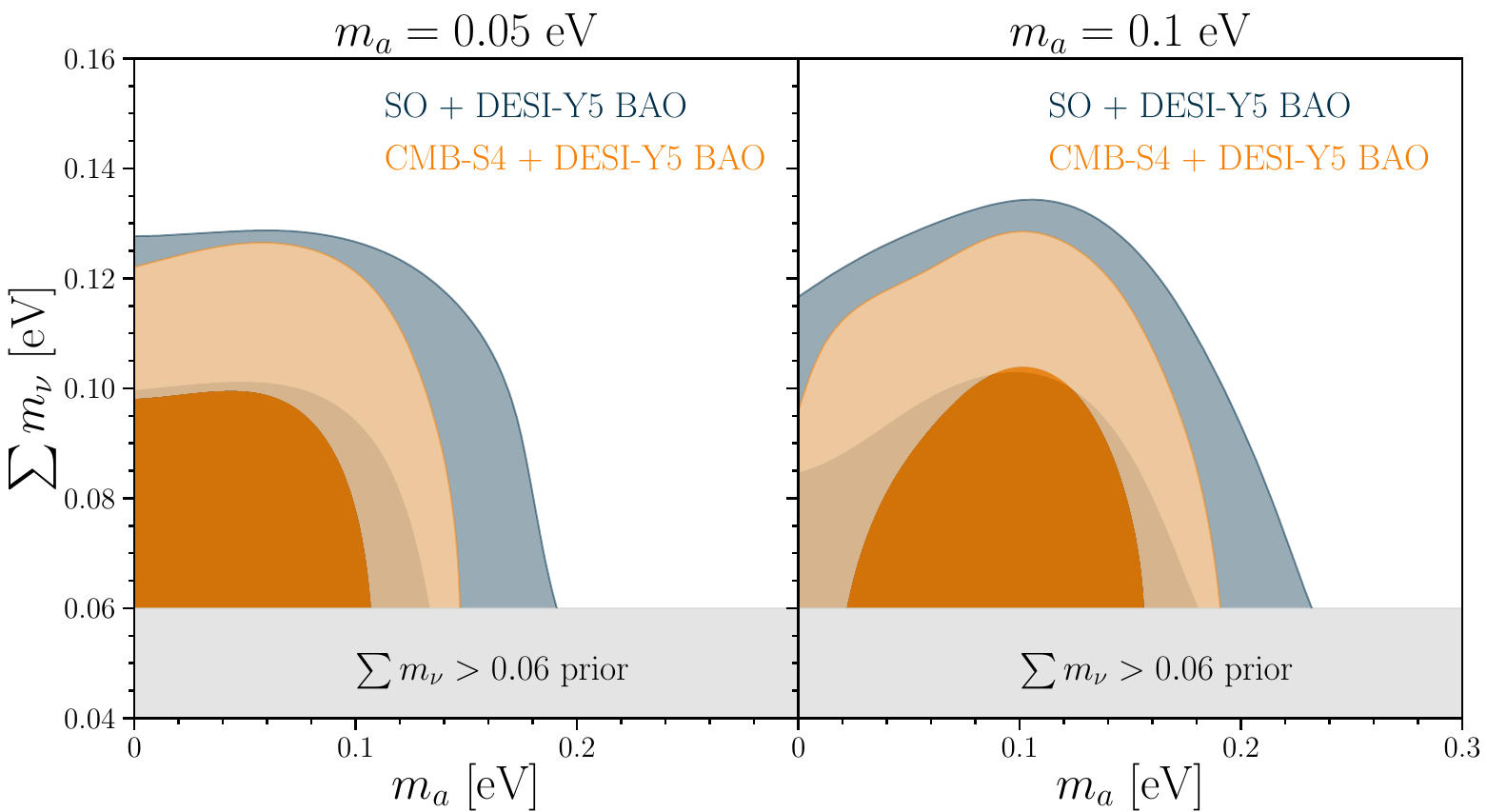}  \caption{$68\%$ and $95\%$ credible region for the axion mass $m_a$ and the sum of neutrino masses from future CMB/LSS surveys for an underlying cosmology with $\sum m_\nu=0.06$ eV and $m_a=0.05$ eV (left panel) and $m_a=0.1$ eV (right panel).
  }
\label{fig:sup_forecast}
\end{figure}

We conclude this section showing in  Fig.~\ref{fig:sup_forecast} the forecasts for the joint $m_{a}$-$\sum m_{\nu}$ 68\% and 95\% credible region obtained for a fiducial cosmology that assumes three equally massive neutrinos with total neutrino mass $\sum m_\nu=0.06$ eV and a QCD axion mass $m_a=0.05$, $0.1$ eV. This investigation explores the conditions under which future cosmological surveys might detect the QCD axion. As shown in Fig.~\ref{fig:sup_forecast}, it is evident that for QCD axion masses much smaller than 0.1 eV, even with CMB-S4 combined with DESI-Y5 BAO data, achieving a 2$\sigma$ detection will be challenging.

\clearpage
\twocolumngrid

\bibliography{biblio} 

\begin{thebibliography}{143}%
\makeatletter
\providecommand \@ifxundefined [1]{%
 \@ifx{#1\undefined}
}%
\providecommand \@ifnum [1]{%
 \ifnum #1\expandafter \@firstoftwo
 \else \expandafter \@secondoftwo
 \fi
}%
\providecommand \@ifx [1]{%
 \ifx #1\expandafter \@firstoftwo
 \else \expandafter \@secondoftwo
 \fi
}%
\providecommand \natexlab [1]{#1}%
\providecommand \enquote  [1]{``#1''}%
\providecommand \bibnamefont  [1]{#1}%
\providecommand \bibfnamefont [1]{#1}%
\providecommand \citenamefont [1]{#1}%
\providecommand \href@noop [0]{\@secondoftwo}%
\providecommand \href [0]{\begingroup \@sanitize@url \@href}%
\providecommand \@href[1]{\@@startlink{#1}\@@href}%
\providecommand \@@href[1]{\endgroup#1\@@endlink}%
\providecommand \@sanitize@url [0]{\catcode `\\12\catcode `\$12\catcode
  `\&12\catcode `\#12\catcode `\^12\catcode `\_12\catcode `\%12\relax}%
\providecommand \@@startlink[1]{}%
\providecommand \@@endlink[0]{}%
\providecommand \url  [0]{\begingroup\@sanitize@url \@url }%
\providecommand \@url [1]{\endgroup\@href {#1}{\urlprefix }}%
\providecommand \urlprefix  [0]{URL }%
\providecommand \Eprint [0]{\href }%
\providecommand \doibase [0]{http://dx.doi.org/}%
\providecommand \selectlanguage [0]{\@gobble}%
\providecommand \bibinfo  [0]{\@secondoftwo}%
\providecommand \bibfield  [0]{\@secondoftwo}%
\providecommand \translation [1]{[#1]}%
\providecommand \BibitemOpen [0]{}%
\providecommand \bibitemStop [0]{}%
\providecommand \bibitemNoStop [0]{.\EOS\space}%
\providecommand \EOS [0]{\spacefactor3000\relax}%
\providecommand \BibitemShut  [1]{\csname bibitem#1\endcsname}%
\let\auto@bib@innerbib\@empty
\bibitem [{\citenamefont {Peccei}\ and\ \citenamefont
  {Quinn}(1977{\natexlab{a}})}]{Peccei:1977hh}%
  \BibitemOpen
  \bibfield  {author} {\bibinfo {author} {\bibfnamefont {R.~D.}\ \bibnamefont
  {Peccei}}\ and\ \bibinfo {author} {\bibfnamefont {H.~R.}\ \bibnamefont
  {Quinn}},\ }\href {\doibase 10.1103/PhysRevLett.38.1440} {\bibfield
  {journal} {\bibinfo  {journal} {Phys. Rev. Lett.}\ }\textbf {\bibinfo
  {volume} {38}},\ \bibinfo {pages} {1440} (\bibinfo {year}
  {1977}{\natexlab{a}})}\BibitemShut {NoStop}%
\bibitem [{\citenamefont {Peccei}\ and\ \citenamefont
  {Quinn}(1977{\natexlab{b}})}]{Peccei:1977ur}%
  \BibitemOpen
  \bibfield  {author} {\bibinfo {author} {\bibfnamefont {R.~D.}\ \bibnamefont
  {Peccei}}\ and\ \bibinfo {author} {\bibfnamefont {H.~R.}\ \bibnamefont
  {Quinn}},\ }\href {\doibase 10.1103/PhysRevD.16.1791} {\bibfield  {journal}
  {\bibinfo  {journal} {Phys. Rev. D}\ }\textbf {\bibinfo {volume} {16}},\
  \bibinfo {pages} {1791} (\bibinfo {year} {1977}{\natexlab{b}})}\BibitemShut
  {NoStop}%
\bibitem [{\citenamefont {Weinberg}(1978)}]{Weinberg:1977ma}%
  \BibitemOpen
  \bibfield  {author} {\bibinfo {author} {\bibfnamefont {S.}~\bibnamefont
  {Weinberg}},\ }\href {\doibase 10.1103/PhysRevLett.40.223} {\bibfield
  {journal} {\bibinfo  {journal} {Phys. Rev. Lett.}\ }\textbf {\bibinfo
  {volume} {40}},\ \bibinfo {pages} {223} (\bibinfo {year} {1978})}\BibitemShut
  {NoStop}%
\bibitem [{\citenamefont {Wilczek}(1978)}]{Wilczek:1977pj}%
  \BibitemOpen
  \bibfield  {author} {\bibinfo {author} {\bibfnamefont {F.}~\bibnamefont
  {Wilczek}},\ }\href {\doibase 10.1103/PhysRevLett.40.279} {\bibfield
  {journal} {\bibinfo  {journal} {Phys. Rev. Lett.}\ }\textbf {\bibinfo
  {volume} {40}},\ \bibinfo {pages} {279} (\bibinfo {year} {1978})}\BibitemShut
  {NoStop}%
\bibitem [{\citenamefont {Baker}\ \emph {et~al.}(2006)\citenamefont {Baker}
  \emph {et~al.}}]{Baker:2006ts}%
  \BibitemOpen
  \bibfield  {author} {\bibinfo {author} {\bibfnamefont {C.~A.}\ \bibnamefont
  {Baker}} \emph {et~al.},\ }\href {\doibase 10.1103/PhysRevLett.97.131801}
  {\bibfield  {journal} {\bibinfo  {journal} {Phys. Rev. Lett.}\ }\textbf
  {\bibinfo {volume} {97}},\ \bibinfo {pages} {131801} (\bibinfo {year}
  {2006})},\ \Eprint {http://arxiv.org/abs/hep-ex/0602020}
  {arXiv:hep-ex/0602020} \BibitemShut {NoStop}%
\bibitem [{\citenamefont {Pendlebury}\ \emph {et~al.}(2015)\citenamefont
  {Pendlebury} \emph {et~al.}}]{Pendlebury:2015lrz}%
  \BibitemOpen
  \bibfield  {author} {\bibinfo {author} {\bibfnamefont {J.~M.}\ \bibnamefont
  {Pendlebury}} \emph {et~al.},\ }\href {\doibase 10.1103/PhysRevD.92.092003}
  {\bibfield  {journal} {\bibinfo  {journal} {Phys. Rev. D}\ }\textbf {\bibinfo
  {volume} {92}},\ \bibinfo {pages} {092003} (\bibinfo {year} {2015})},\
  \Eprint {http://arxiv.org/abs/1509.04411} {arXiv:1509.04411 [hep-ex]}
  \BibitemShut {NoStop}%
\bibitem [{\citenamefont {Abel}\ \emph {et~al.}(2020)\citenamefont {Abel} \emph
  {et~al.}}]{Abel:2020pzs}%
  \BibitemOpen
  \bibfield  {author} {\bibinfo {author} {\bibfnamefont {C.}~\bibnamefont
  {Abel}} \emph {et~al.},\ }\href {\doibase 10.1103/PhysRevLett.124.081803}
  {\bibfield  {journal} {\bibinfo  {journal} {Phys. Rev. Lett.}\ }\textbf
  {\bibinfo {volume} {124}},\ \bibinfo {pages} {081803} (\bibinfo {year}
  {2020})},\ \Eprint {http://arxiv.org/abs/2001.11966} {arXiv:2001.11966
  [hep-ex]} \BibitemShut {NoStop}%
\bibitem [{\citenamefont {Vafa}\ and\ \citenamefont
  {Witten}(1984)}]{Vafa:1984xg}%
  \BibitemOpen
  \bibfield  {author} {\bibinfo {author} {\bibfnamefont {C.}~\bibnamefont
  {Vafa}}\ and\ \bibinfo {author} {\bibfnamefont {E.}~\bibnamefont {Witten}},\
  }\href {\doibase 10.1103/PhysRevLett.53.535} {\bibfield  {journal} {\bibinfo
  {journal} {Phys. Rev. Lett.}\ }\textbf {\bibinfo {volume} {53}},\ \bibinfo
  {pages} {535} (\bibinfo {year} {1984})}\BibitemShut {NoStop}%
\bibitem [{\citenamefont {Dvali}(2022)}]{Dvali:2022fdv}%
  \BibitemOpen
  \bibfield  {author} {\bibinfo {author} {\bibfnamefont {G.}~\bibnamefont
  {Dvali}},\ }\href@noop {} {\  (\bibinfo {year} {2022})},\ \Eprint
  {http://arxiv.org/abs/2209.14219} {arXiv:2209.14219 [hep-ph]} \BibitemShut
  {NoStop}%
\bibitem [{\citenamefont {Preskill}\ \emph {et~al.}(1983)\citenamefont
  {Preskill}, \citenamefont {Wise},\ and\ \citenamefont
  {Wilczek}}]{Preskill:1982cy}%
  \BibitemOpen
  \bibfield  {author} {\bibinfo {author} {\bibfnamefont {J.}~\bibnamefont
  {Preskill}}, \bibinfo {author} {\bibfnamefont {M.~B.}\ \bibnamefont {Wise}},
  \ and\ \bibinfo {author} {\bibfnamefont {F.}~\bibnamefont {Wilczek}},\ }\href
  {\doibase 10.1016/0370-2693(83)90637-8} {\bibfield  {journal} {\bibinfo
  {journal} {Phys. Lett.}\ }\textbf {\bibinfo {volume} {120B}},\ \bibinfo
  {pages} {127} (\bibinfo {year} {1983})}\BibitemShut {NoStop}%
\bibitem [{\citenamefont {Abbott}\ and\ \citenamefont
  {Sikivie}(1983)}]{Abbott:1982af}%
  \BibitemOpen
  \bibfield  {author} {\bibinfo {author} {\bibfnamefont {L.~F.}\ \bibnamefont
  {Abbott}}\ and\ \bibinfo {author} {\bibfnamefont {P.}~\bibnamefont
  {Sikivie}},\ }\href {\doibase 10.1016/0370-2693(83)90638-X} {\bibfield
  {journal} {\bibinfo  {journal} {Phys. Lett.}\ }\textbf {\bibinfo {volume}
  {120B}},\ \bibinfo {pages} {133} (\bibinfo {year} {1983})}\BibitemShut
  {NoStop}%
\bibitem [{\citenamefont {Dine}\ and\ \citenamefont
  {Fischler}(1983)}]{Dine:1982ah}%
  \BibitemOpen
  \bibfield  {author} {\bibinfo {author} {\bibfnamefont {M.}~\bibnamefont
  {Dine}}\ and\ \bibinfo {author} {\bibfnamefont {W.}~\bibnamefont
  {Fischler}},\ }\href {\doibase 10.1016/0370-2693(83)90639-1} {\bibfield
  {journal} {\bibinfo  {journal} {Phys. Lett.}\ }\textbf {\bibinfo {volume}
  {120B}},\ \bibinfo {pages} {137} (\bibinfo {year} {1983})}\BibitemShut
  {NoStop}%
\bibitem [{\citenamefont {Davis}(1986)}]{Davis:1986xc}%
  \BibitemOpen
  \bibfield  {author} {\bibinfo {author} {\bibfnamefont {R.~L.}\ \bibnamefont
  {Davis}},\ }\href {\doibase 10.1016/0370-2693(86)90300-X} {\bibfield
  {journal} {\bibinfo  {journal} {Phys. Lett. B}\ }\textbf {\bibinfo {volume}
  {180}},\ \bibinfo {pages} {225} (\bibinfo {year} {1986})}\BibitemShut
  {NoStop}%
\bibitem [{\citenamefont {Turner}(1987)}]{Turner:1986tb}%
  \BibitemOpen
  \bibfield  {author} {\bibinfo {author} {\bibfnamefont {M.~S.}\ \bibnamefont
  {Turner}},\ }\href {\doibase 10.1103/PhysRevLett.59.2489} {\bibfield
  {journal} {\bibinfo  {journal} {Phys. Rev. Lett.}\ }\textbf {\bibinfo
  {volume} {59}},\ \bibinfo {pages} {2489} (\bibinfo {year} {1987})},\ \bibinfo
  {note} {[Erratum: Phys.Rev.Lett. 60, 1101 (1988)]}\BibitemShut {NoStop}%
\bibitem [{\citenamefont {Kolb}\ and\ \citenamefont
  {Turner}(1990)}]{Kolb:1990vq}%
  \BibitemOpen
  \bibfield  {author} {\bibinfo {author} {\bibfnamefont {E.~W.}\ \bibnamefont
  {Kolb}}\ and\ \bibinfo {author} {\bibfnamefont {M.~S.}\ \bibnamefont
  {Turner}},\ }\href@noop {} {\emph {\bibinfo {title} {{The Early
  Universe}}}},\ Vol.~\bibinfo {volume} {69}\ (\bibinfo {year}
  {1990})\BibitemShut {NoStop}%
\bibitem [{\citenamefont {Berezhiani}\ \emph {et~al.}(1992)\citenamefont
  {Berezhiani}, \citenamefont {Sakharov},\ and\ \citenamefont
  {Khlopov}}]{Berezhiani:1992rk}%
  \BibitemOpen
  \bibfield  {author} {\bibinfo {author} {\bibfnamefont {Z.}~\bibnamefont
  {Berezhiani}}, \bibinfo {author} {\bibfnamefont {A.}~\bibnamefont
  {Sakharov}}, \ and\ \bibinfo {author} {\bibfnamefont {M.}~\bibnamefont
  {Khlopov}},\ }\href@noop {} {\bibfield  {journal} {\bibinfo  {journal} {Sov.
  J. Nucl. Phys.}\ }\textbf {\bibinfo {volume} {55}},\ \bibinfo {pages} {1063}
  (\bibinfo {year} {1992})}\BibitemShut {NoStop}%
\bibitem [{\citenamefont {Chang}\ and\ \citenamefont
  {Choi}(1993)}]{Chang:1993gm}%
  \BibitemOpen
  \bibfield  {author} {\bibinfo {author} {\bibfnamefont {S.}~\bibnamefont
  {Chang}}\ and\ \bibinfo {author} {\bibfnamefont {K.}~\bibnamefont {Choi}},\
  }\href {\doibase 10.1016/0370-2693(93)90656-3} {\bibfield  {journal}
  {\bibinfo  {journal} {Phys. Lett. B}\ }\textbf {\bibinfo {volume} {316}},\
  \bibinfo {pages} {51} (\bibinfo {year} {1993})},\ \Eprint
  {http://arxiv.org/abs/hep-ph/9306216} {arXiv:hep-ph/9306216} \BibitemShut
  {NoStop}%
\bibitem [{\citenamefont {Baumann}\ \emph {et~al.}(2016)\citenamefont
  {Baumann}, \citenamefont {Green},\ and\ \citenamefont
  {Wallisch}}]{Baumann:2016wac}%
  \BibitemOpen
  \bibfield  {author} {\bibinfo {author} {\bibfnamefont {D.}~\bibnamefont
  {Baumann}}, \bibinfo {author} {\bibfnamefont {D.}~\bibnamefont {Green}}, \
  and\ \bibinfo {author} {\bibfnamefont {B.}~\bibnamefont {Wallisch}},\ }\href
  {\doibase 10.1103/PhysRevLett.117.171301} {\bibfield  {journal} {\bibinfo
  {journal} {Phys. Rev. Lett.}\ }\textbf {\bibinfo {volume} {117}},\ \bibinfo
  {pages} {171301} (\bibinfo {year} {2016})},\ \Eprint
  {http://arxiv.org/abs/1604.08614} {arXiv:1604.08614 [astro-ph.CO]}
  \BibitemShut {NoStop}%
\bibitem [{\citenamefont {D'Eramo}\ \emph
  {et~al.}(2022{\natexlab{a}})\citenamefont {D'Eramo}, \citenamefont
  {Di~Valentino}, \citenamefont {Giar\`e}, \citenamefont {Hajkarim},
  \citenamefont {Melchiorri}, \citenamefont {Mena}, \citenamefont {Renzi},\
  and\ \citenamefont {Yun}}]{DEramo:2022nvb}%
  \BibitemOpen
  \bibfield  {author} {\bibinfo {author} {\bibfnamefont {F.}~\bibnamefont
  {D'Eramo}}, \bibinfo {author} {\bibfnamefont {E.}~\bibnamefont
  {Di~Valentino}}, \bibinfo {author} {\bibfnamefont {W.}~\bibnamefont
  {Giar\`e}}, \bibinfo {author} {\bibfnamefont {F.}~\bibnamefont {Hajkarim}},
  \bibinfo {author} {\bibfnamefont {A.}~\bibnamefont {Melchiorri}}, \bibinfo
  {author} {\bibfnamefont {O.}~\bibnamefont {Mena}}, \bibinfo {author}
  {\bibfnamefont {F.}~\bibnamefont {Renzi}}, \ and\ \bibinfo {author}
  {\bibfnamefont {S.}~\bibnamefont {Yun}},\ }\href {\doibase
  10.1088/1475-7516/2022/09/022} {\bibfield  {journal} {\bibinfo  {journal}
  {JCAP}\ }\textbf {\bibinfo {volume} {09}},\ \bibinfo {pages} {022} (\bibinfo
  {year} {2022}{\natexlab{a}})},\ \Eprint {http://arxiv.org/abs/2205.07849}
  {arXiv:2205.07849 [astro-ph.CO]} \BibitemShut {NoStop}%
\bibitem [{\citenamefont {Aoki}\ \emph {et~al.}(2006)\citenamefont {Aoki},
  \citenamefont {Fodor}, \citenamefont {Katz},\ and\ \citenamefont
  {Szabo}}]{Aoki:2006br}%
  \BibitemOpen
  \bibfield  {author} {\bibinfo {author} {\bibfnamefont {Y.}~\bibnamefont
  {Aoki}}, \bibinfo {author} {\bibfnamefont {Z.}~\bibnamefont {Fodor}},
  \bibinfo {author} {\bibfnamefont {S.~D.}\ \bibnamefont {Katz}}, \ and\
  \bibinfo {author} {\bibfnamefont {K.~K.}\ \bibnamefont {Szabo}},\ }\href
  {\doibase 10.1016/j.physletb.2006.10.021} {\bibfield  {journal} {\bibinfo
  {journal} {Phys. Lett. B}\ }\textbf {\bibinfo {volume} {643}},\ \bibinfo
  {pages} {46} (\bibinfo {year} {2006})},\ \Eprint
  {http://arxiv.org/abs/hep-lat/0609068} {arXiv:hep-lat/0609068} \BibitemShut
  {NoStop}%
\bibitem [{\citenamefont {Borsanyi}\ \emph {et~al.}(2010)\citenamefont
  {Borsanyi}, \citenamefont {Fodor}, \citenamefont {Hoelbling}, \citenamefont
  {Katz}, \citenamefont {Krieg}, \citenamefont {Ratti},\ and\ \citenamefont
  {Szabo}}]{Borsanyi:2010bp}%
  \BibitemOpen
  \bibfield  {author} {\bibinfo {author} {\bibfnamefont {S.}~\bibnamefont
  {Borsanyi}}, \bibinfo {author} {\bibfnamefont {Z.}~\bibnamefont {Fodor}},
  \bibinfo {author} {\bibfnamefont {C.}~\bibnamefont {Hoelbling}}, \bibinfo
  {author} {\bibfnamefont {S.~D.}\ \bibnamefont {Katz}}, \bibinfo {author}
  {\bibfnamefont {S.}~\bibnamefont {Krieg}}, \bibinfo {author} {\bibfnamefont
  {C.}~\bibnamefont {Ratti}}, \ and\ \bibinfo {author} {\bibfnamefont {K.~K.}\
  \bibnamefont {Szabo}} (\bibinfo {collaboration} {Wuppertal-Budapest}),\
  }\href {\doibase 10.1007/JHEP09(2010)073} {\bibfield  {journal} {\bibinfo
  {journal} {JHEP}\ }\textbf {\bibinfo {volume} {09}},\ \bibinfo {pages} {073}
  (\bibinfo {year} {2010})},\ \Eprint {http://arxiv.org/abs/1005.3508}
  {arXiv:1005.3508 [hep-lat]} \BibitemShut {NoStop}%
\bibitem [{\citenamefont {Bazavov}\ \emph {et~al.}(2012)\citenamefont {Bazavov}
  \emph {et~al.}}]{Bazavov:2011nk}%
  \BibitemOpen
  \bibfield  {author} {\bibinfo {author} {\bibfnamefont {A.}~\bibnamefont
  {Bazavov}} \emph {et~al.},\ }\href {\doibase 10.1103/PhysRevD.85.054503}
  {\bibfield  {journal} {\bibinfo  {journal} {Phys. Rev. D}\ }\textbf {\bibinfo
  {volume} {85}},\ \bibinfo {pages} {054503} (\bibinfo {year} {2012})},\
  \Eprint {http://arxiv.org/abs/1111.1710} {arXiv:1111.1710 [hep-lat]}
  \BibitemShut {NoStop}%
\bibitem [{\citenamefont {Grilli~di Cortona}\ \emph {et~al.}(2016)\citenamefont
  {Grilli~di Cortona}, \citenamefont {Hardy}, \citenamefont {Pardo~Vega},\ and\
  \citenamefont {Villadoro}}]{diCortona:2015ldu}%
  \BibitemOpen
  \bibfield  {author} {\bibinfo {author} {\bibfnamefont {G.}~\bibnamefont
  {Grilli~di Cortona}}, \bibinfo {author} {\bibfnamefont {E.}~\bibnamefont
  {Hardy}}, \bibinfo {author} {\bibfnamefont {J.}~\bibnamefont {Pardo~Vega}}, \
  and\ \bibinfo {author} {\bibfnamefont {G.}~\bibnamefont {Villadoro}},\ }\href
  {\doibase 10.1007/JHEP01(2016)034} {\bibfield  {journal} {\bibinfo  {journal}
  {JHEP}\ }\textbf {\bibinfo {volume} {01}},\ \bibinfo {pages} {034} (\bibinfo
  {year} {2016})},\ \Eprint {http://arxiv.org/abs/1511.02867} {arXiv:1511.02867
  [hep-ph]} \BibitemShut {NoStop}%
\bibitem [{\citenamefont {Gorghetto}\ and\ \citenamefont
  {Villadoro}(2019)}]{Gorghetto:2018ocs}%
  \BibitemOpen
  \bibfield  {author} {\bibinfo {author} {\bibfnamefont {M.}~\bibnamefont
  {Gorghetto}}\ and\ \bibinfo {author} {\bibfnamefont {G.}~\bibnamefont
  {Villadoro}},\ }\href {\doibase 10.1007/JHEP03(2019)033} {\bibfield
  {journal} {\bibinfo  {journal} {JHEP}\ }\textbf {\bibinfo {volume} {03}},\
  \bibinfo {pages} {033} (\bibinfo {year} {2019})},\ \Eprint
  {http://arxiv.org/abs/1812.01008} {arXiv:1812.01008 [hep-ph]} \BibitemShut
  {NoStop}%
\bibitem [{\citenamefont {Aghanim}\ \emph
  {et~al.}(2020{\natexlab{a}})\citenamefont {Aghanim} \emph
  {et~al.}}]{Aghanim:2018eyx}%
  \BibitemOpen
  \bibfield  {author} {\bibinfo {author} {\bibfnamefont {N.}~\bibnamefont
  {Aghanim}} \emph {et~al.} (\bibinfo {collaboration} {Planck}),\ }\href
  {\doibase 10.1051/0004-6361/201833910} {\bibfield  {journal} {\bibinfo
  {journal} {Astron. Astrophys.}\ }\textbf {\bibinfo {volume} {641}},\ \bibinfo
  {pages} {A6} (\bibinfo {year} {2020}{\natexlab{a}})},\ \Eprint
  {http://arxiv.org/abs/1807.06209} {arXiv:1807.06209 [astro-ph.CO]}
  \BibitemShut {NoStop}%
\bibitem [{\citenamefont {D'Eramo}\ \emph {et~al.}(2021)\citenamefont
  {D'Eramo}, \citenamefont {Hajkarim},\ and\ \citenamefont
  {Yun}}]{DEramo:2021lgb}%
  \BibitemOpen
  \bibfield  {author} {\bibinfo {author} {\bibfnamefont {F.}~\bibnamefont
  {D'Eramo}}, \bibinfo {author} {\bibfnamefont {F.}~\bibnamefont {Hajkarim}}, \
  and\ \bibinfo {author} {\bibfnamefont {S.}~\bibnamefont {Yun}},\ }\href
  {\doibase 10.1007/JHEP10(2021)224} {\bibfield  {journal} {\bibinfo  {journal}
  {JHEP}\ }\textbf {\bibinfo {volume} {10}},\ \bibinfo {pages} {224} (\bibinfo
  {year} {2021})},\ \Eprint {http://arxiv.org/abs/2108.05371} {arXiv:2108.05371
  [hep-ph]} \BibitemShut {NoStop}%
\bibitem [{\citenamefont {Abazajian}\ \emph
  {et~al.}(2019{\natexlab{a}})\citenamefont {Abazajian} \emph
  {et~al.}}]{Abazajian:2019eic}%
  \BibitemOpen
  \bibfield  {author} {\bibinfo {author} {\bibfnamefont {K.}~\bibnamefont
  {Abazajian}} \emph {et~al.},\ }\href@noop {} {\  (\bibinfo {year}
  {2019}{\natexlab{a}})},\ \Eprint {http://arxiv.org/abs/1907.04473}
  {arXiv:1907.04473 [astro-ph.IM]} \BibitemShut {NoStop}%
\bibitem [{\citenamefont {Aiola}\ \emph {et~al.}(2022)\citenamefont {Aiola}
  \emph {et~al.}}]{CMB-HD:2022bsz}%
  \BibitemOpen
  \bibfield  {author} {\bibinfo {author} {\bibfnamefont {S.}~\bibnamefont
  {Aiola}} \emph {et~al.} (\bibinfo {collaboration} {CMB-HD}),\ }\href@noop {}
  {\  (\bibinfo {year} {2022})},\ \Eprint {http://arxiv.org/abs/2203.05728}
  {arXiv:2203.05728 [astro-ph.CO]} \BibitemShut {NoStop}%
\bibitem [{\citenamefont {Carenza}\ \emph {et~al.}(2019)\citenamefont
  {Carenza}, \citenamefont {Fischer}, \citenamefont {Giannotti}, \citenamefont
  {Guo}, \citenamefont {Mart\'\i{}nez-Pinedo},\ and\ \citenamefont
  {Mirizzi}}]{Carenza:2019pxu}%
  \BibitemOpen
  \bibfield  {author} {\bibinfo {author} {\bibfnamefont {P.}~\bibnamefont
  {Carenza}}, \bibinfo {author} {\bibfnamefont {T.}~\bibnamefont {Fischer}},
  \bibinfo {author} {\bibfnamefont {M.}~\bibnamefont {Giannotti}}, \bibinfo
  {author} {\bibfnamefont {G.}~\bibnamefont {Guo}}, \bibinfo {author}
  {\bibfnamefont {G.}~\bibnamefont {Mart\'\i{}nez-Pinedo}}, \ and\ \bibinfo
  {author} {\bibfnamefont {A.}~\bibnamefont {Mirizzi}},\ }\href {\doibase
  10.1088/1475-7516/2019/10/016} {\bibfield  {journal} {\bibinfo  {journal}
  {JCAP}\ }\textbf {\bibinfo {volume} {10}},\ \bibinfo {pages} {016} (\bibinfo
  {year} {2019})},\ \bibinfo {note} {[Erratum: JCAP 05, E01 (2020)]},\ \Eprint
  {http://arxiv.org/abs/1906.11844} {arXiv:1906.11844 [hep-ph]} \BibitemShut
  {NoStop}%
\bibitem [{\citenamefont {Leinson}(2021)}]{Leinson:2021ety}%
  \BibitemOpen
  \bibfield  {author} {\bibinfo {author} {\bibfnamefont {L.~B.}\ \bibnamefont
  {Leinson}},\ }\href {\doibase 10.1088/1475-7516/2021/09/001} {\bibfield
  {journal} {\bibinfo  {journal} {JCAP}\ }\textbf {\bibinfo {volume} {09}},\
  \bibinfo {pages} {001} (\bibinfo {year} {2021})},\ \Eprint
  {http://arxiv.org/abs/2105.14745} {arXiv:2105.14745 [hep-ph]} \BibitemShut
  {NoStop}%
\bibitem [{\citenamefont {Raffelt}(2008)}]{Raffelt:2006cw}%
  \BibitemOpen
  \bibfield  {author} {\bibinfo {author} {\bibfnamefont {G.~G.}\ \bibnamefont
  {Raffelt}},\ }\href {\doibase 10.1007/978-3-540-73518-2_3} {\bibfield
  {journal} {\bibinfo  {journal} {Lect. Notes Phys.}\ }\textbf {\bibinfo
  {volume} {741}},\ \bibinfo {pages} {51} (\bibinfo {year} {2008})},\ \Eprint
  {http://arxiv.org/abs/hep-ph/0611350} {arXiv:hep-ph/0611350} \BibitemShut
  {NoStop}%
\bibitem [{\citenamefont {Miller~Bertolami}\ \emph {et~al.}(2014)\citenamefont
  {Miller~Bertolami}, \citenamefont {Melendez}, \citenamefont {Althaus},\ and\
  \citenamefont {Isern}}]{MillerBertolami:2014rka}%
  \BibitemOpen
  \bibfield  {author} {\bibinfo {author} {\bibfnamefont {M.~M.}\ \bibnamefont
  {Miller~Bertolami}}, \bibinfo {author} {\bibfnamefont {B.~E.}\ \bibnamefont
  {Melendez}}, \bibinfo {author} {\bibfnamefont {L.~G.}\ \bibnamefont
  {Althaus}}, \ and\ \bibinfo {author} {\bibfnamefont {J.}~\bibnamefont
  {Isern}},\ }\href {\doibase 10.1088/1475-7516/2014/10/069} {\bibfield
  {journal} {\bibinfo  {journal} {JCAP}\ }\textbf {\bibinfo {volume} {10}},\
  \bibinfo {pages} {069} (\bibinfo {year} {2014})},\ \Eprint
  {http://arxiv.org/abs/1406.7712} {arXiv:1406.7712 [hep-ph]} \BibitemShut
  {NoStop}%
\bibitem [{\citenamefont {Ayala}\ \emph {et~al.}(2014)\citenamefont {Ayala},
  \citenamefont {Dominguez}, \citenamefont {Giannotti}, \citenamefont
  {Mirizzi},\ and\ \citenamefont {Straniero}}]{Ayala:2014pea}%
  \BibitemOpen
  \bibfield  {author} {\bibinfo {author} {\bibfnamefont {A.}~\bibnamefont
  {Ayala}}, \bibinfo {author} {\bibfnamefont {I.}~\bibnamefont {Dominguez}},
  \bibinfo {author} {\bibfnamefont {M.}~\bibnamefont {Giannotti}}, \bibinfo
  {author} {\bibfnamefont {A.}~\bibnamefont {Mirizzi}}, \ and\ \bibinfo
  {author} {\bibfnamefont {O.}~\bibnamefont {Straniero}},\ }\href {\doibase
  10.1103/PhysRevLett.113.191302} {\bibfield  {journal} {\bibinfo  {journal}
  {Phys. Rev. Lett.}\ }\textbf {\bibinfo {volume} {113}},\ \bibinfo {pages}
  {191302} (\bibinfo {year} {2014})},\ \Eprint {http://arxiv.org/abs/1406.6053}
  {arXiv:1406.6053 [astro-ph.SR]} \BibitemShut {NoStop}%
\bibitem [{\citenamefont {Chang}\ \emph {et~al.}(2018)\citenamefont {Chang},
  \citenamefont {Essig},\ and\ \citenamefont {McDermott}}]{Chang:2018rso}%
  \BibitemOpen
  \bibfield  {author} {\bibinfo {author} {\bibfnamefont {J.~H.}\ \bibnamefont
  {Chang}}, \bibinfo {author} {\bibfnamefont {R.}~\bibnamefont {Essig}}, \ and\
  \bibinfo {author} {\bibfnamefont {S.~D.}\ \bibnamefont {McDermott}},\ }\href
  {\doibase 10.1007/JHEP09(2018)051} {\bibfield  {journal} {\bibinfo  {journal}
  {JHEP}\ }\textbf {\bibinfo {volume} {09}},\ \bibinfo {pages} {051} (\bibinfo
  {year} {2018})},\ \Eprint {http://arxiv.org/abs/1803.00993} {arXiv:1803.00993
  [hep-ph]} \BibitemShut {NoStop}%
\bibitem [{\citenamefont {Bar}\ \emph {et~al.}(2020)\citenamefont {Bar},
  \citenamefont {Blum},\ and\ \citenamefont {D'Amico}}]{Bar:2019ifz}%
  \BibitemOpen
  \bibfield  {author} {\bibinfo {author} {\bibfnamefont {N.}~\bibnamefont
  {Bar}}, \bibinfo {author} {\bibfnamefont {K.}~\bibnamefont {Blum}}, \ and\
  \bibinfo {author} {\bibfnamefont {G.}~\bibnamefont {D'Amico}},\ }\href
  {\doibase 10.1103/PhysRevD.101.123025} {\bibfield  {journal} {\bibinfo
  {journal} {Phys. Rev. D}\ }\textbf {\bibinfo {volume} {101}},\ \bibinfo
  {pages} {123025} (\bibinfo {year} {2020})},\ \Eprint
  {http://arxiv.org/abs/1907.05020} {arXiv:1907.05020 [hep-ph]} \BibitemShut
  {NoStop}%
\bibitem [{\citenamefont {Carenza}\ \emph {et~al.}(2021)\citenamefont
  {Carenza}, \citenamefont {Fore}, \citenamefont {Giannotti}, \citenamefont
  {Mirizzi},\ and\ \citenamefont {Reddy}}]{Carenza:2020cis}%
  \BibitemOpen
  \bibfield  {author} {\bibinfo {author} {\bibfnamefont {P.}~\bibnamefont
  {Carenza}}, \bibinfo {author} {\bibfnamefont {B.}~\bibnamefont {Fore}},
  \bibinfo {author} {\bibfnamefont {M.}~\bibnamefont {Giannotti}}, \bibinfo
  {author} {\bibfnamefont {A.}~\bibnamefont {Mirizzi}}, \ and\ \bibinfo
  {author} {\bibfnamefont {S.}~\bibnamefont {Reddy}},\ }\href {\doibase
  10.1103/PhysRevLett.126.071102} {\bibfield  {journal} {\bibinfo  {journal}
  {Phys. Rev. Lett.}\ }\textbf {\bibinfo {volume} {126}},\ \bibinfo {pages}
  {071102} (\bibinfo {year} {2021})},\ \Eprint
  {http://arxiv.org/abs/2010.02943} {arXiv:2010.02943 [hep-ph]} \BibitemShut
  {NoStop}%
\bibitem [{\citenamefont {Buschmann}\ \emph {et~al.}(2022)\citenamefont
  {Buschmann}, \citenamefont {Dessert}, \citenamefont {Foster}, \citenamefont
  {Long},\ and\ \citenamefont {Safdi}}]{Buschmann:2021juv}%
  \BibitemOpen
  \bibfield  {author} {\bibinfo {author} {\bibfnamefont {M.}~\bibnamefont
  {Buschmann}}, \bibinfo {author} {\bibfnamefont {C.}~\bibnamefont {Dessert}},
  \bibinfo {author} {\bibfnamefont {J.~W.}\ \bibnamefont {Foster}}, \bibinfo
  {author} {\bibfnamefont {A.~J.}\ \bibnamefont {Long}}, \ and\ \bibinfo
  {author} {\bibfnamefont {B.~R.}\ \bibnamefont {Safdi}},\ }\href {\doibase
  10.1103/PhysRevLett.128.091102} {\bibfield  {journal} {\bibinfo  {journal}
  {Phys. Rev. Lett.}\ }\textbf {\bibinfo {volume} {128}},\ \bibinfo {pages}
  {091102} (\bibinfo {year} {2022})},\ \Eprint
  {http://arxiv.org/abs/2111.09892} {arXiv:2111.09892 [hep-ph]} \BibitemShut
  {NoStop}%
\bibitem [{\citenamefont {Dolan}\ \emph {et~al.}(2022)\citenamefont {Dolan},
  \citenamefont {Hiskens},\ and\ \citenamefont {Volkas}}]{Dolan:2022kul}%
  \BibitemOpen
  \bibfield  {author} {\bibinfo {author} {\bibfnamefont {M.~J.}\ \bibnamefont
  {Dolan}}, \bibinfo {author} {\bibfnamefont {F.~J.}\ \bibnamefont {Hiskens}},
  \ and\ \bibinfo {author} {\bibfnamefont {R.~R.}\ \bibnamefont {Volkas}},\
  }\href {\doibase 10.1088/1475-7516/2022/10/096} {\bibfield  {journal}
  {\bibinfo  {journal} {JCAP}\ }\textbf {\bibinfo {volume} {10}},\ \bibinfo
  {pages} {096} (\bibinfo {year} {2022})},\ \Eprint
  {http://arxiv.org/abs/2207.03102} {arXiv:2207.03102 [hep-ph]} \BibitemShut
  {NoStop}%
\bibitem [{\citenamefont {Lella}\ \emph {et~al.}(2023)\citenamefont {Lella},
  \citenamefont {Carenza}, \citenamefont {Co'}, \citenamefont {Lucente},
  \citenamefont {Giannotti}, \citenamefont {Mirizzi},\ and\ \citenamefont
  {Rauscher}}]{Lella:2023bfb}%
  \BibitemOpen
  \bibfield  {author} {\bibinfo {author} {\bibfnamefont {A.}~\bibnamefont
  {Lella}}, \bibinfo {author} {\bibfnamefont {P.}~\bibnamefont {Carenza}},
  \bibinfo {author} {\bibfnamefont {G.}~\bibnamefont {Co'}}, \bibinfo {author}
  {\bibfnamefont {G.}~\bibnamefont {Lucente}}, \bibinfo {author} {\bibfnamefont
  {M.}~\bibnamefont {Giannotti}}, \bibinfo {author} {\bibfnamefont
  {A.}~\bibnamefont {Mirizzi}}, \ and\ \bibinfo {author} {\bibfnamefont
  {T.}~\bibnamefont {Rauscher}},\ }\href@noop {} {\  (\bibinfo {year}
  {2023})},\ \Eprint {http://arxiv.org/abs/2306.01048} {arXiv:2306.01048
  [hep-ph]} \BibitemShut {NoStop}%
\bibitem [{\citenamefont {Hannestad}\ \emph {et~al.}(2005)\citenamefont
  {Hannestad}, \citenamefont {Mirizzi},\ and\ \citenamefont
  {Raffelt}}]{Hannestad:2005df}%
  \BibitemOpen
  \bibfield  {author} {\bibinfo {author} {\bibfnamefont {S.}~\bibnamefont
  {Hannestad}}, \bibinfo {author} {\bibfnamefont {A.}~\bibnamefont {Mirizzi}},
  \ and\ \bibinfo {author} {\bibfnamefont {G.}~\bibnamefont {Raffelt}},\ }\href
  {\doibase 10.1088/1475-7516/2005/07/002} {\bibfield  {journal} {\bibinfo
  {journal} {JCAP}\ }\textbf {\bibinfo {volume} {07}},\ \bibinfo {pages} {002}
  (\bibinfo {year} {2005})},\ \Eprint {http://arxiv.org/abs/hep-ph/0504059}
  {arXiv:hep-ph/0504059} \BibitemShut {NoStop}%
\bibitem [{\citenamefont {Melchiorri}\ \emph {et~al.}(2007)\citenamefont
  {Melchiorri}, \citenamefont {Mena},\ and\ \citenamefont
  {Slosar}}]{Melchiorri:2007cd}%
  \BibitemOpen
  \bibfield  {author} {\bibinfo {author} {\bibfnamefont {A.}~\bibnamefont
  {Melchiorri}}, \bibinfo {author} {\bibfnamefont {O.}~\bibnamefont {Mena}}, \
  and\ \bibinfo {author} {\bibfnamefont {A.}~\bibnamefont {Slosar}},\ }\href
  {\doibase 10.1103/PhysRevD.76.041303} {\bibfield  {journal} {\bibinfo
  {journal} {Phys. Rev. D}\ }\textbf {\bibinfo {volume} {76}},\ \bibinfo
  {pages} {041303} (\bibinfo {year} {2007})},\ \Eprint
  {http://arxiv.org/abs/0705.2695} {arXiv:0705.2695 [astro-ph]} \BibitemShut
  {NoStop}%
\bibitem [{\citenamefont {Hannestad}\ \emph {et~al.}(2007)\citenamefont
  {Hannestad}, \citenamefont {Mirizzi}, \citenamefont {Raffelt},\ and\
  \citenamefont {Wong}}]{Hannestad:2007dd}%
  \BibitemOpen
  \bibfield  {author} {\bibinfo {author} {\bibfnamefont {S.}~\bibnamefont
  {Hannestad}}, \bibinfo {author} {\bibfnamefont {A.}~\bibnamefont {Mirizzi}},
  \bibinfo {author} {\bibfnamefont {G.~G.}\ \bibnamefont {Raffelt}}, \ and\
  \bibinfo {author} {\bibfnamefont {Y.~Y.~Y.}\ \bibnamefont {Wong}},\ }\href
  {\doibase 10.1088/1475-7516/2007/08/015} {\bibfield  {journal} {\bibinfo
  {journal} {JCAP}\ }\textbf {\bibinfo {volume} {08}},\ \bibinfo {pages} {015}
  (\bibinfo {year} {2007})},\ \Eprint {http://arxiv.org/abs/0706.4198}
  {arXiv:0706.4198 [astro-ph]} \BibitemShut {NoStop}%
\bibitem [{\citenamefont {Hannestad}\ \emph {et~al.}(2008)\citenamefont
  {Hannestad}, \citenamefont {Mirizzi}, \citenamefont {Raffelt},\ and\
  \citenamefont {Wong}}]{Hannestad:2008js}%
  \BibitemOpen
  \bibfield  {author} {\bibinfo {author} {\bibfnamefont {S.}~\bibnamefont
  {Hannestad}}, \bibinfo {author} {\bibfnamefont {A.}~\bibnamefont {Mirizzi}},
  \bibinfo {author} {\bibfnamefont {G.~G.}\ \bibnamefont {Raffelt}}, \ and\
  \bibinfo {author} {\bibfnamefont {Y.~Y.~Y.}\ \bibnamefont {Wong}},\ }\href
  {\doibase 10.1088/1475-7516/2008/04/019} {\bibfield  {journal} {\bibinfo
  {journal} {JCAP}\ }\textbf {\bibinfo {volume} {04}},\ \bibinfo {pages} {019}
  (\bibinfo {year} {2008})},\ \Eprint {http://arxiv.org/abs/0803.1585}
  {arXiv:0803.1585 [astro-ph]} \BibitemShut {NoStop}%
\bibitem [{\citenamefont {Hannestad}\ \emph {et~al.}(2010)\citenamefont
  {Hannestad}, \citenamefont {Mirizzi}, \citenamefont {Raffelt},\ and\
  \citenamefont {Wong}}]{Hannestad:2010yi}%
  \BibitemOpen
  \bibfield  {author} {\bibinfo {author} {\bibfnamefont {S.}~\bibnamefont
  {Hannestad}}, \bibinfo {author} {\bibfnamefont {A.}~\bibnamefont {Mirizzi}},
  \bibinfo {author} {\bibfnamefont {G.~G.}\ \bibnamefont {Raffelt}}, \ and\
  \bibinfo {author} {\bibfnamefont {Y.~Y.~Y.}\ \bibnamefont {Wong}},\ }\href
  {\doibase 10.1088/1475-7516/2010/08/001} {\bibfield  {journal} {\bibinfo
  {journal} {JCAP}\ }\textbf {\bibinfo {volume} {08}},\ \bibinfo {pages} {001}
  (\bibinfo {year} {2010})},\ \Eprint {http://arxiv.org/abs/1004.0695}
  {arXiv:1004.0695 [astro-ph.CO]} \BibitemShut {NoStop}%
\bibitem [{\citenamefont {Archidiacono}\ \emph {et~al.}(2013)\citenamefont
  {Archidiacono}, \citenamefont {Hannestad}, \citenamefont {Mirizzi},
  \citenamefont {Raffelt},\ and\ \citenamefont {Wong}}]{Archidiacono:2013cha}%
  \BibitemOpen
  \bibfield  {author} {\bibinfo {author} {\bibfnamefont {M.}~\bibnamefont
  {Archidiacono}}, \bibinfo {author} {\bibfnamefont {S.}~\bibnamefont
  {Hannestad}}, \bibinfo {author} {\bibfnamefont {A.}~\bibnamefont {Mirizzi}},
  \bibinfo {author} {\bibfnamefont {G.}~\bibnamefont {Raffelt}}, \ and\
  \bibinfo {author} {\bibfnamefont {Y.~Y.~Y.}\ \bibnamefont {Wong}},\ }\href
  {\doibase 10.1088/1475-7516/2013/10/020} {\bibfield  {journal} {\bibinfo
  {journal} {JCAP}\ }\textbf {\bibinfo {volume} {10}},\ \bibinfo {pages} {020}
  (\bibinfo {year} {2013})},\ \Eprint {http://arxiv.org/abs/1307.0615}
  {arXiv:1307.0615 [astro-ph.CO]} \BibitemShut {NoStop}%
\bibitem [{\citenamefont {Giusarma}\ \emph {et~al.}(2014)\citenamefont
  {Giusarma}, \citenamefont {Di~Valentino}, \citenamefont {Lattanzi},
  \citenamefont {Melchiorri},\ and\ \citenamefont {Mena}}]{Giusarma:2014zza}%
  \BibitemOpen
  \bibfield  {author} {\bibinfo {author} {\bibfnamefont {E.}~\bibnamefont
  {Giusarma}}, \bibinfo {author} {\bibfnamefont {E.}~\bibnamefont
  {Di~Valentino}}, \bibinfo {author} {\bibfnamefont {M.}~\bibnamefont
  {Lattanzi}}, \bibinfo {author} {\bibfnamefont {A.}~\bibnamefont
  {Melchiorri}}, \ and\ \bibinfo {author} {\bibfnamefont {O.}~\bibnamefont
  {Mena}},\ }\href {\doibase 10.1103/PhysRevD.90.043507} {\bibfield  {journal}
  {\bibinfo  {journal} {Phys. Rev. D}\ }\textbf {\bibinfo {volume} {90}},\
  \bibinfo {pages} {043507} (\bibinfo {year} {2014})},\ \Eprint
  {http://arxiv.org/abs/1403.4852} {arXiv:1403.4852 [astro-ph.CO]} \BibitemShut
  {NoStop}%
\bibitem [{\citenamefont {Di~Valentino}\ \emph {et~al.}(2015)\citenamefont
  {Di~Valentino}, \citenamefont {Gariazzo}, \citenamefont {Giusarma},\ and\
  \citenamefont {Mena}}]{DiValentino:2015zta}%
  \BibitemOpen
  \bibfield  {author} {\bibinfo {author} {\bibfnamefont {E.}~\bibnamefont
  {Di~Valentino}}, \bibinfo {author} {\bibfnamefont {S.}~\bibnamefont
  {Gariazzo}}, \bibinfo {author} {\bibfnamefont {E.}~\bibnamefont {Giusarma}},
  \ and\ \bibinfo {author} {\bibfnamefont {O.}~\bibnamefont {Mena}},\ }\href
  {\doibase 10.1103/PhysRevD.91.123505} {\bibfield  {journal} {\bibinfo
  {journal} {Phys. Rev. D}\ }\textbf {\bibinfo {volume} {91}},\ \bibinfo
  {pages} {123505} (\bibinfo {year} {2015})},\ \Eprint
  {http://arxiv.org/abs/1503.00911} {arXiv:1503.00911 [astro-ph.CO]}
  \BibitemShut {NoStop}%
\bibitem [{\citenamefont {Di~Valentino}\ \emph {et~al.}(2016)\citenamefont
  {Di~Valentino}, \citenamefont {Giusarma}, \citenamefont {Lattanzi},
  \citenamefont {Mena}, \citenamefont {Melchiorri},\ and\ \citenamefont
  {Silk}}]{DiValentino:2015wba}%
  \BibitemOpen
  \bibfield  {author} {\bibinfo {author} {\bibfnamefont {E.}~\bibnamefont
  {Di~Valentino}}, \bibinfo {author} {\bibfnamefont {E.}~\bibnamefont
  {Giusarma}}, \bibinfo {author} {\bibfnamefont {M.}~\bibnamefont {Lattanzi}},
  \bibinfo {author} {\bibfnamefont {O.}~\bibnamefont {Mena}}, \bibinfo {author}
  {\bibfnamefont {A.}~\bibnamefont {Melchiorri}}, \ and\ \bibinfo {author}
  {\bibfnamefont {J.}~\bibnamefont {Silk}},\ }\href {\doibase
  10.1016/j.physletb.2015.11.025} {\bibfield  {journal} {\bibinfo  {journal}
  {Phys. Lett. B}\ }\textbf {\bibinfo {volume} {752}},\ \bibinfo {pages} {182}
  (\bibinfo {year} {2016})},\ \Eprint {http://arxiv.org/abs/1507.08665}
  {arXiv:1507.08665 [astro-ph.CO]} \BibitemShut {NoStop}%
\bibitem [{\citenamefont {Archidiacono}\ \emph {et~al.}(2015)\citenamefont
  {Archidiacono}, \citenamefont {Basse}, \citenamefont {Hamann}, \citenamefont
  {Hannestad}, \citenamefont {Raffelt},\ and\ \citenamefont
  {Wong}}]{Archidiacono:2015mda}%
  \BibitemOpen
  \bibfield  {author} {\bibinfo {author} {\bibfnamefont {M.}~\bibnamefont
  {Archidiacono}}, \bibinfo {author} {\bibfnamefont {T.}~\bibnamefont {Basse}},
  \bibinfo {author} {\bibfnamefont {J.}~\bibnamefont {Hamann}}, \bibinfo
  {author} {\bibfnamefont {S.}~\bibnamefont {Hannestad}}, \bibinfo {author}
  {\bibfnamefont {G.}~\bibnamefont {Raffelt}}, \ and\ \bibinfo {author}
  {\bibfnamefont {Y.~Y.~Y.}\ \bibnamefont {Wong}},\ }\href {\doibase
  10.1088/1475-7516/2015/05/050} {\bibfield  {journal} {\bibinfo  {journal}
  {JCAP}\ }\textbf {\bibinfo {volume} {05}},\ \bibinfo {pages} {050} (\bibinfo
  {year} {2015})},\ \Eprint {http://arxiv.org/abs/1502.03325} {arXiv:1502.03325
  [astro-ph.CO]} \BibitemShut {NoStop}%
\bibitem [{\citenamefont {Ferreira}\ \emph {et~al.}(2020)\citenamefont
  {Ferreira}, \citenamefont {Notari},\ and\ \citenamefont
  {Rompineve}}]{Ferreira:2020bpb}%
  \BibitemOpen
  \bibfield  {author} {\bibinfo {author} {\bibfnamefont {R.~Z.}\ \bibnamefont
  {Ferreira}}, \bibinfo {author} {\bibfnamefont {A.}~\bibnamefont {Notari}}, \
  and\ \bibinfo {author} {\bibfnamefont {F.}~\bibnamefont {Rompineve}},\
  }\href@noop {} {\  (\bibinfo {year} {2020})},\ \Eprint
  {http://arxiv.org/abs/2012.06566} {arXiv:2012.06566 [hep-ph]} \BibitemShut
  {NoStop}%
\bibitem [{\citenamefont {Arias-Arag\'on}\ \emph {et~al.}(2021)\citenamefont
  {Arias-Arag\'on}, \citenamefont {D'Eramo}, \citenamefont {Ferreira},
  \citenamefont {Merlo},\ and\ \citenamefont {Notari}}]{Arias-Aragon:2020shv}%
  \BibitemOpen
  \bibfield  {author} {\bibinfo {author} {\bibfnamefont {F.}~\bibnamefont
  {Arias-Arag\'on}}, \bibinfo {author} {\bibfnamefont {F.}~\bibnamefont
  {D'Eramo}}, \bibinfo {author} {\bibfnamefont {R.~Z.}\ \bibnamefont
  {Ferreira}}, \bibinfo {author} {\bibfnamefont {L.}~\bibnamefont {Merlo}}, \
  and\ \bibinfo {author} {\bibfnamefont {A.}~\bibnamefont {Notari}},\ }\href
  {\doibase 10.1088/1475-7516/2021/03/090} {\bibfield  {journal} {\bibinfo
  {journal} {JCAP}\ }\textbf {\bibinfo {volume} {03}},\ \bibinfo {pages} {090}
  (\bibinfo {year} {2021})},\ \Eprint {http://arxiv.org/abs/2012.04736}
  {arXiv:2012.04736 [hep-ph]} \BibitemShut {NoStop}%
\bibitem [{\citenamefont {D'Eramo}\ \emph
  {et~al.}(2022{\natexlab{b}})\citenamefont {D'Eramo}, \citenamefont
  {Hajkarim},\ and\ \citenamefont {Yun}}]{DEramo:2021psx}%
  \BibitemOpen
  \bibfield  {author} {\bibinfo {author} {\bibfnamefont {F.}~\bibnamefont
  {D'Eramo}}, \bibinfo {author} {\bibfnamefont {F.}~\bibnamefont {Hajkarim}}, \
  and\ \bibinfo {author} {\bibfnamefont {S.}~\bibnamefont {Yun}},\ }\href
  {\doibase 10.1103/PhysRevLett.128.152001} {\bibfield  {journal} {\bibinfo
  {journal} {Phys. Rev. Lett.}\ }\textbf {\bibinfo {volume} {128}},\ \bibinfo
  {pages} {152001} (\bibinfo {year} {2022}{\natexlab{b}})},\ \Eprint
  {http://arxiv.org/abs/2108.04259} {arXiv:2108.04259 [hep-ph]} \BibitemShut
  {NoStop}%
\bibitem [{\citenamefont {Di~Valentino}\ \emph {et~al.}(2023)\citenamefont
  {Di~Valentino}, \citenamefont {Gariazzo}, \citenamefont {Giar\`e},
  \citenamefont {Melchiorri}, \citenamefont {Mena},\ and\ \citenamefont
  {Renzi}}]{DiValentino:2022edq}%
  \BibitemOpen
  \bibfield  {author} {\bibinfo {author} {\bibfnamefont {E.}~\bibnamefont
  {Di~Valentino}}, \bibinfo {author} {\bibfnamefont {S.}~\bibnamefont
  {Gariazzo}}, \bibinfo {author} {\bibfnamefont {W.}~\bibnamefont {Giar\`e}},
  \bibinfo {author} {\bibfnamefont {A.}~\bibnamefont {Melchiorri}}, \bibinfo
  {author} {\bibfnamefont {O.}~\bibnamefont {Mena}}, \ and\ \bibinfo {author}
  {\bibfnamefont {F.}~\bibnamefont {Renzi}},\ }\href {\doibase
  10.1103/PhysRevD.107.103528} {\bibfield  {journal} {\bibinfo  {journal}
  {Phys. Rev. D}\ }\textbf {\bibinfo {volume} {107}},\ \bibinfo {pages}
  {103528} (\bibinfo {year} {2023})},\ \Eprint
  {http://arxiv.org/abs/2212.11926} {arXiv:2212.11926 [astro-ph.CO]}
  \BibitemShut {NoStop}%
\bibitem [{\citenamefont {Georgi}\ \emph {et~al.}(1986)\citenamefont {Georgi},
  \citenamefont {Kaplan},\ and\ \citenamefont {Randall}}]{Georgi:1986df}%
  \BibitemOpen
  \bibfield  {author} {\bibinfo {author} {\bibfnamefont {H.}~\bibnamefont
  {Georgi}}, \bibinfo {author} {\bibfnamefont {D.~B.}\ \bibnamefont {Kaplan}},
  \ and\ \bibinfo {author} {\bibfnamefont {L.}~\bibnamefont {Randall}},\ }\href
  {\doibase 10.1016/0370-2693(86)90688-X} {\bibfield  {journal} {\bibinfo
  {journal} {Phys. Lett. B}\ }\textbf {\bibinfo {volume} {169}},\ \bibinfo
  {pages} {73} (\bibinfo {year} {1986})}\BibitemShut {NoStop}%
\bibitem [{\citenamefont {Masso}\ \emph {et~al.}(2002)\citenamefont {Masso},
  \citenamefont {Rota},\ and\ \citenamefont {Zsembinszki}}]{Masso:2002np}%
  \BibitemOpen
  \bibfield  {author} {\bibinfo {author} {\bibfnamefont {E.}~\bibnamefont
  {Masso}}, \bibinfo {author} {\bibfnamefont {F.}~\bibnamefont {Rota}}, \ and\
  \bibinfo {author} {\bibfnamefont {G.}~\bibnamefont {Zsembinszki}},\ }\href
  {\doibase 10.1103/PhysRevD.66.023004} {\bibfield  {journal} {\bibinfo
  {journal} {Phys. Rev. D}\ }\textbf {\bibinfo {volume} {66}},\ \bibinfo
  {pages} {023004} (\bibinfo {year} {2002})},\ \Eprint
  {http://arxiv.org/abs/hep-ph/0203221} {arXiv:hep-ph/0203221} \BibitemShut
  {NoStop}%
\bibitem [{\citenamefont {Di~Luzio}\ \emph {et~al.}(2021)\citenamefont
  {Di~Luzio}, \citenamefont {Martinelli},\ and\ \citenamefont
  {Piazza}}]{DiLuzio:2021vjd}%
  \BibitemOpen
  \bibfield  {author} {\bibinfo {author} {\bibfnamefont {L.}~\bibnamefont
  {Di~Luzio}}, \bibinfo {author} {\bibfnamefont {G.}~\bibnamefont
  {Martinelli}}, \ and\ \bibinfo {author} {\bibfnamefont {G.}~\bibnamefont
  {Piazza}},\ }\href {\doibase 10.1103/PhysRevLett.126.241801} {\bibfield
  {journal} {\bibinfo  {journal} {Phys. Rev. Lett.}\ }\textbf {\bibinfo
  {volume} {126}},\ \bibinfo {pages} {241801} (\bibinfo {year} {2021})},\
  \Eprint {http://arxiv.org/abs/2101.10330} {arXiv:2101.10330 [hep-ph]}
  \BibitemShut {NoStop}%
\bibitem [{\citenamefont {Di~Luzio}\ \emph {et~al.}(2022)\citenamefont
  {Di~Luzio}, \citenamefont {Martin~Camalich}, \citenamefont {Martinelli},
  \citenamefont {Oller},\ and\ \citenamefont {Piazza}}]{DiLuzio:2022gsc}%
  \BibitemOpen
  \bibfield  {author} {\bibinfo {author} {\bibfnamefont {L.}~\bibnamefont
  {Di~Luzio}}, \bibinfo {author} {\bibfnamefont {J.}~\bibnamefont
  {Martin~Camalich}}, \bibinfo {author} {\bibfnamefont {G.}~\bibnamefont
  {Martinelli}}, \bibinfo {author} {\bibfnamefont {J.~A.}\ \bibnamefont
  {Oller}}, \ and\ \bibinfo {author} {\bibfnamefont {G.}~\bibnamefont
  {Piazza}},\ }\href@noop {} {\  (\bibinfo {year} {2022})},\ \Eprint
  {http://arxiv.org/abs/2211.05073} {arXiv:2211.05073 [hep-ph]} \BibitemShut
  {NoStop}%
\bibitem [{\citenamefont {Notari}\ \emph {et~al.}(2023)\citenamefont {Notari},
  \citenamefont {Rompineve},\ and\ \citenamefont {Villadoro}}]{Notari:2022zxo}%
  \BibitemOpen
  \bibfield  {author} {\bibinfo {author} {\bibfnamefont {A.}~\bibnamefont
  {Notari}}, \bibinfo {author} {\bibfnamefont {F.}~\bibnamefont {Rompineve}}, \
  and\ \bibinfo {author} {\bibfnamefont {G.}~\bibnamefont {Villadoro}},\ }\href
  {\doibase 10.1103/PhysRevLett.131.011004} {\bibfield  {journal} {\bibinfo
  {journal} {Phys. Rev. Lett.}\ }\textbf {\bibinfo {volume} {131}},\ \bibinfo
  {pages} {011004} (\bibinfo {year} {2023})},\ \Eprint
  {http://arxiv.org/abs/2211.03799} {arXiv:2211.03799 [hep-ph]} \BibitemShut
  {NoStop}%
\bibitem [{\citenamefont {Graf}\ and\ \citenamefont
  {Steffen}(2011)}]{Graf:2010tv}%
  \BibitemOpen
  \bibfield  {author} {\bibinfo {author} {\bibfnamefont {P.}~\bibnamefont
  {Graf}}\ and\ \bibinfo {author} {\bibfnamefont {F.~D.}\ \bibnamefont
  {Steffen}},\ }\href {\doibase 10.1103/PhysRevD.83.075011} {\bibfield
  {journal} {\bibinfo  {journal} {Phys. Rev. D}\ }\textbf {\bibinfo {volume}
  {83}},\ \bibinfo {pages} {075011} (\bibinfo {year} {2011})},\ \Eprint
  {http://arxiv.org/abs/1008.4528} {arXiv:1008.4528 [hep-ph]} \BibitemShut
  {NoStop}%
\bibitem [{\citenamefont {Salvio}\ \emph {et~al.}(2014)\citenamefont {Salvio},
  \citenamefont {Strumia},\ and\ \citenamefont {Xue}}]{Salvio:2013iaa}%
  \BibitemOpen
  \bibfield  {author} {\bibinfo {author} {\bibfnamefont {A.}~\bibnamefont
  {Salvio}}, \bibinfo {author} {\bibfnamefont {A.}~\bibnamefont {Strumia}}, \
  and\ \bibinfo {author} {\bibfnamefont {W.}~\bibnamefont {Xue}},\ }\href
  {\doibase 10.1088/1475-7516/2014/01/011} {\bibfield  {journal} {\bibinfo
  {journal} {JCAP}\ }\textbf {\bibinfo {volume} {01}},\ \bibinfo {pages} {011}
  (\bibinfo {year} {2014})},\ \Eprint {http://arxiv.org/abs/1310.6982}
  {arXiv:1310.6982 [hep-ph]} \BibitemShut {NoStop}%
\bibitem [{\citenamefont {Laine}\ and\ \citenamefont
  {Vuorinen}(2016)}]{Laine:2016hma}%
  \BibitemOpen
  \bibfield  {author} {\bibinfo {author} {\bibfnamefont {M.}~\bibnamefont
  {Laine}}\ and\ \bibinfo {author} {\bibfnamefont {A.}~\bibnamefont
  {Vuorinen}},\ }\href {\doibase 10.1007/978-3-319-31933-9} {\emph {\bibinfo
  {title} {{Basics of Thermal Field Theory}}}},\ Vol.\ \bibinfo {volume} {925}\
  (\bibinfo  {publisher} {Springer},\ \bibinfo {year} {2016})\ \Eprint
  {http://arxiv.org/abs/1701.01554} {arXiv:1701.01554 [hep-ph]} \BibitemShut
  {NoStop}%
\bibitem [{\citenamefont {Adame}\ \emph {et~al.}(2024)\citenamefont {Adame}
  \emph {et~al.}}]{DESI:2024mwx}%
  \BibitemOpen
  \bibfield  {author} {\bibinfo {author} {\bibfnamefont {A.~G.}\ \bibnamefont
  {Adame}} \emph {et~al.} (\bibinfo {collaboration} {DESI}),\ }\href@noop {} {\
   (\bibinfo {year} {2024})},\ \Eprint {http://arxiv.org/abs/2404.03002}
  {arXiv:2404.03002 [astro-ph.CO]} \BibitemShut {NoStop}%
\bibitem [{\citenamefont {Aiola}\ \emph {et~al.}(2020)\citenamefont {Aiola}
  \emph {et~al.}}]{ACT:2020gnv}%
  \BibitemOpen
  \bibfield  {author} {\bibinfo {author} {\bibfnamefont {S.}~\bibnamefont
  {Aiola}} \emph {et~al.} (\bibinfo {collaboration} {ACT}),\ }\href {\doibase
  10.1088/1475-7516/2020/12/047} {\bibfield  {journal} {\bibinfo  {journal}
  {JCAP}\ }\textbf {\bibinfo {volume} {12}},\ \bibinfo {pages} {047} (\bibinfo
  {year} {2020})},\ \Eprint {http://arxiv.org/abs/2007.07288} {arXiv:2007.07288
  [astro-ph.CO]} \BibitemShut {NoStop}%
\bibitem [{\citenamefont {{Balkenhol}}\ and\ \citenamefont {{SPT-3G
  Collaboration}}(2023)}]{Balkenhol23}%
  \BibitemOpen
  \bibfield  {author} {\bibinfo {author} {\bibfnamefont {L.}~\bibnamefont
  {{Balkenhol}}}\ and\ \bibinfo {author} {\bibnamefont {{SPT-3G
  Collaboration}}},\ }\href {\doibase 10.1103/PhysRevD.108.023510} {\bibfield
  {journal} {\bibinfo  {journal} {\prd}\ }\textbf {\bibinfo {volume} {108}},\
  \bibinfo {eid} {023510} (\bibinfo {year} {2023})},\ \Eprint
  {http://arxiv.org/abs/2212.05642} {arXiv:2212.05642 [astro-ph.CO]}
  \BibitemShut {NoStop}%
\bibitem [{\citenamefont {Di~Luzio}\ and\ \citenamefont
  {Piazza}(2022)}]{DiLuzio:2022tbb}%
  \BibitemOpen
  \bibfield  {author} {\bibinfo {author} {\bibfnamefont {L.}~\bibnamefont
  {Di~Luzio}}\ and\ \bibinfo {author} {\bibfnamefont {G.}~\bibnamefont
  {Piazza}},\ }\href@noop {} {\  (\bibinfo {year} {2022})},\ \Eprint
  {http://arxiv.org/abs/2206.04061} {arXiv:2206.04061 [hep-ph]} \BibitemShut
  {NoStop}%
\bibitem [{\citenamefont {Bonanno}\ \emph
  {et~al.}(2023{\natexlab{a}})\citenamefont {Bonanno}, \citenamefont
  {D'Angelo}, \citenamefont {D'Elia}, \citenamefont {Maio},\ and\ \citenamefont
  {Naviglio}}]{Bonanno:2023thi}%
  \BibitemOpen
  \bibfield  {author} {\bibinfo {author} {\bibfnamefont {C.}~\bibnamefont
  {Bonanno}}, \bibinfo {author} {\bibfnamefont {F.}~\bibnamefont {D'Angelo}},
  \bibinfo {author} {\bibfnamefont {M.}~\bibnamefont {D'Elia}}, \bibinfo
  {author} {\bibfnamefont {L.}~\bibnamefont {Maio}}, \ and\ \bibinfo {author}
  {\bibfnamefont {M.}~\bibnamefont {Naviglio}},\ }\href@noop {} {\  (\bibinfo
  {year} {2023}{\natexlab{a}})},\ \Eprint {http://arxiv.org/abs/2308.01287}
  {arXiv:2308.01287 [hep-lat]} \BibitemShut {NoStop}%
\bibitem [{\citenamefont {Bonanno}\ \emph
  {et~al.}(2023{\natexlab{b}})\citenamefont {Bonanno}, \citenamefont
  {D'Angelo}, \citenamefont {D'Elia}, \citenamefont {Maio},\ and\ \citenamefont
  {Naviglio}}]{Bonanno:2023xfv}%
  \BibitemOpen
  \bibfield  {author} {\bibinfo {author} {\bibfnamefont {C.}~\bibnamefont
  {Bonanno}}, \bibinfo {author} {\bibfnamefont {F.}~\bibnamefont {D'Angelo}},
  \bibinfo {author} {\bibfnamefont {M.}~\bibnamefont {D'Elia}}, \bibinfo
  {author} {\bibfnamefont {L.}~\bibnamefont {Maio}}, \ and\ \bibinfo {author}
  {\bibfnamefont {M.}~\bibnamefont {Naviglio}},\ }in\ \href@noop {} {\emph
  {\bibinfo {booktitle} {{26th High-Energy Physics International Conference in
  QCD}}}}\ (\bibinfo {year} {2023})\ \Eprint {http://arxiv.org/abs/2309.13327}
  {arXiv:2309.13327 [hep-lat]} \BibitemShut {NoStop}%
\bibitem [{\citenamefont {Ade}\ \emph {et~al.}(2019)\citenamefont {Ade} \emph
  {et~al.}}]{SimonsObservatory:2018koc}%
  \BibitemOpen
  \bibfield  {author} {\bibinfo {author} {\bibfnamefont {P.}~\bibnamefont
  {Ade}} \emph {et~al.} (\bibinfo {collaboration} {Simons Observatory}),\
  }\href {\doibase 10.1088/1475-7516/2019/02/056} {\bibfield  {journal}
  {\bibinfo  {journal} {JCAP}\ }\textbf {\bibinfo {volume} {02}},\ \bibinfo
  {pages} {056} (\bibinfo {year} {2019})},\ \Eprint
  {http://arxiv.org/abs/1808.07445} {arXiv:1808.07445 [astro-ph.CO]}
  \BibitemShut {NoStop}%
\bibitem [{\citenamefont {Abazajian}\ \emph
  {et~al.}(2019{\natexlab{b}})\citenamefont {Abazajian} \emph
  {et~al.}}]{abazajian2019cmbs4}%
  \BibitemOpen
  \bibfield  {author} {\bibinfo {author} {\bibfnamefont {K.~N.}\ \bibnamefont
  {Abazajian}} \emph {et~al.},\ }\href@noop {} {\enquote {\bibinfo {title}
  {Cmb-s4 decadal survey apc white paper},}\ } (\bibinfo {year}
  {2019}{\natexlab{b}}),\ \Eprint {http://arxiv.org/abs/1908.01062}
  {arXiv:1908.01062 [astro-ph.IM]} \BibitemShut {NoStop}%
\bibitem [{\citenamefont {Abazajian}\ \emph {et~al.}(2022)\citenamefont
  {Abazajian} \emph {et~al.}}]{CMB-S4:2022ght}%
  \BibitemOpen
  \bibfield  {author} {\bibinfo {author} {\bibfnamefont {K.}~\bibnamefont
  {Abazajian}} \emph {et~al.} (\bibinfo {collaboration} {CMB-S4}),\ }\href@noop
  {} {\  (\bibinfo {year} {2022})},\ \Eprint {http://arxiv.org/abs/2203.08024}
  {arXiv:2203.08024 [astro-ph.CO]} \BibitemShut {NoStop}%
\bibitem [{\citenamefont {Aghamousa}\ \emph {et~al.}(2016)\citenamefont
  {Aghamousa} \emph {et~al.}}]{DESI:2016fyo}%
  \BibitemOpen
  \bibfield  {author} {\bibinfo {author} {\bibfnamefont {A.}~\bibnamefont
  {Aghamousa}} \emph {et~al.} (\bibinfo {collaboration} {DESI}),\ }\href@noop
  {} {\  (\bibinfo {year} {2016})},\ \Eprint {http://arxiv.org/abs/1611.00036}
  {arXiv:1611.00036 [astro-ph.IM]} \BibitemShut {NoStop}%
\bibitem [{\citenamefont {Di~Luzio}\ \emph {et~al.}(2020)\citenamefont
  {Di~Luzio}, \citenamefont {Giannotti}, \citenamefont {Nardi},\ and\
  \citenamefont {Visinelli}}]{DiLuzio:2020wdo}%
  \BibitemOpen
  \bibfield  {author} {\bibinfo {author} {\bibfnamefont {L.}~\bibnamefont
  {Di~Luzio}}, \bibinfo {author} {\bibfnamefont {M.}~\bibnamefont {Giannotti}},
  \bibinfo {author} {\bibfnamefont {E.}~\bibnamefont {Nardi}}, \ and\ \bibinfo
  {author} {\bibfnamefont {L.}~\bibnamefont {Visinelli}},\ }\href {\doibase
  10.1016/j.physrep.2020.06.002} {\bibfield  {journal} {\bibinfo  {journal}
  {Phys. Rept.}\ }\textbf {\bibinfo {volume} {870}},\ \bibinfo {pages} {1}
  (\bibinfo {year} {2020})},\ \Eprint {http://arxiv.org/abs/2003.01100}
  {arXiv:2003.01100 [hep-ph]} \BibitemShut {NoStop}%
\bibitem [{\citenamefont {Aydemir}\ \emph {et~al.}(2012)\citenamefont
  {Aydemir}, \citenamefont {Anber},\ and\ \citenamefont
  {Donoghue}}]{Aydemir:2012nz}%
  \BibitemOpen
  \bibfield  {author} {\bibinfo {author} {\bibfnamefont {U.}~\bibnamefont
  {Aydemir}}, \bibinfo {author} {\bibfnamefont {M.~M.}\ \bibnamefont {Anber}},
  \ and\ \bibinfo {author} {\bibfnamefont {J.~F.}\ \bibnamefont {Donoghue}},\
  }\href {\doibase 10.1103/PhysRevD.86.014025} {\bibfield  {journal} {\bibinfo
  {journal} {Phys. Rev. D}\ }\textbf {\bibinfo {volume} {86}},\ \bibinfo
  {pages} {014025} (\bibinfo {year} {2012})},\ \Eprint
  {http://arxiv.org/abs/1203.5153} {arXiv:1203.5153 [hep-ph]} \BibitemShut
  {NoStop}%
\bibitem [{\citenamefont {Schenk}(1993)}]{Schenk:1993ru}%
  \BibitemOpen
  \bibfield  {author} {\bibinfo {author} {\bibfnamefont {A.}~\bibnamefont
  {Schenk}},\ }\href {\doibase 10.1103/PhysRevD.47.5138} {\bibfield  {journal}
  {\bibinfo  {journal} {Phys. Rev. D}\ }\textbf {\bibinfo {volume} {47}},\
  \bibinfo {pages} {5138} (\bibinfo {year} {1993})}\BibitemShut {NoStop}%
\bibitem [{\citenamefont {Truong}(1988)}]{Truong:1988zp}%
  \BibitemOpen
  \bibfield  {author} {\bibinfo {author} {\bibfnamefont {T.~N.}\ \bibnamefont
  {Truong}},\ }\href {\doibase 10.1103/PhysRevLett.61.2526} {\bibfield
  {journal} {\bibinfo  {journal} {Phys. Rev. Lett.}\ }\textbf {\bibinfo
  {volume} {61}},\ \bibinfo {pages} {2526} (\bibinfo {year}
  {1988})}\BibitemShut {NoStop}%
\bibitem [{\citenamefont {Salas-Bern\'ardez}\ \emph {et~al.}(2021)\citenamefont
  {Salas-Bern\'ardez}, \citenamefont {Llanes-Estrada}, \citenamefont
  {Escudero-Pedrosa},\ and\ \citenamefont {Oller}}]{Salas-Bernardez:2020hua}%
  \BibitemOpen
  \bibfield  {author} {\bibinfo {author} {\bibfnamefont {A.}~\bibnamefont
  {Salas-Bern\'ardez}}, \bibinfo {author} {\bibfnamefont {F.~J.}\ \bibnamefont
  {Llanes-Estrada}}, \bibinfo {author} {\bibfnamefont {J.}~\bibnamefont
  {Escudero-Pedrosa}}, \ and\ \bibinfo {author} {\bibfnamefont {J.~A.}\
  \bibnamefont {Oller}},\ }\href {\doibase 10.21468/SciPostPhys.11.2.020}
  {\bibfield  {journal} {\bibinfo  {journal} {SciPost Phys.}\ }\textbf
  {\bibinfo {volume} {11}},\ \bibinfo {pages} {020} (\bibinfo {year} {2021})},\
  \Eprint {http://arxiv.org/abs/2010.13709} {arXiv:2010.13709 [hep-ph]}
  \BibitemShut {NoStop}%
\bibitem [{\citenamefont {Watson}(1952)}]{PhysRev.88.1163}%
  \BibitemOpen
  \bibfield  {author} {\bibinfo {author} {\bibfnamefont {K.~M.}\ \bibnamefont
  {Watson}},\ }\href {\doibase 10.1103/PhysRev.88.1163} {\bibfield  {journal}
  {\bibinfo  {journal} {Phys. Rev.}\ }\textbf {\bibinfo {volume} {88}},\
  \bibinfo {pages} {1163} (\bibinfo {year} {1952})}\BibitemShut {NoStop}%
\bibitem [{\citenamefont {Ma}\ and\ \citenamefont
  {Bertschinger}(1995)}]{Ma:1995ey}%
  \BibitemOpen
  \bibfield  {author} {\bibinfo {author} {\bibfnamefont {C.-P.}\ \bibnamefont
  {Ma}}\ and\ \bibinfo {author} {\bibfnamefont {E.}~\bibnamefont
  {Bertschinger}},\ }\href {\doibase 10.1086/176550} {\bibfield  {journal}
  {\bibinfo  {journal} {Astrophys. J.}\ }\textbf {\bibinfo {volume} {455}},\
  \bibinfo {pages} {7} (\bibinfo {year} {1995})},\ \Eprint
  {http://arxiv.org/abs/astro-ph/9506072} {arXiv:astro-ph/9506072} \BibitemShut
  {NoStop}%
\bibitem [{\citenamefont {Lesgourgues}\ and\ \citenamefont
  {Tram}(2011)}]{Lesgourgues:2011rh}%
  \BibitemOpen
  \bibfield  {author} {\bibinfo {author} {\bibfnamefont {J.}~\bibnamefont
  {Lesgourgues}}\ and\ \bibinfo {author} {\bibfnamefont {T.}~\bibnamefont
  {Tram}},\ }\href {\doibase 10.1088/1475-7516/2011/09/032} {\bibfield
  {journal} {\bibinfo  {journal} {JCAP}\ }\textbf {\bibinfo {volume} {09}},\
  \bibinfo {pages} {032} (\bibinfo {year} {2011})},\ \Eprint
  {http://arxiv.org/abs/1104.2935} {arXiv:1104.2935 [astro-ph.CO]} \BibitemShut
  {NoStop}%
\bibitem [{\citenamefont {Lesgourgues}\ and\ \citenamefont
  {Pastor}(2006)}]{Lesgourgues:2006nd}%
  \BibitemOpen
  \bibfield  {author} {\bibinfo {author} {\bibfnamefont {J.}~\bibnamefont
  {Lesgourgues}}\ and\ \bibinfo {author} {\bibfnamefont {S.}~\bibnamefont
  {Pastor}},\ }\href {\doibase 10.1016/j.physrep.2006.04.001} {\bibfield
  {journal} {\bibinfo  {journal} {Phys. Rept.}\ }\textbf {\bibinfo {volume}
  {429}},\ \bibinfo {pages} {307} (\bibinfo {year} {2006})},\ \Eprint
  {http://arxiv.org/abs/astro-ph/0603494} {arXiv:astro-ph/0603494} \BibitemShut
  {NoStop}%
\bibitem [{\citenamefont {Aghanim}\ \emph
  {et~al.}(2020{\natexlab{b}})\citenamefont {Aghanim} \emph
  {et~al.}}]{Planck:2018vyg}%
  \BibitemOpen
  \bibfield  {author} {\bibinfo {author} {\bibfnamefont {N.}~\bibnamefont
  {Aghanim}} \emph {et~al.} (\bibinfo {collaboration} {Planck}),\ }\href
  {\doibase 10.1051/0004-6361/201833910} {\bibfield  {journal} {\bibinfo
  {journal} {Astron. Astrophys.}\ }\textbf {\bibinfo {volume} {641}},\ \bibinfo
  {pages} {A6} (\bibinfo {year} {2020}{\natexlab{b}})},\ \bibinfo {note}
  {[Erratum: Astron.Astrophys. 652, C4 (2021)]},\ \Eprint
  {http://arxiv.org/abs/1807.06209} {arXiv:1807.06209 [astro-ph.CO]}
  \BibitemShut {NoStop}%
\bibitem [{\citenamefont {Aghanim}\ \emph
  {et~al.}(2020{\natexlab{c}})\citenamefont {Aghanim} \emph
  {et~al.}}]{Aghanim:2019ame}%
  \BibitemOpen
  \bibfield  {author} {\bibinfo {author} {\bibfnamefont {N.}~\bibnamefont
  {Aghanim}} \emph {et~al.} (\bibinfo {collaboration} {Planck}),\ }\href
  {\doibase 10.1051/0004-6361/201936386} {\bibfield  {journal} {\bibinfo
  {journal} {Astron. Astrophys.}\ }\textbf {\bibinfo {volume} {641}},\ \bibinfo
  {pages} {A5} (\bibinfo {year} {2020}{\natexlab{c}})},\ \Eprint
  {http://arxiv.org/abs/1907.12875} {arXiv:1907.12875 [astro-ph.CO]}
  \BibitemShut {NoStop}%
\bibitem [{\citenamefont {Alam}\ \emph {et~al.}(2017)\citenamefont {Alam} \emph
  {et~al.}}]{Alam:2016hwk}%
  \BibitemOpen
  \bibfield  {author} {\bibinfo {author} {\bibfnamefont {S.}~\bibnamefont
  {Alam}} \emph {et~al.} (\bibinfo {collaboration} {BOSS}),\ }\href {\doibase
  10.1093/mnras/stx721} {\bibfield  {journal} {\bibinfo  {journal} {Mon. Not.
  Roy. Astron. Soc.}\ }\textbf {\bibinfo {volume} {470}},\ \bibinfo {pages}
  {2617} (\bibinfo {year} {2017})},\ \Eprint {http://arxiv.org/abs/1607.03155}
  {arXiv:1607.03155 [astro-ph.CO]} \BibitemShut {NoStop}%
\bibitem [{\citenamefont {Beutler}\ \emph {et~al.}(2011)\citenamefont
  {Beutler}, \citenamefont {Blake}, \citenamefont {Colless}, \citenamefont
  {Jones}, \citenamefont {Staveley-Smith} \emph {et~al.}}]{Beutler:2011hx}%
  \BibitemOpen
  \bibfield  {author} {\bibinfo {author} {\bibfnamefont {F.}~\bibnamefont
  {Beutler}}, \bibinfo {author} {\bibfnamefont {C.}~\bibnamefont {Blake}},
  \bibinfo {author} {\bibfnamefont {M.}~\bibnamefont {Colless}}, \bibinfo
  {author} {\bibfnamefont {D.~H.}\ \bibnamefont {Jones}}, \bibinfo {author}
  {\bibfnamefont {L.}~\bibnamefont {Staveley-Smith}},  \emph {et~al.},\ }\href
  {\doibase 10.1111/j.1365-2966.2011.19250.x} {\bibfield  {journal} {\bibinfo
  {journal} {Mon. Not. Roy. Astron. Soc.}\ }\textbf {\bibinfo {volume} {416}},\
  \bibinfo {pages} {3017} (\bibinfo {year} {2011})},\ \Eprint
  {http://arxiv.org/abs/1106.3366} {arXiv:1106.3366 [astro-ph.CO]} \BibitemShut
  {NoStop}%
\bibitem [{\citenamefont {Ross}\ \emph {et~al.}(2015)\citenamefont {Ross},
  \citenamefont {Samushia}, \citenamefont {Howlett}, \citenamefont {Percival},
  \citenamefont {Burden},\ and\ \citenamefont {Manera}}]{Ross:2014qpa}%
  \BibitemOpen
  \bibfield  {author} {\bibinfo {author} {\bibfnamefont {A.~J.}\ \bibnamefont
  {Ross}}, \bibinfo {author} {\bibfnamefont {L.}~\bibnamefont {Samushia}},
  \bibinfo {author} {\bibfnamefont {C.}~\bibnamefont {Howlett}}, \bibinfo
  {author} {\bibfnamefont {W.~J.}\ \bibnamefont {Percival}}, \bibinfo {author}
  {\bibfnamefont {A.}~\bibnamefont {Burden}}, \ and\ \bibinfo {author}
  {\bibfnamefont {M.}~\bibnamefont {Manera}},\ }\href {\doibase
  10.1093/mnras/stv154} {\bibfield  {journal} {\bibinfo  {journal} {Mon. Not.
  Roy. Astron. Soc.}\ }\textbf {\bibinfo {volume} {449}},\ \bibinfo {pages}
  {835} (\bibinfo {year} {2015})},\ \Eprint {http://arxiv.org/abs/1409.3242}
  {arXiv:1409.3242 [astro-ph.CO]} \BibitemShut {NoStop}%
\bibitem [{\citenamefont {Alam}\ \emph {et~al.}(2021)\citenamefont {Alam} \emph
  {et~al.}}]{Alam:2020sor}%
  \BibitemOpen
  \bibfield  {author} {\bibinfo {author} {\bibfnamefont {S.}~\bibnamefont
  {Alam}} \emph {et~al.} (\bibinfo {collaboration} {eBOSS}),\ }\href {\doibase
  10.1103/PhysRevD.103.083533} {\bibfield  {journal} {\bibinfo  {journal}
  {Phys. Rev. D}\ }\textbf {\bibinfo {volume} {103}},\ \bibinfo {pages}
  {083533} (\bibinfo {year} {2021})},\ \Eprint
  {http://arxiv.org/abs/2007.08991} {arXiv:2007.08991 [astro-ph.CO]}
  \BibitemShut {NoStop}%
\bibitem [{\citenamefont {Scolnic}\ \emph {et~al.}(2018)\citenamefont {Scolnic}
  \emph {et~al.}}]{Scolnic:2017caz}%
  \BibitemOpen
  \bibfield  {author} {\bibinfo {author} {\bibfnamefont {D.~M.}\ \bibnamefont
  {Scolnic}} \emph {et~al.},\ }\href {\doibase 10.3847/1538-4357/aab9bb}
  {\bibfield  {journal} {\bibinfo  {journal} {Astrophys. J.}\ }\textbf
  {\bibinfo {volume} {859}},\ \bibinfo {pages} {101} (\bibinfo {year}
  {2018})},\ \Eprint {http://arxiv.org/abs/1710.00845} {arXiv:1710.00845
  [astro-ph.CO]} \BibitemShut {NoStop}%
\bibitem [{\citenamefont {Qu}\ \emph {et~al.}(2023)\citenamefont {Qu} \emph
  {et~al.}}]{ACT:2023dou}%
  \BibitemOpen
  \bibfield  {author} {\bibinfo {author} {\bibfnamefont {F.~J.}\ \bibnamefont
  {Qu}} \emph {et~al.} (\bibinfo {collaboration} {ACT}),\ }\href@noop {} {\
  (\bibinfo {year} {2023})},\ \Eprint {http://arxiv.org/abs/2304.05202}
  {arXiv:2304.05202 [astro-ph.CO]} \BibitemShut {NoStop}%
\bibitem [{\citenamefont {Carron}\ \emph {et~al.}(2022)\citenamefont {Carron},
  \citenamefont {Mirmelstein},\ and\ \citenamefont {Lewis}}]{Carron_2022}%
  \BibitemOpen
  \bibfield  {author} {\bibinfo {author} {\bibfnamefont {J.}~\bibnamefont
  {Carron}}, \bibinfo {author} {\bibfnamefont {M.}~\bibnamefont {Mirmelstein}},
  \ and\ \bibinfo {author} {\bibfnamefont {A.}~\bibnamefont {Lewis}},\ }\href
  {\doibase 10.1088/1475-7516/2022/09/039} {\bibfield  {journal} {\bibinfo
  {journal} {Journal of Cosmology and Astroparticle Physics}\ }\textbf
  {\bibinfo {volume} {2022}},\ \bibinfo {pages} {039} (\bibinfo {year}
  {2022})}\BibitemShut {NoStop}%
\bibitem [{\citenamefont {{Lesgourgues}}(2011)}]{2011arXiv1104.2932L}%
  \BibitemOpen
  \bibfield  {author} {\bibinfo {author} {\bibfnamefont {J.}~\bibnamefont
  {{Lesgourgues}}},\ }\href {\doibase 10.48550/arXiv.1104.2932} {\bibfield
  {journal} {\bibinfo  {journal} {arXiv e-prints}\ ,\ \bibinfo {eid}
  {arXiv:1104.2932}} (\bibinfo {year} {2011})},\ \Eprint
  {http://arxiv.org/abs/1104.2932} {arXiv:1104.2932 [astro-ph.IM]} \BibitemShut
  {NoStop}%
\bibitem [{\citenamefont {{Blas}}\ \emph {et~al.}(2011)\citenamefont {{Blas}},
  \citenamefont {{Lesgourgues}},\ and\ \citenamefont
  {{Tram}}}]{2011JCAP07034B}%
  \BibitemOpen
  \bibfield  {author} {\bibinfo {author} {\bibfnamefont {D.}~\bibnamefont
  {{Blas}}}, \bibinfo {author} {\bibfnamefont {J.}~\bibnamefont
  {{Lesgourgues}}}, \ and\ \bibinfo {author} {\bibfnamefont {T.}~\bibnamefont
  {{Tram}}},\ }\href {\doibase 10.1088/1475-7516/2011/07/034} {\bibfield
  {journal} {\bibinfo  {journal} {JCAP}\ }\textbf {\bibinfo {volume} {2011}},\
  \bibinfo {eid} {034} (\bibinfo {year} {2011})},\ \Eprint
  {http://arxiv.org/abs/1104.2933} {arXiv:1104.2933 [astro-ph.CO]} \BibitemShut
  {NoStop}%
\bibitem [{\citenamefont {Capozzi}\ \emph {et~al.}(2017)\citenamefont
  {Capozzi}, \citenamefont {Di~Valentino}, \citenamefont {Lisi}, \citenamefont
  {Marrone}, \citenamefont {Melchiorri},\ and\ \citenamefont
  {Palazzo}}]{Capozzi:2017ipn}%
  \BibitemOpen
  \bibfield  {author} {\bibinfo {author} {\bibfnamefont {F.}~\bibnamefont
  {Capozzi}}, \bibinfo {author} {\bibfnamefont {E.}~\bibnamefont
  {Di~Valentino}}, \bibinfo {author} {\bibfnamefont {E.}~\bibnamefont {Lisi}},
  \bibinfo {author} {\bibfnamefont {A.}~\bibnamefont {Marrone}}, \bibinfo
  {author} {\bibfnamefont {A.}~\bibnamefont {Melchiorri}}, \ and\ \bibinfo
  {author} {\bibfnamefont {A.}~\bibnamefont {Palazzo}},\ }\href {\doibase
  10.1103/PhysRevD.95.096014} {\bibfield  {journal} {\bibinfo  {journal} {Phys.
  Rev. D}\ }\textbf {\bibinfo {volume} {95}},\ \bibinfo {pages} {096014}
  (\bibinfo {year} {2017})},\ \bibinfo {note} {[Addendum: Phys.Rev.D 101,
  116013 (2020)]},\ \Eprint {http://arxiv.org/abs/2003.08511} {arXiv:2003.08511
  [hep-ph]} \BibitemShut {NoStop}%
\bibitem [{\citenamefont {de~Salas}\ \emph {et~al.}(2021)\citenamefont
  {de~Salas}, \citenamefont {Forero}, \citenamefont {Gariazzo}, \citenamefont
  {Mart{\'\i}nez-Mirav{\'e}}, \citenamefont {Mena}, \citenamefont {Ternes},
  \citenamefont {T{\'o}rtola},\ and\ \citenamefont {Valle}}]{deSalas:2020pgw}%
  \BibitemOpen
  \bibfield  {author} {\bibinfo {author} {\bibfnamefont {P.~F.}\ \bibnamefont
  {de~Salas}}, \bibinfo {author} {\bibfnamefont {D.~V.}\ \bibnamefont
  {Forero}}, \bibinfo {author} {\bibfnamefont {S.}~\bibnamefont {Gariazzo}},
  \bibinfo {author} {\bibfnamefont {P.}~\bibnamefont
  {Mart{\'\i}nez-Mirav{\'e}}}, \bibinfo {author} {\bibfnamefont
  {O.}~\bibnamefont {Mena}}, \bibinfo {author} {\bibfnamefont {C.~A.}\
  \bibnamefont {Ternes}}, \bibinfo {author} {\bibfnamefont {M.}~\bibnamefont
  {T{\'o}rtola}}, \ and\ \bibinfo {author} {\bibfnamefont {J.~W.~F.}\
  \bibnamefont {Valle}},\ }\href {\doibase 10.1007/JHEP02(2021)071} {\bibfield
  {journal} {\bibinfo  {journal} {JHEP}\ }\textbf {\bibinfo {volume} {02}},\
  \bibinfo {pages} {071} (\bibinfo {year} {2021})},\ \Eprint
  {http://arxiv.org/abs/2006.11237} {arXiv:2006.11237 [hep-ph]} \BibitemShut
  {NoStop}%
\bibitem [{\citenamefont {Gonzalez-Garcia}\ \emph {et~al.}(2021)\citenamefont
  {Gonzalez-Garcia}, \citenamefont {Maltoni},\ and\ \citenamefont
  {Schwetz}}]{Gonzalez-Garcia:2021dve}%
  \BibitemOpen
  \bibfield  {author} {\bibinfo {author} {\bibfnamefont {M.~C.}\ \bibnamefont
  {Gonzalez-Garcia}}, \bibinfo {author} {\bibfnamefont {M.}~\bibnamefont
  {Maltoni}}, \ and\ \bibinfo {author} {\bibfnamefont {T.}~\bibnamefont
  {Schwetz}},\ }\href {\doibase 10.3390/universe7120459} {\bibfield  {journal}
  {\bibinfo  {journal} {Universe}\ }\textbf {\bibinfo {volume} {7}},\ \bibinfo
  {pages} {459} (\bibinfo {year} {2021})},\ \Eprint
  {http://arxiv.org/abs/2111.03086} {arXiv:2111.03086 [hep-ph]} \BibitemShut
  {NoStop}%
\bibitem [{\citenamefont {Torrado}\ and\ \citenamefont
  {Lewis}(2021)}]{Torrado:2020dgo}%
  \BibitemOpen
  \bibfield  {author} {\bibinfo {author} {\bibfnamefont {J.}~\bibnamefont
  {Torrado}}\ and\ \bibinfo {author} {\bibfnamefont {A.}~\bibnamefont
  {Lewis}},\ }\href {\doibase 10.1088/1475-7516/2021/05/057} {\bibfield
  {journal} {\bibinfo  {journal} {JCAP}\ }\textbf {\bibinfo {volume} {05}},\
  \bibinfo {pages} {057} (\bibinfo {year} {2021})},\ \Eprint
  {http://arxiv.org/abs/2005.05290} {arXiv:2005.05290 [astro-ph.IM]}
  \BibitemShut {NoStop}%
\bibitem [{\citenamefont {Lewis}(2019)}]{Lewis:2019xzd}%
  \BibitemOpen
  \bibfield  {author} {\bibinfo {author} {\bibfnamefont {A.}~\bibnamefont
  {Lewis}},\ }\href@noop {} {\  (\bibinfo {year} {2019})},\ \Eprint
  {http://arxiv.org/abs/1910.13970} {arXiv:1910.13970 [astro-ph.IM]}
  \BibitemShut {NoStop}%
\bibitem [{\citenamefont {Sch\"oneberg}\ \emph {et~al.}(2019)\citenamefont
  {Sch\"oneberg}, \citenamefont {Lesgourgues},\ and\ \citenamefont
  {Hooper}}]{Schoneberg:2019wmt}%
  \BibitemOpen
  \bibfield  {author} {\bibinfo {author} {\bibfnamefont {N.}~\bibnamefont
  {Sch\"oneberg}}, \bibinfo {author} {\bibfnamefont {J.}~\bibnamefont
  {Lesgourgues}}, \ and\ \bibinfo {author} {\bibfnamefont {D.~C.}\ \bibnamefont
  {Hooper}},\ }\href {\doibase 10.1088/1475-7516/2019/10/029} {\bibfield
  {journal} {\bibinfo  {journal} {JCAP}\ }\textbf {\bibinfo {volume} {10}},\
  \bibinfo {pages} {029} (\bibinfo {year} {2019})},\ \Eprint
  {http://arxiv.org/abs/1907.11594} {arXiv:1907.11594 [astro-ph.CO]}
  \BibitemShut {NoStop}%
\bibitem [{\citenamefont {Lee}\ and\ \citenamefont
  {Ali-Ha\"\i{}moud}(2020)}]{Lee:2020obi}%
  \BibitemOpen
  \bibfield  {author} {\bibinfo {author} {\bibfnamefont {N.}~\bibnamefont
  {Lee}}\ and\ \bibinfo {author} {\bibfnamefont {Y.}~\bibnamefont
  {Ali-Ha\"\i{}moud}},\ }\href {\doibase 10.1103/PhysRevD.102.083517}
  {\bibfield  {journal} {\bibinfo  {journal} {Phys. Rev. D}\ }\textbf {\bibinfo
  {volume} {102}},\ \bibinfo {pages} {083517} (\bibinfo {year} {2020})},\
  \Eprint {http://arxiv.org/abs/2007.14114} {arXiv:2007.14114 [astro-ph.CO]}
  \BibitemShut {NoStop}%
\bibitem [{\citenamefont {Steigman}(2010)}]{Steigman:2010pa}%
  \BibitemOpen
  \bibfield  {author} {\bibinfo {author} {\bibfnamefont {G.}~\bibnamefont
  {Steigman}},\ }\href {\doibase 10.1088/1475-7516/2010/04/029} {\bibfield
  {journal} {\bibinfo  {journal} {JCAP}\ }\textbf {\bibinfo {volume} {04}},\
  \bibinfo {pages} {029} (\bibinfo {year} {2010})},\ \Eprint
  {http://arxiv.org/abs/1002.3604} {arXiv:1002.3604 [astro-ph.CO]} \BibitemShut
  {NoStop}%
\bibitem [{\citenamefont {Sarkar}(1996)}]{Sarkar:1995dd}%
  \BibitemOpen
  \bibfield  {author} {\bibinfo {author} {\bibfnamefont {S.}~\bibnamefont
  {Sarkar}},\ }\href {\doibase 10.1088/0034-4885/59/12/001} {\bibfield
  {journal} {\bibinfo  {journal} {Rept. Prog. Phys.}\ }\textbf {\bibinfo
  {volume} {59}},\ \bibinfo {pages} {1493} (\bibinfo {year} {1996})},\ \Eprint
  {http://arxiv.org/abs/hep-ph/9602260} {arXiv:hep-ph/9602260} \BibitemShut
  {NoStop}%
\bibitem [{\citenamefont {Olive}\ \emph {et~al.}(2000)\citenamefont {Olive},
  \citenamefont {Steigman},\ and\ \citenamefont {Walker}}]{Olive:1999ij}%
  \BibitemOpen
  \bibfield  {author} {\bibinfo {author} {\bibfnamefont {K.~A.}\ \bibnamefont
  {Olive}}, \bibinfo {author} {\bibfnamefont {G.}~\bibnamefont {Steigman}}, \
  and\ \bibinfo {author} {\bibfnamefont {T.~P.}\ \bibnamefont {Walker}},\
  }\href {\doibase 10.1016/S0370-1573(00)00031-4} {\bibfield  {journal}
  {\bibinfo  {journal} {Phys. Rept.}\ }\textbf {\bibinfo {volume} {333}},\
  \bibinfo {pages} {389} (\bibinfo {year} {2000})},\ \Eprint
  {http://arxiv.org/abs/https://arxiv.org/abs/astro-ph/9905320}
  {arXiv:https://arxiv.org/abs/astro-ph/9905320} \BibitemShut {NoStop}%
\bibitem [{\citenamefont {Pospelov}\ and\ \citenamefont
  {Pradler}(2010)}]{Pospelov:2010hj}%
  \BibitemOpen
  \bibfield  {author} {\bibinfo {author} {\bibfnamefont {M.}~\bibnamefont
  {Pospelov}}\ and\ \bibinfo {author} {\bibfnamefont {J.}~\bibnamefont
  {Pradler}},\ }\href {\doibase 10.1146/annurev.nucl.012809.104521} {\bibfield
  {journal} {\bibinfo  {journal} {Ann. Rev. Nucl. Part. Sci.}\ }\textbf
  {\bibinfo {volume} {60}},\ \bibinfo {pages} {539} (\bibinfo {year} {2010})},\
  \Eprint {http://arxiv.org/abs/1011.1054} {arXiv:1011.1054 [hep-ph]}
  \BibitemShut {NoStop}%
\bibitem [{\citenamefont {{Hsyu}}\ \emph {et~al.}(2020)\citenamefont {{Hsyu}},
  \citenamefont {{Cooke}}, \citenamefont {{Prochaska}},\ and\ \citenamefont
  {{Bolte}}}]{2020ApJ89677H}%
  \BibitemOpen
  \bibfield  {author} {\bibinfo {author} {\bibfnamefont {T.}~\bibnamefont
  {{Hsyu}}}, \bibinfo {author} {\bibfnamefont {R.~J.}\ \bibnamefont {{Cooke}}},
  \bibinfo {author} {\bibfnamefont {J.~X.}\ \bibnamefont {{Prochaska}}}, \ and\
  \bibinfo {author} {\bibfnamefont {M.}~\bibnamefont {{Bolte}}},\ }\href
  {\doibase 10.3847/1538-4357/ab91af} {\bibfield  {journal} {\bibinfo
  {journal} {\apj}\ }\textbf {\bibinfo {volume} {896}},\ \bibinfo {eid} {77}
  (\bibinfo {year} {2020})},\ \Eprint
  {http://arxiv.org/abs/https://arxiv.org/abs/2005.12290}
  {arXiv:https://arxiv.org/abs/2005.12290 [astro-ph.GA]} \BibitemShut {NoStop}%
\bibitem [{\citenamefont {Kurichin}\ \emph {et~al.}(2021)\citenamefont
  {Kurichin}, \citenamefont {Kislitsyn}, \citenamefont {Klimenko},
  \citenamefont {Balashev},\ and\ \citenamefont {Ivanchik}}]{Kurichin:2021ppm}%
  \BibitemOpen
  \bibfield  {author} {\bibinfo {author} {\bibfnamefont {O.~A.}\ \bibnamefont
  {Kurichin}}, \bibinfo {author} {\bibfnamefont {P.~A.}\ \bibnamefont
  {Kislitsyn}}, \bibinfo {author} {\bibfnamefont {V.~V.}\ \bibnamefont
  {Klimenko}}, \bibinfo {author} {\bibfnamefont {S.~A.}\ \bibnamefont
  {Balashev}}, \ and\ \bibinfo {author} {\bibfnamefont {A.~V.}\ \bibnamefont
  {Ivanchik}},\ }\href {\doibase 10.1093/mnras/stab215} {\bibfield  {journal}
  {\bibinfo  {journal} {Mon. Not. Roy. Astron. Soc.}\ }\textbf {\bibinfo
  {volume} {502}},\ \bibinfo {pages} {3045} (\bibinfo {year} {2021})},\ \Eprint
  {http://arxiv.org/abs/2101.09127} {arXiv:2101.09127 [astro-ph.CO]}
  \BibitemShut {NoStop}%
\bibitem [{\citenamefont {Pitrou}\ \emph {et~al.}(2018)\citenamefont {Pitrou},
  \citenamefont {Coc}, \citenamefont {Uzan},\ and\ \citenamefont
  {Vangioni}}]{Pitrou18}%
  \BibitemOpen
  \bibfield  {author} {\bibinfo {author} {\bibfnamefont {C.}~\bibnamefont
  {Pitrou}}, \bibinfo {author} {\bibfnamefont {A.}~\bibnamefont {Coc}},
  \bibinfo {author} {\bibfnamefont {J.-P.}\ \bibnamefont {Uzan}}, \ and\
  \bibinfo {author} {\bibfnamefont {E.}~\bibnamefont {Vangioni}},\ }\href
  {\doibase 10.1016/j.physrep.2018.04.005} {\bibfield  {journal} {\bibinfo
  {journal} {Physics Reports}\ }\textbf {\bibinfo {volume} {754}},\ \bibinfo
  {pages} {1–66} (\bibinfo {year} {2018})}\BibitemShut {NoStop}%
\bibitem [{\citenamefont {Riemer-S\o{}rensen}\ \emph
  {et~al.}(2017)\citenamefont {Riemer-S\o{}rensen}, \citenamefont {Kotu\v{s}},
  \citenamefont {Webb}, \citenamefont {Ali}, \citenamefont {Dumont},
  \citenamefont {Murphy},\ and\ \citenamefont
  {Carswell}}]{Riemer-Sorensen:2017pey}%
  \BibitemOpen
  \bibfield  {author} {\bibinfo {author} {\bibfnamefont {S.}~\bibnamefont
  {Riemer-S\o{}rensen}}, \bibinfo {author} {\bibfnamefont {S.}~\bibnamefont
  {Kotu\v{s}}}, \bibinfo {author} {\bibfnamefont {J.~K.}\ \bibnamefont {Webb}},
  \bibinfo {author} {\bibfnamefont {K.}~\bibnamefont {Ali}}, \bibinfo {author}
  {\bibfnamefont {V.}~\bibnamefont {Dumont}}, \bibinfo {author} {\bibfnamefont
  {M.~T.}\ \bibnamefont {Murphy}}, \ and\ \bibinfo {author} {\bibfnamefont
  {R.~F.}\ \bibnamefont {Carswell}},\ }\href {\doibase 10.1093/mnras/stx681}
  {\bibfield  {journal} {\bibinfo  {journal} {Mon. Not. Roy. Astron. Soc.}\
  }\textbf {\bibinfo {volume} {468}},\ \bibinfo {pages} {3239} (\bibinfo {year}
  {2017})},\ \Eprint {http://arxiv.org/abs/1703.06656} {arXiv:1703.06656
  [astro-ph.CO]} \BibitemShut {NoStop}%
\bibitem [{\citenamefont {Cooke}\ \emph {et~al.}(2018)\citenamefont {Cooke},
  \citenamefont {Pettini},\ and\ \citenamefont {Steidel}}]{Cooke:2017cwo}%
  \BibitemOpen
  \bibfield  {author} {\bibinfo {author} {\bibfnamefont {R.~J.}\ \bibnamefont
  {Cooke}}, \bibinfo {author} {\bibfnamefont {M.}~\bibnamefont {Pettini}}, \
  and\ \bibinfo {author} {\bibfnamefont {C.~C.}\ \bibnamefont {Steidel}},\
  }\href {\doibase 10.3847/1538-4357/aaab53} {\bibfield  {journal} {\bibinfo
  {journal} {Astrophys. J.}\ }\textbf {\bibinfo {volume} {855}},\ \bibinfo
  {pages} {102} (\bibinfo {year} {2018})},\ \Eprint
  {http://arxiv.org/abs/1710.11129} {arXiv:1710.11129 [astro-ph.CO]}
  \BibitemShut {NoStop}%
\bibitem [{\citenamefont {Pitrou}\ \emph {et~al.}(2021)\citenamefont {Pitrou},
  \citenamefont {Coc}, \citenamefont {Uzan},\ and\ \citenamefont
  {Vangioni}}]{Pitrou:2021vqr}%
  \BibitemOpen
  \bibfield  {author} {\bibinfo {author} {\bibfnamefont {C.}~\bibnamefont
  {Pitrou}}, \bibinfo {author} {\bibfnamefont {A.}~\bibnamefont {Coc}},
  \bibinfo {author} {\bibfnamefont {J.-P.}\ \bibnamefont {Uzan}}, \ and\
  \bibinfo {author} {\bibfnamefont {E.}~\bibnamefont {Vangioni}},\ }\href
  {\doibase 10.1038/s42254-021-00294-6} {\bibfield  {journal} {\bibinfo
  {journal} {Nature Rev. Phys.}\ }\textbf {\bibinfo {volume} {3}},\ \bibinfo
  {pages} {231} (\bibinfo {year} {2021})},\ \Eprint
  {http://arxiv.org/abs/2104.11148} {arXiv:2104.11148 [astro-ph.CO]}
  \BibitemShut {NoStop}%
\bibitem [{\citenamefont {Burns}\ \emph
  {et~al.}(2023{\natexlab{a}})\citenamefont {Burns}, \citenamefont {Tait},\
  and\ \citenamefont {Valli}}]{Burns:2022hkq}%
  \BibitemOpen
  \bibfield  {author} {\bibinfo {author} {\bibfnamefont {A.-K.}\ \bibnamefont
  {Burns}}, \bibinfo {author} {\bibfnamefont {T.~M.~P.}\ \bibnamefont {Tait}},
  \ and\ \bibinfo {author} {\bibfnamefont {M.}~\bibnamefont {Valli}},\ }\href
  {\doibase 10.1103/PhysRevLett.130.131001} {\bibfield  {journal} {\bibinfo
  {journal} {Phys. Rev. Lett.}\ }\textbf {\bibinfo {volume} {130}},\ \bibinfo
  {pages} {131001} (\bibinfo {year} {2023}{\natexlab{a}})},\ \Eprint
  {http://arxiv.org/abs/2206.00693} {arXiv:2206.00693 [hep-ph]} \BibitemShut
  {NoStop}%
\bibitem [{\citenamefont {Yeh}\ \emph {et~al.}(2022)\citenamefont {Yeh},
  \citenamefont {Shelton}, \citenamefont {Olive},\ and\ \citenamefont
  {Fields}}]{Yeh:2022heq}%
  \BibitemOpen
  \bibfield  {author} {\bibinfo {author} {\bibfnamefont {T.-H.}\ \bibnamefont
  {Yeh}}, \bibinfo {author} {\bibfnamefont {J.}~\bibnamefont {Shelton}},
  \bibinfo {author} {\bibfnamefont {K.~A.}\ \bibnamefont {Olive}}, \ and\
  \bibinfo {author} {\bibfnamefont {B.~D.}\ \bibnamefont {Fields}},\ }\href
  {\doibase 10.1088/1475-7516/2022/10/046} {\bibfield  {journal} {\bibinfo
  {journal} {JCAP}\ }\textbf {\bibinfo {volume} {10}},\ \bibinfo {pages} {046}
  (\bibinfo {year} {2022})},\ \Eprint {http://arxiv.org/abs/2207.13133}
  {arXiv:2207.13133 [astro-ph.CO]} \BibitemShut {NoStop}%
\bibitem [{\citenamefont {Burns}\ \emph
  {et~al.}(2023{\natexlab{b}})\citenamefont {Burns}, \citenamefont {Tait},\
  and\ \citenamefont {Valli}}]{Burns:2023sgx}%
  \BibitemOpen
  \bibfield  {author} {\bibinfo {author} {\bibfnamefont {A.-K.}\ \bibnamefont
  {Burns}}, \bibinfo {author} {\bibfnamefont {T.~M.~P.}\ \bibnamefont {Tait}},
  \ and\ \bibinfo {author} {\bibfnamefont {M.}~\bibnamefont {Valli}},\
  }\href@noop {} {\  (\bibinfo {year} {2023}{\natexlab{b}})},\ \Eprint
  {http://arxiv.org/abs/2307.07061} {arXiv:2307.07061 [hep-ph]} \BibitemShut
  {NoStop}%
\bibitem [{\citenamefont {Xu}\ \emph {et~al.}(2013)\citenamefont {Xu},
  \citenamefont {Takahashi}, \citenamefont {Goriely}, \citenamefont {Arnould},
  \citenamefont {Ohta},\ and\ \citenamefont {Utsunomiya}}]{Xu:2013fha}%
  \BibitemOpen
  \bibfield  {author} {\bibinfo {author} {\bibfnamefont {Y.}~\bibnamefont
  {Xu}}, \bibinfo {author} {\bibfnamefont {K.}~\bibnamefont {Takahashi}},
  \bibinfo {author} {\bibfnamefont {S.}~\bibnamefont {Goriely}}, \bibinfo
  {author} {\bibfnamefont {M.}~\bibnamefont {Arnould}}, \bibinfo {author}
  {\bibfnamefont {M.}~\bibnamefont {Ohta}}, \ and\ \bibinfo {author}
  {\bibfnamefont {H.}~\bibnamefont {Utsunomiya}},\ }\href {\doibase
  10.1016/j.nuclphysa.2013.09.007} {\bibfield  {journal} {\bibinfo  {journal}
  {Nucl. Phys. A}\ }\textbf {\bibinfo {volume} {918}},\ \bibinfo {pages} {61}
  (\bibinfo {year} {2013})},\ \Eprint {http://arxiv.org/abs/1310.7099}
  {arXiv:1310.7099 [nucl-th]} \BibitemShut {NoStop}%
\bibitem [{\citenamefont {Workman}\ \emph
  {et~al.}(2022{\natexlab{a}})\citenamefont {Workman} \emph
  {et~al.}}]{PDG:2022pth}%
  \BibitemOpen
  \bibfield  {author} {\bibinfo {author} {\bibfnamefont {R.~L.}\ \bibnamefont
  {Workman}} \emph {et~al.} (\bibinfo {collaboration} {Particle Data Group}),\
  }\href {\doibase 10.1093/ptep/ptac097} {\bibfield  {journal} {\bibinfo
  {journal} {PTEP}\ }\textbf {\bibinfo {volume} {2022}},\ \bibinfo {pages}
  {083C01} (\bibinfo {year} {2022}{\natexlab{a}})}\BibitemShut {NoStop}%
\bibitem [{\citenamefont {Motloch}\ and\ \citenamefont {Hu}(2020)}]{motloch20}%
  \BibitemOpen
  \bibfield  {author} {\bibinfo {author} {\bibfnamefont {P.}~\bibnamefont
  {Motloch}}\ and\ \bibinfo {author} {\bibfnamefont {W.}~\bibnamefont {Hu}},\
  }\href {\doibase 10.1103/PhysRevD.101.083515} {\bibfield  {journal} {\bibinfo
   {journal} {Phys. Rev. D}\ }\textbf {\bibinfo {volume} {101}},\ \bibinfo
  {pages} {083515} (\bibinfo {year} {2020})}\BibitemShut {NoStop}%
\bibitem [{\citenamefont {Bianchini}\ \emph {et~al.}(2020)\citenamefont
  {Bianchini} \emph {et~al.}}]{bianchini20a}%
  \BibitemOpen
  \bibfield  {author} {\bibinfo {author} {\bibfnamefont {F.}~\bibnamefont
  {Bianchini}} \emph {et~al.} (\bibinfo {collaboration} {SPTpol}),\ }\href
  {\doibase 10.3847/1538-4357/ab6082} {\bibfield  {journal} {\bibinfo
  {journal} {\apj}\ }\textbf {\bibinfo {volume} {888}},\ \bibinfo {eid} {119}
  (\bibinfo {year} {2020})},\ \Eprint {http://arxiv.org/abs/1910.07157}
  {arXiv:1910.07157 [astro-ph.CO]} \BibitemShut {NoStop}%
\bibitem [{\citenamefont {Lesgourgues}\ and\ \citenamefont
  {Pastor}(1999)}]{Lesgourgues:1999wu}%
  \BibitemOpen
  \bibfield  {author} {\bibinfo {author} {\bibfnamefont {J.}~\bibnamefont
  {Lesgourgues}}\ and\ \bibinfo {author} {\bibfnamefont {S.}~\bibnamefont
  {Pastor}},\ }\href {\doibase 10.1103/PhysRevD.60.103521} {\bibfield
  {journal} {\bibinfo  {journal} {Phys. Rev. D}\ }\textbf {\bibinfo {volume}
  {60}},\ \bibinfo {pages} {103521} (\bibinfo {year} {1999})},\ \Eprint
  {http://arxiv.org/abs/hep-ph/9904411} {arXiv:hep-ph/9904411} \BibitemShut
  {NoStop}%
\bibitem [{\citenamefont {Arias}\ \emph {et~al.}(2023)\citenamefont {Arias},
  \citenamefont {Bernal}, \citenamefont {Osi\'nski}, \citenamefont
  {Roszkowski},\ and\ \citenamefont {Venegas}}]{Arias:2023wyg}%
  \BibitemOpen
  \bibfield  {author} {\bibinfo {author} {\bibfnamefont {P.}~\bibnamefont
  {Arias}}, \bibinfo {author} {\bibfnamefont {N.}~\bibnamefont {Bernal}},
  \bibinfo {author} {\bibfnamefont {J.~K.}\ \bibnamefont {Osi\'nski}}, \bibinfo
  {author} {\bibfnamefont {L.}~\bibnamefont {Roszkowski}}, \ and\ \bibinfo
  {author} {\bibfnamefont {M.}~\bibnamefont {Venegas}},\ }\href@noop {} {\
  (\bibinfo {year} {2023})},\ \Eprint {http://arxiv.org/abs/2308.01352}
  {arXiv:2308.01352 [hep-ph]} \BibitemShut {NoStop}%
\bibitem [{\citenamefont {McLerran}\ \emph {et~al.}(1991)\citenamefont
  {McLerran}, \citenamefont {Mottola},\ and\ \citenamefont
  {Shaposhnikov}}]{McLerran:1990de}%
  \BibitemOpen
  \bibfield  {author} {\bibinfo {author} {\bibfnamefont {L.~D.}\ \bibnamefont
  {McLerran}}, \bibinfo {author} {\bibfnamefont {E.}~\bibnamefont {Mottola}}, \
  and\ \bibinfo {author} {\bibfnamefont {M.~E.}\ \bibnamefont {Shaposhnikov}},\
  }\href {\doibase 10.1103/PhysRevD.43.2027} {\bibfield  {journal} {\bibinfo
  {journal} {Phys. Rev. D}\ }\textbf {\bibinfo {volume} {43}},\ \bibinfo
  {pages} {2027} (\bibinfo {year} {1991})}\BibitemShut {NoStop}%
\bibitem [{\citenamefont {Giudice}\ and\ \citenamefont
  {Shaposhnikov}(1994)}]{Giudice:1993bb}%
  \BibitemOpen
  \bibfield  {author} {\bibinfo {author} {\bibfnamefont {G.~F.}\ \bibnamefont
  {Giudice}}\ and\ \bibinfo {author} {\bibfnamefont {M.~E.}\ \bibnamefont
  {Shaposhnikov}},\ }\href {\doibase 10.1016/0370-2693(94)91202-5} {\bibfield
  {journal} {\bibinfo  {journal} {Phys. Lett. B}\ }\textbf {\bibinfo {volume}
  {326}},\ \bibinfo {pages} {118} (\bibinfo {year} {1994})},\ \Eprint
  {http://arxiv.org/abs/hep-ph/9311367} {arXiv:hep-ph/9311367} \BibitemShut
  {NoStop}%
\bibitem [{\citenamefont {Fukushima}\ \emph {et~al.}(2008)\citenamefont
  {Fukushima}, \citenamefont {Kharzeev},\ and\ \citenamefont
  {Warringa}}]{Fukushima:2008xe}%
  \BibitemOpen
  \bibfield  {author} {\bibinfo {author} {\bibfnamefont {K.}~\bibnamefont
  {Fukushima}}, \bibinfo {author} {\bibfnamefont {D.~E.}\ \bibnamefont
  {Kharzeev}}, \ and\ \bibinfo {author} {\bibfnamefont {H.~J.}\ \bibnamefont
  {Warringa}},\ }\href {\doibase 10.1103/PhysRevD.78.074033} {\bibfield
  {journal} {\bibinfo  {journal} {Phys. Rev. D}\ }\textbf {\bibinfo {volume}
  {78}},\ \bibinfo {pages} {074033} (\bibinfo {year} {2008})},\ \Eprint
  {http://arxiv.org/abs/0808.3382} {arXiv:0808.3382 [hep-ph]} \BibitemShut
  {NoStop}%
\bibitem [{\citenamefont {Moore}\ and\ \citenamefont
  {Tassler}(2011)}]{Moore:2010jd}%
  \BibitemOpen
  \bibfield  {author} {\bibinfo {author} {\bibfnamefont {G.~D.}\ \bibnamefont
  {Moore}}\ and\ \bibinfo {author} {\bibfnamefont {M.}~\bibnamefont
  {Tassler}},\ }\href {\doibase 10.1007/JHEP02(2011)105} {\bibfield  {journal}
  {\bibinfo  {journal} {JHEP}\ }\textbf {\bibinfo {volume} {02}},\ \bibinfo
  {pages} {105} (\bibinfo {year} {2011})},\ \Eprint
  {http://arxiv.org/abs/1011.1167} {arXiv:1011.1167 [hep-ph]} \BibitemShut
  {NoStop}%
\bibitem [{\citenamefont {Schlegel}\ \emph {et~al.}(2022)\citenamefont
  {Schlegel} \emph {et~al.}}]{Schlegel:2022vrv}%
  \BibitemOpen
  \bibfield  {author} {\bibinfo {author} {\bibfnamefont {D.~J.}\ \bibnamefont
  {Schlegel}} \emph {et~al.},\ }\href@noop {} {\  (\bibinfo {year} {2022})},\
  \Eprint {http://arxiv.org/abs/2209.04322} {arXiv:2209.04322 [astro-ph.IM]}
  \BibitemShut {NoStop}%
\bibitem [{\citenamefont {MacInnis}\ \emph {et~al.}(2023)\citenamefont
  {MacInnis}, \citenamefont {Sehgal},\ and\ \citenamefont
  {Rothermel}}]{MacInnis:2023vif}%
  \BibitemOpen
  \bibfield  {author} {\bibinfo {author} {\bibfnamefont {A.}~\bibnamefont
  {MacInnis}}, \bibinfo {author} {\bibfnamefont {N.}~\bibnamefont {Sehgal}}, \
  and\ \bibinfo {author} {\bibfnamefont {M.}~\bibnamefont {Rothermel}},\
  }\href@noop {} {\  (\bibinfo {year} {2023})},\ \Eprint
  {http://arxiv.org/abs/2309.03021} {arXiv:2309.03021 [astro-ph.CO]}
  \BibitemShut {NoStop}%
\bibitem [{\citenamefont {Ir\v{s}i\v{c}}\ \emph {et~al.}(2023)\citenamefont
  {Ir\v{s}i\v{c}} \emph {et~al.}}]{Irsic:2023equ}%
  \BibitemOpen
  \bibfield  {author} {\bibinfo {author} {\bibfnamefont {V.}~\bibnamefont
  {Ir\v{s}i\v{c}}} \emph {et~al.},\ }\href@noop {} {\  (\bibinfo {year}
  {2023})},\ \Eprint {http://arxiv.org/abs/2309.04533} {arXiv:2309.04533
  [astro-ph.CO]} \BibitemShut {NoStop}%
\bibitem [{\citenamefont {Chudaykin}\ \emph {et~al.}(2020)\citenamefont
  {Chudaykin}, \citenamefont {Ivanov}, \citenamefont {Philcox},\ and\
  \citenamefont {Simonovi\'c}}]{Chudaykin:2020aoj}%
  \BibitemOpen
  \bibfield  {author} {\bibinfo {author} {\bibfnamefont {A.}~\bibnamefont
  {Chudaykin}}, \bibinfo {author} {\bibfnamefont {M.~M.}\ \bibnamefont
  {Ivanov}}, \bibinfo {author} {\bibfnamefont {O.~H.~E.}\ \bibnamefont
  {Philcox}}, \ and\ \bibinfo {author} {\bibfnamefont {M.}~\bibnamefont
  {Simonovi\'c}},\ }\href {\doibase 10.1103/PhysRevD.102.063533} {\bibfield
  {journal} {\bibinfo  {journal} {Phys. Rev. D}\ }\textbf {\bibinfo {volume}
  {102}},\ \bibinfo {pages} {063533} (\bibinfo {year} {2020})},\ \Eprint
  {http://arxiv.org/abs/2004.10607} {arXiv:2004.10607 [astro-ph.CO]}
  \BibitemShut {NoStop}%
\bibitem [{\citenamefont {D'Amico}\ \emph {et~al.}(2021)\citenamefont
  {D'Amico}, \citenamefont {Senatore},\ and\ \citenamefont
  {Zhang}}]{DAmico:2020kxu}%
  \BibitemOpen
  \bibfield  {author} {\bibinfo {author} {\bibfnamefont {G.}~\bibnamefont
  {D'Amico}}, \bibinfo {author} {\bibfnamefont {L.}~\bibnamefont {Senatore}}, \
  and\ \bibinfo {author} {\bibfnamefont {P.}~\bibnamefont {Zhang}},\ }\href
  {\doibase 10.1088/1475-7516/2021/01/006} {\bibfield  {journal} {\bibinfo
  {journal} {JCAP}\ }\textbf {\bibinfo {volume} {01}},\ \bibinfo {pages} {006}
  (\bibinfo {year} {2021})},\ \Eprint {http://arxiv.org/abs/2003.07956}
  {arXiv:2003.07956 [astro-ph.CO]} \BibitemShut {NoStop}%
\bibitem [{\citenamefont {Karkare}\ \emph {et~al.}(2022)\citenamefont
  {Karkare}, \citenamefont {Dizgah}, \citenamefont {Keating}, \citenamefont
  {Breysse},\ and\ \citenamefont {Chung}}]{Karkare:2022bai}%
  \BibitemOpen
  \bibfield  {author} {\bibinfo {author} {\bibfnamefont {K.~S.}\ \bibnamefont
  {Karkare}}, \bibinfo {author} {\bibfnamefont {A.~M.}\ \bibnamefont {Dizgah}},
  \bibinfo {author} {\bibfnamefont {G.~K.}\ \bibnamefont {Keating}}, \bibinfo
  {author} {\bibfnamefont {P.}~\bibnamefont {Breysse}}, \ and\ \bibinfo
  {author} {\bibfnamefont {D.~T.}\ \bibnamefont {Chung}} (\bibinfo
  {collaboration} {Snowmass Cosmic Frontier 5 Topical Group}),\ }in\ \href@noop
  {} {\emph {\bibinfo {booktitle} {{Snowmass 2021}}}}\ (\bibinfo {year}
  {2022})\ \Eprint {http://arxiv.org/abs/2203.07258} {arXiv:2203.07258
  [astro-ph.CO]} \BibitemShut {NoStop}%
\bibitem [{\citenamefont {{Yueh}}\ and\ \citenamefont
  {{Buchler}}(1976)}]{1976Ap&SS..39..429Y}%
  \BibitemOpen
  \bibfield  {author} {\bibinfo {author} {\bibfnamefont {W.~R.}\ \bibnamefont
  {{Yueh}}}\ and\ \bibinfo {author} {\bibfnamefont {J.~R.}\ \bibnamefont
  {{Buchler}}},\ }\href {\doibase 10.1007/BF00648341} {\bibfield  {journal}
  {\bibinfo  {journal} {{\it Astrophys. Space Sci.}}\ }\textbf {\bibinfo
  {volume} {39}},\ \bibinfo {pages} {429} (\bibinfo {year} {1976})}\BibitemShut
  {NoStop}%
\bibitem [{\citenamefont {Hannestad}\ and\ \citenamefont
  {Madsen}(1995)}]{Hannestad:1995rs}%
  \BibitemOpen
  \bibfield  {author} {\bibinfo {author} {\bibfnamefont {S.}~\bibnamefont
  {Hannestad}}\ and\ \bibinfo {author} {\bibfnamefont {J.}~\bibnamefont
  {Madsen}},\ }\href {\doibase 10.1103/PhysRevD.52.1764} {\bibfield  {journal}
  {\bibinfo  {journal} {Phys. Rev. D}\ }\textbf {\bibinfo {volume} {52}},\
  \bibinfo {pages} {1764} (\bibinfo {year} {1995})},\ \Eprint
  {http://arxiv.org/abs/astro-ph/9506015} {arXiv:astro-ph/9506015} \BibitemShut
  {NoStop}%
\bibitem [{\citenamefont {Workman}\ \emph
  {et~al.}(2022{\natexlab{b}})\citenamefont {Workman} \emph
  {et~al.}}]{ParticleDataGroup:2022pth}%
  \BibitemOpen
  \bibfield  {author} {\bibinfo {author} {\bibfnamefont {R.~L.}\ \bibnamefont
  {Workman}} \emph {et~al.} (\bibinfo {collaboration} {Particle Data Group}),\
  }\href {\doibase 10.1093/ptep/ptac097} {\bibfield  {journal} {\bibinfo
  {journal} {PTEP}\ }\textbf {\bibinfo {volume} {2022}},\ \bibinfo {pages}
  {083C01} (\bibinfo {year} {2022}{\natexlab{b}})}\BibitemShut {NoStop}%
\bibitem [{\citenamefont {Lepage}(2021)}]{Lepage:2020tgj}%
  \BibitemOpen
  \bibfield  {author} {\bibinfo {author} {\bibfnamefont {G.~P.}\ \bibnamefont
  {Lepage}},\ }\href {\doibase 10.1016/j.jcp.2021.110386} {\bibfield  {journal}
  {\bibinfo  {journal} {J. Comput. Phys.}\ }\textbf {\bibinfo {volume} {439}},\
  \bibinfo {pages} {110386} (\bibinfo {year} {2021})},\ \Eprint
  {http://arxiv.org/abs/2009.05112} {arXiv:2009.05112 [physics.comp-ph]}
  \BibitemShut {NoStop}%
\bibitem [{\citenamefont {Dobado}\ and\ \citenamefont
  {Pelaez}(1997)}]{Dobado:1996ps}%
  \BibitemOpen
  \bibfield  {author} {\bibinfo {author} {\bibfnamefont {A.}~\bibnamefont
  {Dobado}}\ and\ \bibinfo {author} {\bibfnamefont {J.~R.}\ \bibnamefont
  {Pelaez}},\ }\href {\doibase 10.1103/PhysRevD.56.3057} {\bibfield  {journal}
  {\bibinfo  {journal} {Phys. Rev. D}\ }\textbf {\bibinfo {volume} {56}},\
  \bibinfo {pages} {3057} (\bibinfo {year} {1997})},\ \Eprint
  {http://arxiv.org/abs/hep-ph/9604416} {arXiv:hep-ph/9604416} \BibitemShut
  {NoStop}%
\bibitem [{\citenamefont {Aoki}\ \emph {et~al.}(2022)\citenamefont {Aoki} \emph
  {et~al.}}]{FlavourLatticeAveragingGroupFLAG:2021npn}%
  \BibitemOpen
  \bibfield  {author} {\bibinfo {author} {\bibfnamefont {Y.}~\bibnamefont
  {Aoki}} \emph {et~al.} (\bibinfo {collaboration} {Flavour Lattice Averaging
  Group (FLAG)}),\ }\href {\doibase 10.1140/epjc/s10052-022-10536-1} {\bibfield
   {journal} {\bibinfo  {journal} {Eur. Phys. J. C}\ }\textbf {\bibinfo
  {volume} {82}},\ \bibinfo {pages} {869} (\bibinfo {year} {2022})},\ \Eprint
  {http://arxiv.org/abs/2111.09849} {arXiv:2111.09849 [hep-lat]} \BibitemShut
  {NoStop}%
\bibitem [{\citenamefont {Gasser}\ and\ \citenamefont
  {Leutwyler}(1984)}]{Gasser:1983yg}%
  \BibitemOpen
  \bibfield  {author} {\bibinfo {author} {\bibfnamefont {J.}~\bibnamefont
  {Gasser}}\ and\ \bibinfo {author} {\bibfnamefont {H.}~\bibnamefont
  {Leutwyler}},\ }\href {\doibase 10.1016/0003-4916(84)90242-2} {\bibfield
  {journal} {\bibinfo  {journal} {Annals Phys.}\ }\textbf {\bibinfo {volume}
  {158}},\ \bibinfo {pages} {142} (\bibinfo {year} {1984})}\BibitemShut
  {NoStop}%
\bibitem [{\citenamefont {Garcia-Martin}\ \emph {et~al.}(2011)\citenamefont
  {Garcia-Martin}, \citenamefont {Kaminski}, \citenamefont {Pelaez},
  \citenamefont {Ruiz~de Elvira},\ and\ \citenamefont
  {Yndurain}}]{Garcia-Martin:2011iqs}%
  \BibitemOpen
  \bibfield  {author} {\bibinfo {author} {\bibfnamefont {R.}~\bibnamefont
  {Garcia-Martin}}, \bibinfo {author} {\bibfnamefont {R.}~\bibnamefont
  {Kaminski}}, \bibinfo {author} {\bibfnamefont {J.~R.}\ \bibnamefont
  {Pelaez}}, \bibinfo {author} {\bibfnamefont {J.}~\bibnamefont {Ruiz~de
  Elvira}}, \ and\ \bibinfo {author} {\bibfnamefont {F.~J.}\ \bibnamefont
  {Yndurain}},\ }\href {\doibase 10.1103/PhysRevD.83.074004} {\bibfield
  {journal} {\bibinfo  {journal} {Phys. Rev. D}\ }\textbf {\bibinfo {volume}
  {83}},\ \bibinfo {pages} {074004} (\bibinfo {year} {2011})},\ \Eprint
  {http://arxiv.org/abs/1102.2183} {arXiv:1102.2183 [hep-ph]} \BibitemShut
  {NoStop}%
\bibitem [{\citenamefont {Aghanim}\ \emph {et~al.}(2019)\citenamefont {Aghanim}
  \emph {et~al.}}]{Planck18res}%
  \BibitemOpen
  \bibfield  {author} {\bibinfo {author} {\bibfnamefont {N.}~\bibnamefont
  {Aghanim}} \emph {et~al.} (\bibinfo {collaboration} {Planck}),\ }\href@noop
  {} {}\bibinfo {howpublished}
  {\url{https://wiki.cosmos.esa.int/planck-legacy-archive/images/4/43/Baseline_params_table_2018_68pc_v2.pdf}}
  (\bibinfo {year} {2019})\BibitemShut {NoStop}%
\bibitem [{\citenamefont {Hamimeche}\ and\ \citenamefont
  {Lewis}(2008)}]{Hamimeche:2008ai}%
  \BibitemOpen
  \bibfield  {author} {\bibinfo {author} {\bibfnamefont {S.}~\bibnamefont
  {Hamimeche}}\ and\ \bibinfo {author} {\bibfnamefont {A.}~\bibnamefont
  {Lewis}},\ }\href {\doibase 10.1103/PhysRevD.77.103013} {\bibfield  {journal}
  {\bibinfo  {journal} {Phys. Rev. D}\ }\textbf {\bibinfo {volume} {77}},\
  \bibinfo {pages} {103013} (\bibinfo {year} {2008})},\ \Eprint
  {http://arxiv.org/abs/0801.0554} {arXiv:0801.0554 [astro-ph]} \BibitemShut
  {NoStop}%
\bibitem [{\citenamefont {Coc}\ \emph {et~al.}(2014)\citenamefont {Coc},
  \citenamefont {Uzan},\ and\ \citenamefont {Vangioni}}]{Coc:2014oia}%
  \BibitemOpen
  \bibfield  {author} {\bibinfo {author} {\bibfnamefont {A.}~\bibnamefont
  {Coc}}, \bibinfo {author} {\bibfnamefont {J.-P.}\ \bibnamefont {Uzan}}, \
  and\ \bibinfo {author} {\bibfnamefont {E.}~\bibnamefont {Vangioni}},\ }\href
  {\doibase 10.1088/1475-7516/2014/10/050} {\bibfield  {journal} {\bibinfo
  {journal} {JCAP}\ }\textbf {\bibinfo {volume} {10}},\ \bibinfo {pages} {050}
  (\bibinfo {year} {2014})},\ \Eprint {http://arxiv.org/abs/1403.6694}
  {arXiv:1403.6694 [astro-ph.CO]} \BibitemShut {NoStop}%
\bibitem [{\citenamefont {Aver}\ \emph {et~al.}(2015)\citenamefont {Aver},
  \citenamefont {Olive},\ and\ \citenamefont {Skillman}}]{Aver:2015iza}%
  \BibitemOpen
  \bibfield  {author} {\bibinfo {author} {\bibfnamefont {E.}~\bibnamefont
  {Aver}}, \bibinfo {author} {\bibfnamefont {K.~A.}\ \bibnamefont {Olive}}, \
  and\ \bibinfo {author} {\bibfnamefont {E.~D.}\ \bibnamefont {Skillman}},\
  }\href {\doibase 10.1088/1475-7516/2015/07/011} {\bibfield  {journal}
  {\bibinfo  {journal} {JCAP}\ }\textbf {\bibinfo {volume} {07}},\ \bibinfo
  {pages} {011} (\bibinfo {year} {2015})},\ \Eprint
  {http://arxiv.org/abs/1503.08146} {arXiv:1503.08146 [astro-ph.CO]}
  \BibitemShut {NoStop}%
\bibitem [{\citenamefont {Aver}\ \emph {et~al.}(2021)\citenamefont {Aver},
  \citenamefont {Berg}, \citenamefont {Olive}, \citenamefont {Pogge},
  \citenamefont {Salzer},\ and\ \citenamefont {Skillman}}]{Aver:2020fon}%
  \BibitemOpen
  \bibfield  {author} {\bibinfo {author} {\bibfnamefont {E.}~\bibnamefont
  {Aver}}, \bibinfo {author} {\bibfnamefont {D.~A.}\ \bibnamefont {Berg}},
  \bibinfo {author} {\bibfnamefont {K.~A.}\ \bibnamefont {Olive}}, \bibinfo
  {author} {\bibfnamefont {R.~W.}\ \bibnamefont {Pogge}}, \bibinfo {author}
  {\bibfnamefont {J.~J.}\ \bibnamefont {Salzer}}, \ and\ \bibinfo {author}
  {\bibfnamefont {E.~D.}\ \bibnamefont {Skillman}},\ }\href {\doibase
  10.1088/1475-7516/2021/03/027} {\bibfield  {journal} {\bibinfo  {journal}
  {JCAP}\ }\textbf {\bibinfo {volume} {03}},\ \bibinfo {pages} {027} (\bibinfo
  {year} {2021})},\ \Eprint {http://arxiv.org/abs/2010.04180} {arXiv:2010.04180
  [astro-ph.CO]} \BibitemShut {NoStop}%
\bibitem [{\citenamefont {Cooke}\ \emph {et~al.}(2016)\citenamefont {Cooke},
  \citenamefont {Pettini}, \citenamefont {Nollett},\ and\ \citenamefont
  {Jorgenson}}]{Cooke:2016rky}%
  \BibitemOpen
  \bibfield  {author} {\bibinfo {author} {\bibfnamefont {R.~J.}\ \bibnamefont
  {Cooke}}, \bibinfo {author} {\bibfnamefont {M.}~\bibnamefont {Pettini}},
  \bibinfo {author} {\bibfnamefont {K.~M.}\ \bibnamefont {Nollett}}, \ and\
  \bibinfo {author} {\bibfnamefont {R.}~\bibnamefont {Jorgenson}},\ }\href
  {\doibase 10.3847/0004-637X/830/2/148} {\bibfield  {journal} {\bibinfo
  {journal} {Astrophys. J.}\ }\textbf {\bibinfo {volume} {830}},\ \bibinfo
  {pages} {148} (\bibinfo {year} {2016})},\ \Eprint
  {http://arxiv.org/abs/1607.03900} {arXiv:1607.03900 [astro-ph.CO]}
  \BibitemShut {NoStop}%
\bibitem [{\citenamefont {Rosenberg}\ \emph {et~al.}(2022)\citenamefont
  {Rosenberg}, \citenamefont {Gratton},\ and\ \citenamefont
  {Efstathiou}}]{Rosenberg:2022sdy}%
  \BibitemOpen
  \bibfield  {author} {\bibinfo {author} {\bibfnamefont {E.}~\bibnamefont
  {Rosenberg}}, \bibinfo {author} {\bibfnamefont {S.}~\bibnamefont {Gratton}},
  \ and\ \bibinfo {author} {\bibfnamefont {G.}~\bibnamefont {Efstathiou}},\
  }\href {\doibase 10.1093/mnras/stac2744} {\bibfield  {journal} {\bibinfo
  {journal} {Mon. Not. Roy. Astron. Soc.}\ }\textbf {\bibinfo {volume} {517}},\
  \bibinfo {pages} {4620} (\bibinfo {year} {2022})},\ \Eprint
  {http://arxiv.org/abs/2205.10869} {arXiv:2205.10869 [astro-ph.CO]}
  \BibitemShut {NoStop}%
\bibitem [{\citenamefont {Tristram}\ \emph {et~al.}(2024)\citenamefont
  {Tristram} \emph {et~al.}}]{Tristram:2023haj}%
  \BibitemOpen
  \bibfield  {author} {\bibinfo {author} {\bibfnamefont {M.}~\bibnamefont
  {Tristram}} \emph {et~al.},\ }\href {\doibase 10.1051/0004-6361/202348015}
  {\bibfield  {journal} {\bibinfo  {journal} {Astron. Astrophys.}\ }\textbf
  {\bibinfo {volume} {682}},\ \bibinfo {pages} {A37} (\bibinfo {year}
  {2024})},\ \Eprint {http://arxiv.org/abs/2309.10034} {arXiv:2309.10034
  [astro-ph.CO]} \BibitemShut {NoStop}%
\end{thebibliography}%

\end{document}